\documentclass{article}
\usepackage{a4wide}
\usepackage{epsfig}

\newcommand{\be}{\begin{equation}}
\newcommand{\ee}{\end{equation}}
\newcommand{\ba}{\begin{eqnarray}}
\newcommand{\ea}{\end{eqnarray}}
\newcommand{\lapprox}{%
\mathrel{%
\setbox0=\hbox{$<$}\raise0.6ex\copy0\kern-\wd0\lower0.65ex\hbox{$\sim$}}}

\newcommand{\gapprox}{%
\mathrel{%
\setbox0=\hbox{$>$}\raise0.6ex\copy0\kern-\wd0\lower0.65ex\hbox{$\sim$}}}

\renewcommand{\theequation}{\arabic{section}.\arabic{equation}}

\begin{document}

\thispagestyle{empty}
\begin{flushright}
LU TP 00-11 \\
hep-ph/0003258\\
March 2000
\end{flushright}
\vspace{2cm}
\begin{center}
\begin{Large}
$K_{\ell 4}$ Form-Factors and $\pi$-$\pi$ Scattering
 \\[1cm]
\end{Large}
G. Amor\'os$^{a,b}$, J. Bijnens$^b$ and P. Talavera$^b$\\[2cm]
${}^a$ Dept. of Physics, Univ. Helsinki, P.O. Box 9, FIN--00014
Helsinki\\
${}^b$ Dept. of Theor. Phys., Univ. Lund, S\"olvegatan 14A, S--22362 Lund
\\[1cm]
{\bf Pacs:} 12.39.Fe, 12.40.Yx, 12.15.Ff, 14.40.-n\\[0.2mm]
\end{center}

\begin{abstract}
\noindent
The $F$ and $G$ form-factors of $K_{\ell4}$ and the
quark condensates are calculated to
${\cal O}(p^6)$ in Chiral Perturbation Theory (CHPT).
Full formulas are presented as much as possible.
A full refit of most of the ${\cal O}(p^4)$ CHPT parameters is done with
a discussion of all inputs and underlying assumptions.
We discuss the consequences for the 
vacuum expectation values, decay constants, pseudoscalar
masses and $\pi$-$\pi$ scattering. 
\end{abstract}
\setcounter{page}{0}

\clearpage
\tableofcontents

\setcounter{equation}{0}
\section{Introduction}
\label{intro}

The low-energy realization of QCD is still disconnected from 
the high-energy structure where perturbative QCD applications have made it
clear beyond doubt that QCD is the theory of the strong interaction.
At low energies it is possible to obtain a family of theories within
a low-energy expansion using only symmetry principles from QCD.
This family of low-energy theories is known as Chiral Perturbation 
Theory (CHPT) \cite{CHPT}.
The perturbation expansion is in terms
of the energy, momenta and meson masses, collectively denoted as $p$.
Since this is a perturbative framework, in order to compare with
experiment it is very important to have calculations performed
to as high an order as possible. CHPT in the three flavour sector
was firmly worked out in \cite{GL1} where all parameters were
determined to ${\cal O}(p^4)$.

The $K_{e 4}$ decay\footnote{The early history of this decay
is reviewed in \cite{Chounet}.}, is one of the several
processes that can be 
calculated in the framework of CHPT. It has been calculated to tree level,
to one-loop \cite{Kl4oneloop}
and to improved one-loop \cite{BCG}. More precise $K_{e4}$ experiments in
progress will go beyond the precision of these calculation.
In addition $K_{e4}$ is the main input in determining three of the CHPT
parameters.
A lot of work has gone into
improving the calculation of $\pi$-$\pi$ scattering in CHPT,
\cite{pipi,pipi2,Knecht}.
The precision of the prediction on the latter process
is now such that these CHPT parameters are the main uncertainty thus
requiring
an extension of the calculation for $K_{e 4}$ as well.

This is the purpose of the present work: one further step in the 
perturbative CHPT series of $K_{e 4}$ or, in other words, to obtain a two-loop 
result. Using previous two-loop results \cite{Amoros}
a complete three-flavour and, as far as possible, a model independent
result is obtained. We discuss 
several model dependent assumptions, mainly standard, to 
increase confidence in the generality of the numerical results.
In addition we also present the calculation of the vacuum expectation values
to two-loops.

There are several papers where a complete two-loop calculation is done. 
Most of them with the two-flavour chiral Lagrangian 
as the effective theory, with the
other light flavour and the excited states contributing only through the 
local constants in the CHPT Lagrangian.
This two-flavour framework seems to be the best one to extract 
the low-energy parameters of 
the pion-pion scattering, with other states starting to modify the behaviour 
at higher energies. Also, it is the two-flavour sector where
 chiral symmetry predictions are expected to work best
because of the clean feature of 
pseudo-Goldstone bosons for the pions, since the relevant
quark mass is so small.
On the other hand, the two-flavour effective Lagrangian 
can be applied in a small region of physical phenomena.
A more interesting Lagrangian 
has to include the third flavour, but in this case 
the convergence is not so obvious and we can ask the question:
is the Kaon mass small enough to allow a perturbative theory? 
To answer this we need a large spectrum of calculations to
two-loop accuracy.
   
At present, besides results to tree and one-loop level, there is
some work to try to estimate the $K_{e4}$ two-loop contribution \cite{BCG} 
and there exist several complete two-loop results \cite{2loop,BCT}
for other processes. 
We presented a short summary of the main results with a slightly
different input in \cite{LEC}. Here we extend the discussion,
present full results and discuss more implications.

In Sect. \ref{kin} we present the kinematics with the description of the 
process and the variables. Sect. \ref{Form-Factors} introduces the standard 
notation for the form-factors followed
by a short overview of CHPT, Sect. \ref{CHPTsection},
and a discussion of the available data in Sect. \ref{data}.
Previous theoretical results are described in Sect. \ref{oldresults}.
Next we  discuss the calculation, method and checks, Sect. \ref{p6}
and in
Sect. \ref{estimates}
some of the ${\cal O}(p^6)$ constants in CHPT
are estimated using resonance saturation. Details about the 
${\cal O}(p^4)$ constants (LEC), fits and errors are in Sect. \ref{Values}.
We also discuss all the inputs and underlying assumptions there
and compare with previous results and a few models, Sect. \ref{models}.
The  resulting theoretical form-factors and partial waves
are shown in Sect. \ref{FFandPWE}. The comparison with a proposed
experimental parametrization is done in Sect. \ref{kl4parametrization}.
Sect. \ref{pipi} updates the $\pi$-$\pi$ scattering threshold parameters 
with our fit for the LEC as well as comparisons
with available low-energy data.
Using the LEC fit, we discuss the behaviour of 
the vacuum expectation values, masses and decay constants
with the ratio of the 
quark masses in Sects. \ref{vev} and \ref{massessect}.
Finally we shortly summarize the results.
The appendices provide explicit formulas
for the form-factors, vacuum expectation values
and the one-loop integrals.

\section{Definitions and the Data}

\subsection{Kinematics}
\label{kin}

\begin{figure}
\begin{center}
\epsfig{file=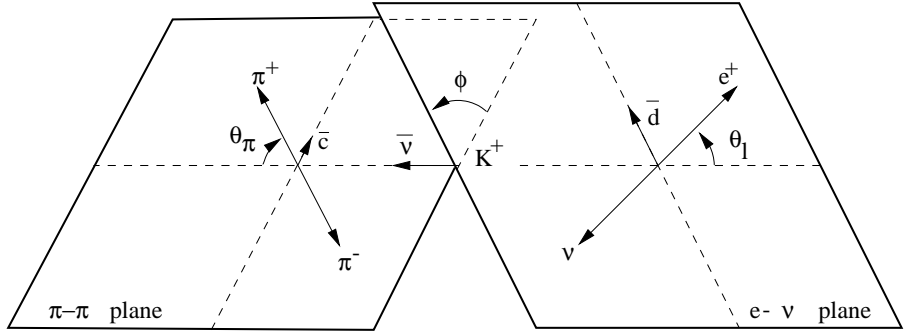,width=12cm}
\end{center}
\caption{\label{figkinematics} Kinematical variables in $K_{\ell 4}$ decays.}
\end{figure}

We shortly review the variables that parametrize the
processes
\ba
\label{processes}
K^+(p)&\to&\pi^+(p_+)\pi^-(p_-)\ell^+(p_\ell)\nu_\ell(p_\nu)\,,
\nonumber\\
K^+(p)&\to&\pi^0(p_+)\pi^0(p_-)\ell^+(p_\ell)\nu_\ell(p_\nu)\,,
\nonumber\\
K^0(p)&\to&\pi^-(p_+)\pi^0(p_-)\ell^+(p_\ell)\nu_\ell(p_\nu)\,.
\ea
In brackets we have labelled the corresponding momentum.
For the last two processes we keep the momentum notation
of the first.
The amplitudes for the three processes we denote
by $T^{+-}$, $T^{00}$ and $T^{-0}$ respectively.
Following
the original work \cite{cabibbo} we consider three reference frames, the 
Kaon rest frame, the dipion center-of-mass system ($\pi-\pi$ plane) and
the dilepton center-of-mass system ($e-\nu$ plane). 
The $K_{\ell 4}$ decays can be parametrized 
in terms of five
kinematical variables
(see Fig. \ref{figkinematics}):\\
i) $s_\pi$, the squared effective mass of the dipion system.\\
ii) $s_{\ell}$, the squared effective mass of the dilepton system.\\
iii) $\theta_\pi$,  the angle between the $\pi^+$ and the
  dipion line of flight with respect to the Kaon rest frame.\\
iv) $\theta_{\ell}$, the angle between the $\ell^+$ 
and the dilepton line of flight with respect to the  Kaon rest frame.\\
v) $\phi$, the angle between the $\pi-\pi$ and $e-\nu$ planes
with respect to the Kaon rest frame.

These variables can vary inside the range
\ba
4 m_\pi^2 \leq s_\pi = (p_++p_-)^2 \leq (m_K-m_\ell )^2\,,
\nonumber\\
m_\ell ^2 \leq s_\ell  = (p_\ell +p_\nu)^2 \leq (m_K-\sqrt{s_\pi})^2\,,
\nonumber\\
0 \leq \theta_\pi, \theta_\ell \leq \pi,\quad 0\leq \phi \leq 2\pi\,.
\ea
Instead of using the previous variables,
and besides $s_\pi$, $\cos\theta_\pi$ and $\phi$,
we also use
\be
t_\pi = (p_+-p)^2 \quad \mbox{and}\quad u_\pi=(p_--p)^2\,,
\ee
related through 
\ba
&&s_\pi+t_\pi+u_\pi = m_K^2+ 2\, m_\pi^2 + s_\ell\,,
\nonumber\\&&
t_\pi-u_\pi=-2\sigma_\pi X \cos\theta_\pi\,,
\ea
with
\ba
X &=& \frac{1}{2} \lambda^{1/2}(m_K^2,s_\pi,s_\ell)\,,
\nonumber\\
\sigma_\pi&=&\sqrt{1-4m_\pi^2/s_\pi}\,,
\nonumber\\
\lambda(m_1,m_2,m_3) &=& m_1^2+m_2^2+m_3^2 - 2\,( m_1 m_2+m_1 m_3+ m_2 m_3)\,.
\ea

\subsection{Form-Factors}
\label{Form-Factors}

With the previous notation the amplitude for the 
decay $K^+ \rightarrow \pi^+ \pi^- \ell^+ \nu_\ell$ is 
\be
\label{matrix}
T^{+-} = \frac{G_F}{\sqrt{2}} V^*_{us} \overline{u}(p_\nu) 
\gamma_\mu (1 - \gamma_5) v (p_\ell) ( V^\mu - A^\mu) \;,
\ee
with
\ba
V_\mu &=& - \frac{H}{m^3_K} \epsilon_{\mu \nu \rho \sigma} 
(p_\ell+p_\nu)^\nu (p_++p_-)^\rho (p_+-p_-)^\sigma \;,\\
A_\mu &=& - \frac{i}{m_K} [ (p_++p_-)_\mu \; F + 
(p_+-p_-)_\mu \; G + (p_\ell+p_\nu)_\mu \; R ]\;.
\ea
$V_{us}$ is the relevant CKM matrix-element. 
The other two amplitudes, $T^{-0}$ and $T^{00}$, are defined similarly.

Here we are interested in the $F$ and $G$ form-factors. 
The $H$ form-factor is already known to ${\cal O}(p^6)$ \cite{Hanomaly} and 
the $R$ form-factor always appears with the $m_\ell^2$ factor and is 
negligible for the electron case. $R$ is known to ${\cal O}(p^4)$ \cite{BCG}.
The form-factors are functions of $s_\pi$, $s_\ell$ and $\cos\theta_\pi$
only, or alternatively of $s_\pi$, $t_\pi$ and $u_\pi$.

The relations between the form-factors and the intensities, easier to 
obtain from the experiment, can be found in \cite{PT} or in 
\cite{BCG}.

The amplitudes for the three processes of Eq. (\ref{processes})
are related using isospin by
\be
T^{+-} = \frac{T^{- 0}}{\sqrt{2}} + T^{0 0}
\ee
with $T^{ij}$ the matrix element defined in Eq. (\ref{matrix}).
$T^{- 0}$ is anti-symmetric under the interchange of the pion momenta while
$T^{0 0}$ is symmetric. This also implies relations between
the form-factors themselves.
Observe the different phase 
convention in the isospin states compared to \cite{BCG} where $M^{0 0}$ 
appears with a minus sign because of the Condon--Shortley phase convention.

\subsection{Chiral Perturbation Theory}
\label{CHPTsection}

CHPT is an expansion in the energy, with the light mesons 
being pseudo-Goldstone 
bosons. 
The main feature is the spontaneously broken chiral symmetry.
The pseudoscalar fields are the lowest mass excitations of
the vacuum expectation values. 
Several realizations are possible, the 
non-linear realization is more useful with the power counting
present explicitly.
Explicit 
chiral symmetry breaking terms are included as perturbations of the symmetry.
For an extensive discussion about CHPT we refer the reader to
\cite{CHPT} and references therein.
In the following we include the basic notation and terms we are working on.

The pseudoscalar fields are included in
\be
U(\phi) = u(\phi)^2 = \exp(i \sqrt{2} \Phi/F_0)\,,
\ee
where
\ba
\Phi (x) \equiv
\frac{ \vec{\lambda}}{ \sqrt 2} \, \vec{\phi}
 = \, \left( \begin{array}{ccc}
\displaystyle\frac{ \pi^0}{ \sqrt 2} \, + \, \frac{ \eta_8}{ \sqrt 6}
 & \pi^+ & K^+ \\
\pi^- &\displaystyle - \frac{\pi^0}{\sqrt 2} \, + \, \frac{ \eta_8}
{\sqrt 6}    & K^0 \\
K^- & \bar K^0 &\displaystyle - \frac{ 2 \, \eta_8}{\sqrt 6}
\end{array}  \right) .
\ea
With the chiral symmetry constraints and the transformation properties under 
this symmetry as well as Lorentz invariance and charge conjugation invariance, 
the Lagrangian starts as
\be
{\cal L}^{\mbox{\small effective}} =  {\cal L}_2 + {\cal L}_4 + {\cal L}_6 + 
\cdots = {\cal L}_2 + \sum_{i=1}^{10} L_i \; O_4^i 
+ \sum_{i=1}^2 H_i \; \tilde{O}_4^i 
+ \sum_{i=1}^{90} C_i \; O_6^i
+ \sum_{i=91}^{94} C_i \;\tilde{O}_6^i + \cdots 
\label{EffLag}
\ee
where the subscripts stand for the power of the low momenta,
energies or masses, collectively denoted as $p$.

The expressions for the first two terms are ($F_0$ is the pion decay constant 
in the chiral limit)
\be
\label{l2}
{\cal L}_2 = \frac{F_0^2}{4} \{\langle D_\mu U^\dagger D^\mu U \rangle 
+\langle \chi^\dagger U+\chi U^\dagger \rangle \}\, ,
\ee
and
\ba
\label{l4}
{\cal L}_4&&\hspace{-0.5cm} = 
L_1 \langle D_\mu U^\dagger D^\mu U \rangle^2
+L_2 \langle D_\mu U^\dagger D_\nu U \rangle 
     \langle D^\mu U^\dagger D^\nu U \rangle \nonumber\\&&\hspace{-0.5cm}
+L_3 \langle D^\mu U^\dagger D_\mu U D^\nu U^\dagger D_\nu U\rangle
+L_4 \langle D^\mu U^\dagger D_\mu U \rangle \langle \chi^\dagger U+\chi U^\dagger \rangle
\nonumber\\&&\hspace{-0.5cm}
+L_5 \langle D^\mu U^\dagger D_\mu U (\chi^\dagger U+U^\dagger \chi ) \rangle
+L_6 \langle \chi^\dagger U+\chi U^\dagger \rangle^2
\nonumber\\&&\hspace{-0.5cm}
+L_7 \langle \chi^\dagger U-\chi U^\dagger \rangle^2
+L_8 \langle \chi^\dagger U \chi^\dagger U + \chi U^\dagger \chi U^\dagger \rangle
\nonumber\\&&\hspace{-0.5cm}
-i L_9 \langle F^R_{\mu\nu} D^\mu U D^\nu U^\dagger +
               F^L_{\mu\nu} D^\mu U^\dagger D^\nu U \rangle
\nonumber\\&&\hspace{-0.5cm}
+L_{10} \langle U^\dagger  F^R_{\mu\nu} U F^{L\mu\nu} \rangle
+H_1 \langle F^R_{\mu\nu} F^{R\mu\nu} + F^L_{\mu\nu} F^{L\mu\nu} \rangle
+H_2 \langle \chi^\dagger \chi \rangle\,.
\ea
The next-to-next-to-leading order ${\cal L}_6$ is a rather
long expression and can be
found in \cite{BCElag}. It contains 90+4 terms compared to
the 10+2 of ${\cal L}_4$.

The notation for the covariant derivative is 
\be
D_\mu U = \partial_\mu U -i r_\mu U + i U l_\mu \,,
\ee
with $l_\mu$ and $r_\mu$ the left and right external currents related with 
the vector and axial-vector external currents
\be
r_\mu = v_\mu + a_\mu, \quad
l_\mu=v_\mu - a_\mu.
\ee
Their field strength tensors are 
\be
\label{fieldstrength}
F_L^{\mu \nu} = \partial^\mu l^\nu -\partial^\nu l^\mu -i [l^\mu,l^\nu], \quad
F_R^{\mu \nu} = \partial^\mu r^\nu -\partial^\nu r^\mu -i [r^\mu,r^\nu]\,.
\ee
Scalar ($s$) and pseudoscalar ($p$) external fields are introduced with 
\be
\chi = 2 B_0 (s+ip)
\ee
where the explicit masses are included through the scalar current $s$
\be 
s= {\cal M} + \cdots \,.
\ee
${\cal M}$ stands for the diagonal quark mass matrix, 
${\cal M} = \mbox{diag}\,(m_u,m_d,m_s)$.
For later use we define the following quantities
\begin{eqnarray}
v^{\mu\nu} &=& \left(F_R^{\mu\nu} + F_L^{\mu\nu}\right)/2\,,
\nonumber\\ 
a^{\mu\nu} &=& \left(F_R^{\mu\nu} - F_L^{\mu\nu}\right)/2\,,
\nonumber\\
u_\mu &=& i\{u^\dagger(\partial_\mu-i r_\mu)u
  -u(\partial_\mu-i l_\mu)u^\dagger\}\,,
\nonumber\\
\Gamma_\mu &=& \frac{1}{2}\{u^\dagger(\partial_\mu-i r_\mu)u
  +u(\partial_\mu-i l_\mu)u^\dagger\}\,,
\nonumber\\
\chi_\pm&=&u^\dagger \chi u^\dagger\pm u \chi^\dagger u\,,
\nonumber\\
\nabla_\mu X &=& \partial_\mu X + \Gamma_\mu X - X \Gamma_\mu\,.
\end{eqnarray}

\subsection{Available $K_{\ell4}$ Experimental Results}
\label{data}

Let us briefly comment on the status of the present data. There are
two experiments \cite{Rosselet,Makoff} on which our analysis is based. 
These two dominate in statistics and precision over previous ones.
Earlier experiments are within errors compatible using isospin relations
with these two and are discussed in \cite{Kl4oneloop}.
In fact our main results rely entirely on \cite{Rosselet}
and \cite{Makoff} is used to 
give an idea on the experimental uncertainties involved. 
In the experiment \cite{Rosselet} the data are analyzed with
the partial wave expansion
\ba
F&=& f_s e^{i\delta_s}+f_p e^{i \delta_p} cos\theta_\pi + 
\mbox{$d$-wave}\,,\nonumber\\
G&=& g e^{i\delta_p}+ \mbox{$d$-wave}\,,
\ea
where $f_s, f_p$ and $g$ are assumed to be real.
Within the  experimental sensitivity no dependence of the form-factors on
$s_\ell$ nor on $d$-waves was observed.
We confirm this using our calculation as shown in
Figs. \ref{figFGsl} and \ref{figFpartial}
in Sect. \ref{FFandPWE}. 
The expected errors
in the planned experiments will
allow to see the $s_\ell$ 
dependence.
The form-factor $f_p$ was found to be compatible with zero
and consequently neglected
when the final value of $g$ was derived.
As we show below, Fig. \ref{figfitparam},
this is a borderline assumption even for the present sensitivity.
Neither was there dependence on $s_\pi$ found for the reduced form-factor
\be
\overline{g} = \frac{g}{f_s}\,,
\ee
thus the latter is assumed to be constant. 

Furthermore $f_s(s_\pi)$ was parametrized as
\be
f_s(s_\pi) = f_s(0) (1+\lambda_f q^2)\,, 
\quad q^2 = (s_\pi-4 m_\pi^2)/4 m_\pi^2\,.
\ee
Notice that $f_s$ is assumed to have only a linear dependence in $s_\pi$.
As shown in Fig.  \ref{figFGsl} future experiments
might detect curvature. The linear dependence can 
\emph{only} be  ensured for $s_\pi < 0.11$~GeV$^2$.
The assumption of constant $\overline{g}$ constrains the 
$g$ form-factor parametrization to be
\be
g(s_\pi) = g(0) (1+\lambda_g q^2)\,,
\ee
with the same slope as the $f_s$ form-factor,
 i.e $\lambda=\lambda_f=\lambda_g$. As is clear,
from  Fig. \ref{figFGsl} the theoretical slopes are different but the effect
still remain within present errors.
Using $V_{us}=0.220$
the values of  \cite{Rosselet} 
yield for the form-factors and slope at threshold
\be
f_s(0)=5.59\pm 0.14,\quad g(0)=4.77\pm 0.27,\quad \lambda=0.08\pm 0.02\,.
\ee
{}From \cite{Makoff} the same assumptions and using isospin give
\be
g(0) = 5.50 \pm 0.50\,,
\ee
thus both experiments are marginally compatible.
Using the PDG averaging methods \cite{PDG} on both results we obtain
\be
g(0)=4.93\pm 0.31\,,
\ee
that we use in fit 9.

\setcounter{equation}{0}
\section{Form-Factors at Next-to-Leading Order }
\label{oldresults}

\begin{figure}
\begin{center}
\epsfig{file=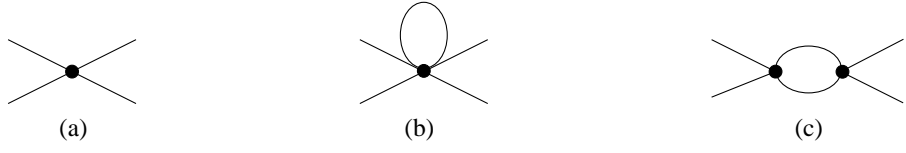,width=12cm}
\end{center}
\caption{\label{fig1loop} 
(a) One-particle irreducible tree level diagram.
(b) One-particle irreducible one-loop diagrams.
Dots refer to strong vertices or
current insertions from ${\cal L}_2$, ${\cal L}_4$ or ${\cal L}_6$.
External legs stand for
pseudoscalar or weak current. Internal lines are pseudoscalars only.}
\end{figure}

For completeness we start with the 
calculation up to one-loop precision. At tree level --see diagram (a)
in Fig. \ref{fig1loop}--
both the $F$ and $G$ form-factors
are equal and given by a single insertion of ${\cal L}_2$ \cite{weinberg}
\be
F=G=\frac{m_K}{\sqrt{2} F_\pi}\,,
\ee
where $m_K$ is the physical Kaon mass. At one-loop \cite{Kl4oneloop}
$F$ and $G$ come through the topologies shown
in Fig. \ref{fig1loop}, containing
two vertices of ${\cal L}_2$ in (b-c) or a single vertex of ${\cal L}_4$
in (a). 
The result can be cast in the general form
\ba
\label{expansion}
F(s_\pi,t_\pi,u_\pi)&=&  \frac{m_K}{\sqrt{2} F_\pi} 
\Big\{1+\frac{F_{1-loop}}{F_\pi^2}
 \Big\}+{\cal O}(p^6)\,,\nonumber\\
G(s_\pi,t_\pi,u_\pi)&=&  \frac{m_K}{\sqrt{2} F_\pi}
 \Big\{1+\frac{G_{1-loop}}{F_\pi^2}
 \Big\}+{\cal O}(p^6)\,.
\ea
$F_{1-loop}$ $(G_{1-loop})$ contain three possible contributions:\\
i) a unitary correction generated by the loop graphs
 --see diagram (c) in Fig. \ref{fig1loop}--
given by the functions $\overline{B}_i$ --see
App. \ref{1loop}-- some of which
develop an imaginary part responsible for
the $I=0$~$s$-wave, $I=1$~{$p$-wave}
$\pi$-$\pi$ phase-shift.\\
ii) A purely logarithmic piece coming from the tad-pole,
(b) diagram in Fig. \ref{fig1loop}.\\
iii) A polynomial part which split in two pieces.
The first one cures the scale dependence
of the one-loop functions, making the full process scale-independent
while the remaining finite 
piece is tuned to match some experimental observables.

A straightforward calculation \cite{Kl4oneloop} leads to
\noindent
\ba
\label{F1loop}
   F_{1-loop}&&\hspace{-0.6cm} =
        4 m^2_\pi ( - 16 L^r_1 + 4L^r_2 - 3
         L^r_3 + 8 L^r_4 + L^r_5 - 
         L^r_9) + 2 m^2_K ( 8L^r_2 + 2L^r_3 - 
         L^r_9)  
\nonumber\\&&\hspace{-0.6cm}
+ 2 s_\pi ( 16 L^r_1 + 4L^r_3 + 
         L^r_9) -  2 t_\pi ( 4L^r_2 + 2L^r_3 - 
         L^r_9 ) + 2u_\pi (- 4L^r_2 + L^r_9 )
\nonumber\\&&\hspace{-0.6cm}
       + 7/24 \Big( \overline{A}(m^2_\pi) 
       + 6 \overline{A}(m^2_K)  
       + 3 \overline{A}(m^2_\eta) \Big)
       + (  - 3/2m^2_\pi + s_\pi )  \, \overline{B}(m^2_\pi,m^2_\pi,s_\pi) 
\nonumber\\&&\hspace{-0.6cm}
       + ( m^2_\pi + m^2_K - 
         t_\pi )/4 \, \overline{B}(m^2_\pi,m^2_K,t_\pi)  
       + ( 2m^2_\pi + m^2_K - 
         u_\pi )/3  \, \overline{B}(m^2_\pi,m^2_K,u_\pi) 
\nonumber\\&&\hspace{-0.6cm}
       + s_\pi/2 \Big( \overline{B}(m^2_K,m^2_K,s_\pi)    
       + \overline{B}(m^2_\eta,m^2_\eta,s_\pi)  \Big)
       +  ( 2m^2_\pi + s_\pi/3 ) \, \overline{B}_1(m^2_\pi,m^2_\pi,s_\pi) 
\nonumber\\&&\hspace{-0.6cm}
       + ( m^2_\pi + m^2_K/2 - 2
         t_\pi )/3 \, \overline{B}_1(m^2_\pi,m^2_K,t_\pi)  
       - u_\pi/3 \, \overline{B}_1(m^2_\pi,m^2_K,u_\pi)  
       + s_\pi/2 \, \overline{B}_1(m^2_K,m^2_K,s_\pi) 
\nonumber\\&&\hspace{-0.6cm}
       +  (  - m^2_\pi/3 + 7m^2_K/3 - 
         t_\pi )/4 \overline{B_1}(m^2_K,m^2_\eta,t_\pi) 
       -2/3 s_\pi  \, \overline{B}_{21}(m^2_\pi,m^2_\pi,s_\pi)
\nonumber\\&&\hspace{-0.6cm}
       + ( 3m^2_\pi - 3m^2_K - 5/3
         t_\pi )/4 \, \overline{B}_{21}(m^2_\pi,m^2_K,t_\pi)  
       +  ( 3m^2_\pi - 3m^2_K - 
         t_\pi )/4 \, \overline{B}_{21}(m^2_K,m^2_\eta,t_\pi) 
\nonumber\\&&\hspace{-0.6cm}
       - 2/3 \, \overline{B}_{22}(m^2_\pi,m^2_\pi,s_\pi) 
       - 8/3 \, \overline{B}_{22}(m^2_\pi,m^2_K,t_\pi)  
       - 5/2 \, \overline{B}_{22}(m^2_K,m^2_\eta,t_\pi)  \,,
\ea
and for the $G$ form-factor to
\ba
\label{G1loop}
   G_{1-loop}&& \hspace{-0.5cm}=
       m^2_\pi 4 (  -  L^r_3 + L^r_5 -   
         L^r_9 ) - 2 m^2_K ( 2 L^r_3 + L^r_9 ) + 2   s_\pi L^r_9 + 2 t_\pi(
         4 L^r_2 + 2  L^r_3 +  L^r_9 ) 
\nonumber\\&&\hspace{-0.5cm}
+ 2 u_\pi (- 4 
           L^r_2 +    L^r_9 )
       + 1/24 \Big( 35\overline{A}(m^2_\pi) 
       +  14 \overline{A}(m^2_K) 
       - 3 \overline{A}(m^2_\eta) \Big)
\nonumber\\&&\hspace{-0.5cm}
       +   (  - m^2_\pi   - m^2_K   + 
            t_\pi )/4\, \overline{B}(m^2_\pi,m^2_K,t_\pi)   
       +   ( 2 m^2_\pi   + m^2_K   -  
           u_\pi )/3\, \overline{B}(m^2_\pi,m^2_K,u_\pi)   
\nonumber\\&&\hspace{-0.5cm}
       +   (  - m^2_\pi   - m^2_K/2   + 2
          t_\pi )/3
 \, \overline{B}_{1}(m^2_\pi,m^2_K,t_\pi)   
       -  u_\pi/3  \, \overline{B}_{1}(m^2_\pi,m^2_K,u_\pi)    
\nonumber\\&&\hspace{-0.5cm}
       +   ( m^2_\pi/3   - 7/3 m^2_K   + 
            t_\pi )/4
 \, \overline{B}_{1}(m^2_K,m^2_\eta,t_\pi)  
       +    3/4 (  - m^2_\pi   + m^2_K   + 5/9 t_\pi )
 \, \overline{B}_{21}(m^2_\pi,m^2_K,t_\pi)  
\nonumber\\&&\hspace{-0.5cm}
       +    (  - 3 m^2_\pi   + 3 m^2_K   + 
         t_\pi )/4
\, \overline{B}_{21}(m^2_K,m^2_\eta,t_\pi)  
       -2  \, \overline{B}_{22}(m^2_\pi,m^2_\pi,s_\pi)  
       - \overline{B}_{22}(m^2_\pi,m^2_K,t_\pi)/3
\nonumber\\&&\hspace{-0.5cm}
       - \overline{B}_{22}(m^2_K,m^2_K,s_\pi)   
       - \overline{B}_{22}(m^2_K,m^2_\eta,t_\pi)/2\,.
\ea

It is worthwhile to point out at this level some relevant features
in Eqs. (\ref{F1loop}), (\ref{G1loop}).
For instance 
they depend on the ${\cal L}_4$ constants $L_1^r, L_2^r,L_3^r,L_4^r,L_5^r$
and $L_9^r$. 
Provided we make some assumptions (large $N_c$ and neglecting $d$-waves)
we can use the low-energy data together
with Eqs. (\ref{F1loop}), (\ref{G1loop}) to obtain values for
$L_1^r, L_2^r$ and $L_3^r$.
Eqs. (\ref{F1loop}), (\ref{G1loop})
 have also acquired a different
energy dependence that brings the one-loop result nearer to the experimental
values \cite{Rosselet}.
In fact these one-loop corrections are rather large making 
the low-energy constants evaluation sensitive to variations of
the treatment of
$F_\pi$ and pseudoscalar masses.

A possible improvement in the $L_1^r, L_2^r$ and $L_3^r$ evaluation 
is to consider an Omn\`es-representation for both form-factors.
In other words, to calculate
the dispersive contribution from the next chiral order considering that no 
intermediate new channels are opened \cite{BCG}.
Even more, renormalization group analysis
allows to fully calculate the double logarithm contribution \cite{BCEdble}.
Both showed that in this
particular case the next order is expected to be quite sizeable.

This together with the discrepancy between the $SU(2)$ CHPT constants
$\bar l_1$ and $\bar l_2$
obtained from the $K_{\ell4}$ fit \cite{BCG}
and Roy equation analysis at ${\cal O}(p^4)$ \cite{Buttiker} and
${\cal O}(p^6)$ \cite{Girlanda:1997ed} motivated us to
calculate the two-loop contribution to $K_{\ell4}$.

\setcounter{equation}{0}
\section{Form-Factors at Next-to-Next-to-Leading Order }
\label{p6}

\subsection{General Technique}
\label{technique}

The calculation is done with dimensional regularization (DR) with 
the extended dimension $d = 4 - 2 \epsilon$. In this scheme the chiral 
symmetry is preserved and because of the mass independent regularization 
the divergences do not mix different orders in the chiral expansion.

Once we have all the mathematical tools to calculate all the integrals, 
the process to obtain the form-factors is quite general. We summarize here 
the main steps and checks.

\begin{description}
\item[One-loop:] We reproduce the one-loop result obtained some 
time ago \cite{Kl4oneloop}. For reasons explained later, in the expansion 
with the DR parameter $\epsilon$ we keep the ${\cal O}(\epsilon^1)$ term.

\item[Masses and decay constants:] A two-loop calculation is a step beyond and
with some differences with the one-loop calculation. One has to be much more
careful in precisely defining quantities.
E.g. in a one-loop calculation we can safely write all the masses as physical 
because the difference is in the next order. This is the reason why we have 
to keep in mind which are the masses to consider in the two-loop result. 
The masses can be written as:
\be
m^2 = m_0^2 + m^{2 (4)} + m^{2 (6)} + ...
\label{masses}
\ee
with $m_0$ the lowest-order mass
and where the superscripts (4) and (6) stand for ${\cal O}(p^4)$ and 
${\cal O}(p^6)$. For any ${\cal O}(p^6)$ contribution the masses $m_0^2$ 
can be used since the difference is ${\cal O}(p^8)$.
However for ${\cal O}(p^4)$
contribution the masses with the first two terms of Eq.~(\ref{masses}) should 
be taken. For the tree level the three terms are needed. In CHPT lowest-order
and 
renormalized masses are finite and then the result can be in principle be 
obtained with lowest-order masses appearing inside the loops and in
the chiral symmetry 
breaking terms, but we have to use the full Eq.~(\ref{masses}) for the squared
momenta. We choose to write all the masses in terms of the physical ones.
The same discussion can be done for the decay constants; we choose also to 
write the physical decay constant $F_\pi$ instead of the lowest-order
one $F_0$. 

\item[Wave-function-renormalization (WFR):] We have to include the WFR 
for each external leg. That means the 
matrix element is multiplied
by $Z^{1/2}_K Z^{1/2}_{\pi_1} Z^{1/2}_{\pi_2}$,
with $Z_{K,\pi}$ the renormalization constant for the fields
including ${\cal O}(p^6)$. This is equivalent to include the 
one-particle-reducible diagrams.

\item[Cancellation of divergences:] There are two kind of divergences to
cancel. 
The first one is pure polynomial; they cancel with the renormalization of the 
$C_i$ constants (see Eq.~(\ref{EffLag})). These divergences 
were already found in \cite{BCEinf}. 

Another kind of divergences are non-local, i.e., can not be expressed as 
contributions from terms in the effective Lagrangian. They appear in
the separate contributions and cancel in any quantum field theory.
But this is a global cancellation, so it taking place, as is the case
for our calculation,  
is a strong check of the result. 

A new feature of the two-loop calculations is that the expansion in the 
DR parameter 
until ${\cal O}(\epsilon^1)$ needs to be kept.
The reason for that is the existence of 
terms as $L_i A(m^2)$ (or as $L_i L_j$, $L_i B_j$, $\cdots$; 
in general products 
of one-loop contributions: one-loop$ \times$ one-loop)
where $A(m^2)$ is a one-loop integral. Once the 
$L_i$ constant is renormalized
\be
L_i = L_i^r + \Gamma_i \lambda
\label{L_shift}
\ee
where $\lambda = \frac{1}{2 \epsilon} + C$ and with $\Gamma$ and $C$ real and 
finite numbers, there appears a new finite term. 

\item[Double logarithms:] Because the nonlocal divergences have to cancel,
the double poles in $\epsilon$ can be obtained from a one-loop calculation.
These can be calculated in general \cite{BCEdble} or from
the $L_i\times$one-loop graphs for a particular process.
We 
obtain the same result as \cite{BCEdble} for the double logarithms and 
the products $L_i \times L_j$ and $L_i \times \log$.

\item[Three-point one-loop integrals:]
The contribution from the first four 
diagrams in the Fig.~\ref{fig2loop1}, (d--g),
could be obtained through the renormalization 
of the meson line in diagrams (b) and (c) of Fig. \ref{fig1loop}.
Therefore the result should be a product of at most two-point 
one-loop integrals and the naive three-point one-loop integrals from
diagrams (e) and (g) must cancel when (c) is rewritten
in terms of physical masses.

\begin{figure}
\begin{center}
\epsfig{file=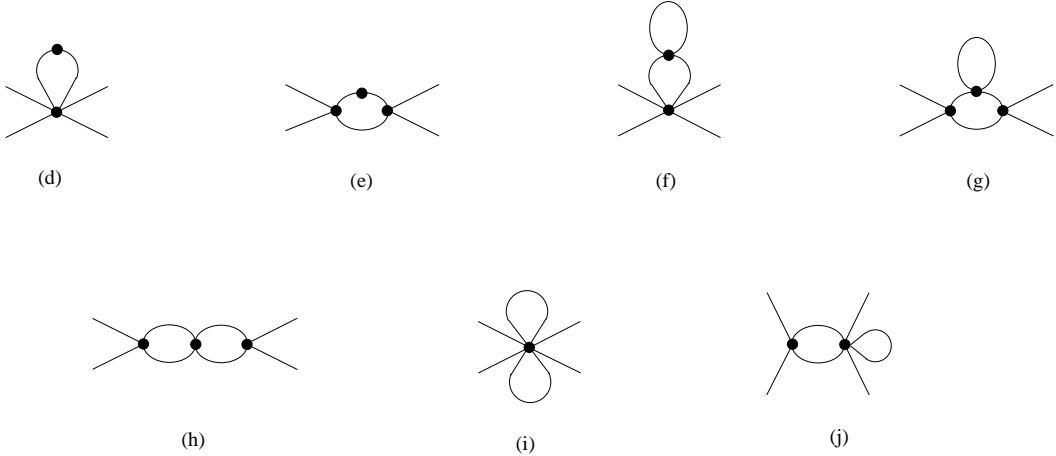,width=14cm}
\end{center}
\caption{\label{fig2loop1} 
One-particle irreducible ${\cal O}(p^6)$ diagrams with only one-loop
integrals. The dots are ${\cal O}(p^2)$ vertices except
the top dot in (d) and (e) which is an ${\cal O}(p^4)$ vertex.
In addition there are the diagrams
(b) and (c) of Fig. \ref{fig1loop} with one
of the dots replaced by a ${\cal O}(p^4)$ vertex.}
\end{figure}

\begin{figure}
\begin{center}
\epsfig{file=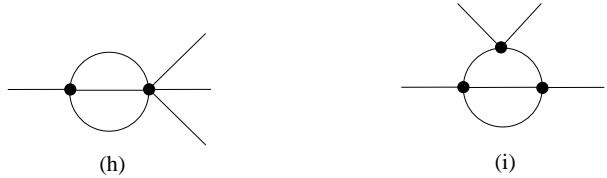,width=8cm}
\end{center}
\caption{\label{fig2loop2} 
One-particle irreducible ${\cal O}(p^6)$ diagrams with irreducible two-loop
integrals. The dots are ${\cal O}(p^2)$ vertices }
\end{figure}

\item[Irreducible two-loop integrals:]
The diagrams in Fig.~\ref{fig2loop2} generate sunset-integrals and
vertex-integrals already known in the literature.
In a previous work \cite{Amoros} 
on the masses and decay constants of the pseudoscalar mesons 
to two-loops, the sunset-integrals are calculated using the results obtained 
in \cite{DT,GS}. For the vertex-integrals we use another technique 
explained in \cite{Ghinculov} and also useful to 
recalculate the sunset-integrals. This provided
another check on our integral calculations.
Additionally, in the vertex-integrals we obtain the same imaginary part 
from the two-particle discontinuities as using the Cutkosky rules. This 
check is done below the three-particle thresholds. We refer
to \cite{Ghinculov} for a longer discussion.

\item[Isospin relations:]
Isospin symmetry relates the decays as discussed in Sect. \ref{Form-Factors}.
Satisfying these relations is another check of our result. 

\item[Gell-Mann-Okubo (GMO) relation:]
The last comment in this subsection is related to effects from ${\cal O}(p^8)$.
As explained in the previous paragraph about masses, the correction to the 
${\cal O}(p^6)$ using the physical masses instead of the bare 
masses is ${\cal O}(p^8)$. Then, to two-loop order it is safe to use 
the GMO relation $3 m^2_\eta = 4 m^2_K - m^2_\pi$.
We can calculate with and without this relation. The result is 
a variation roughly of $10\%$ of the tail of the ${\cal O}(p^6)$ part
of the form-factors 
in Fig.~\ref{figVVfitparam}. The plot in the kinematic regime accessible 
for the experiment ($s_\pi < 0.16 \, \, \mbox{GeV}^2 $) is not modified.
This particular effect is mainly due to terms containing
$(3 m^2_\eta - 4 m^2_K + m^2_\pi)/(u_\pi,t_\pi,s_\pi)$
which appear when reducing
the integrals to a more minimal basis.
Another example of this type of effects is discussed in
Sects. \ref{vev} and \ref{massessect}.
\end{description}

\subsection{Results}
\label{tworesults}

Up to ${\cal O}(p^6)$ we split the contributions as follows
\ba
\label{expansion2}
F(s_\pi,t_\pi,u_\pi)&=& \frac{m_K}{\sqrt{2} F_\pi}
 \Big\{1+\frac{F_{1-loop}}{F_\pi^2}
+\frac{1}{F_\pi^4} ( F_{LL}+F_{V}+F_{ct}) \Big\}+  {\cal O}(p^8)\,,\nonumber\\
G(s_\pi,t_\pi,u_\pi)&=& \frac{m_K}{\sqrt{2} F_\pi} 
\Big\{1+\frac{G_{1-loop}}{F_\pi^2}
+ \frac{1}{F_\pi^4} ( G_{LL}+G_{V}+G_{ct}) \Big\}+{\cal O}(p^8)\,.
\ea
The two first terms have been explained in Sect. \ref{oldresults}.
The ${\cal O}(p^6)$ part is split as follows:\\
$\mathbf{F(G)_V}$: 
the part from diagrams with vertices from only the lowest-order Lagrangian, it 
contains the pure two-loop unitarity contribution.
$F_V$ depends only on the physical masses
of the pseudoscalars.\\
$\mathbf{F(G)_{ct}}$: is the part of the
polynomial contribution depending on the couplings of ${\cal L}_6$.\\
$\mathbf{F(G)_{LL}}$: is 
the remaining part, in particular all dependence at ${\cal O}(p^6)$
on the $L_i^r$ is collected here.

The diagrams in the Fig.~\ref{fig2loop1} can be 
written as product of one-loop integrals and evaluated 
with the standard methods. For the diagrams in the Fig.~\ref{fig2loop2}
we profit 
from previous work with the sunset \cite{Amoros} and 
the vertex-integrals \cite{Ghinculov}. The method to manage 
the sunset-integrals is extensively explained in 
\cite{Amoros} and we refer to this work. For the vertex diagram we refer
to \cite{Ghinculov} since we do not explicitly present any formulas
containing them.
We observe a fairly strong numerical cancellation between the contributions
from irreducible vertex-integrals and the rest of the terms in $F_V$ and $G_V$.

We present the results for  $F(G)_{LL}$ and $F(G)_{ct}$ 
in App. \ref{FGLL} and App. \ref{counter} respectively. For $F(G)_V$ 
a significantly longer expression is found. Since it only depends
on the physical masses, which are not likely to change
significantly, a simple parametrization is sufficient.
We present it for three sets of masses for the three
possible decays. Using $m_{K^+}$ and $m_{\pi^+}$ for
$K^+\to\pi^+\pi^-\ell^+\nu_\ell$;
 $m_{K^+}$ and $m_{\pi^0}$ for
$K^+\to\pi^0\pi^0\ell^+\nu_\ell$ and
 $m_{K^0}$ and $m_{\pi\mbox{\small av}}=(m_{\pi^+}+m_{\pi^0})/2$ for
$K^0\to\pi^-\pi^0\ell^+\nu_\ell$.

Notice that in all cases we quote $F$ and $G$ as for
$K^+\to\pi^+\pi^-\ell^+\nu_\ell$. Taking the even or odd part in
$\cos\theta_\pi$ then gives the form-factors for the other
decays.

\begin{figure}
\begin{center}
\epsfig{file=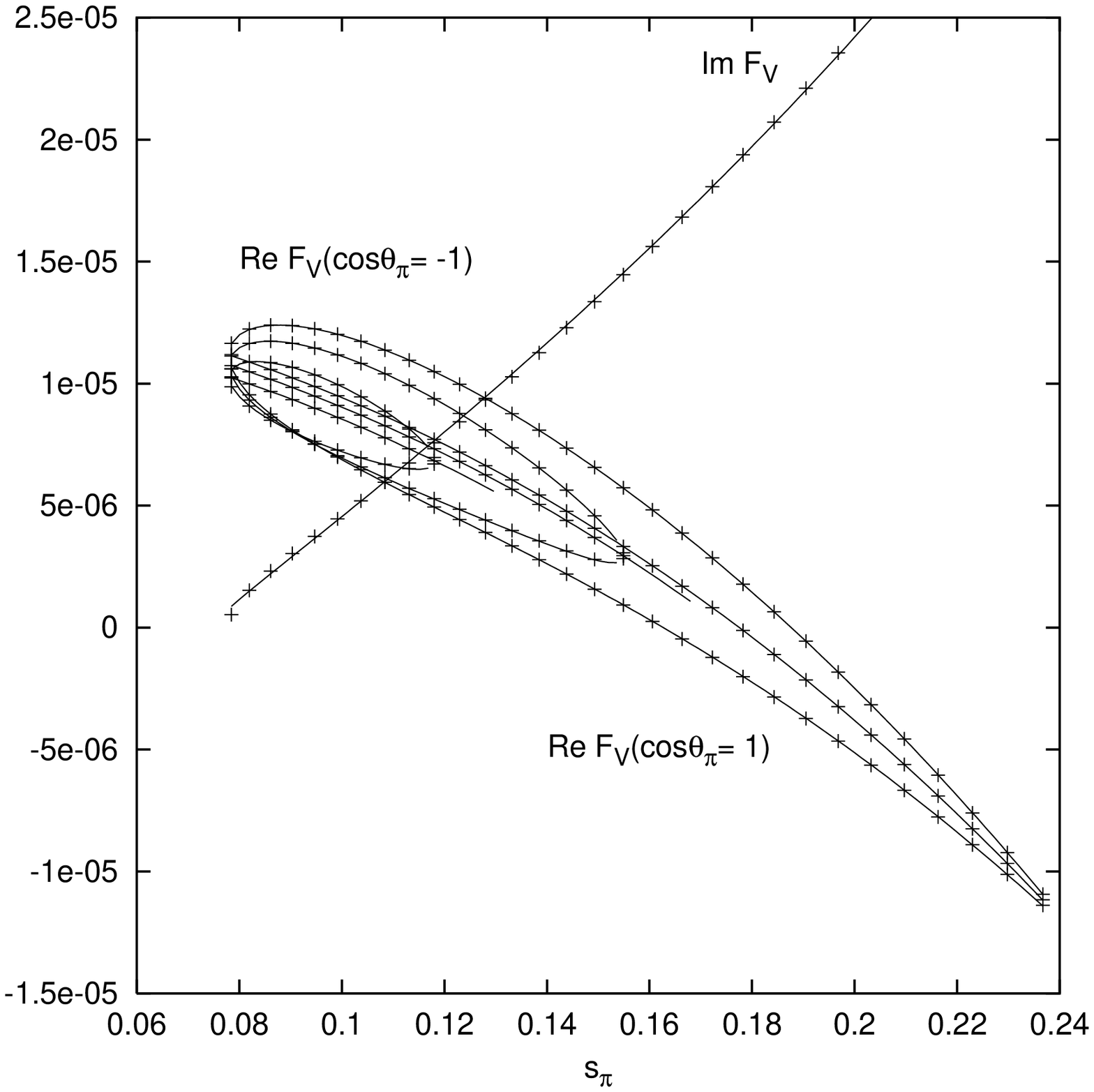,width=0.48\textwidth, angle=-90}
\epsfig{file=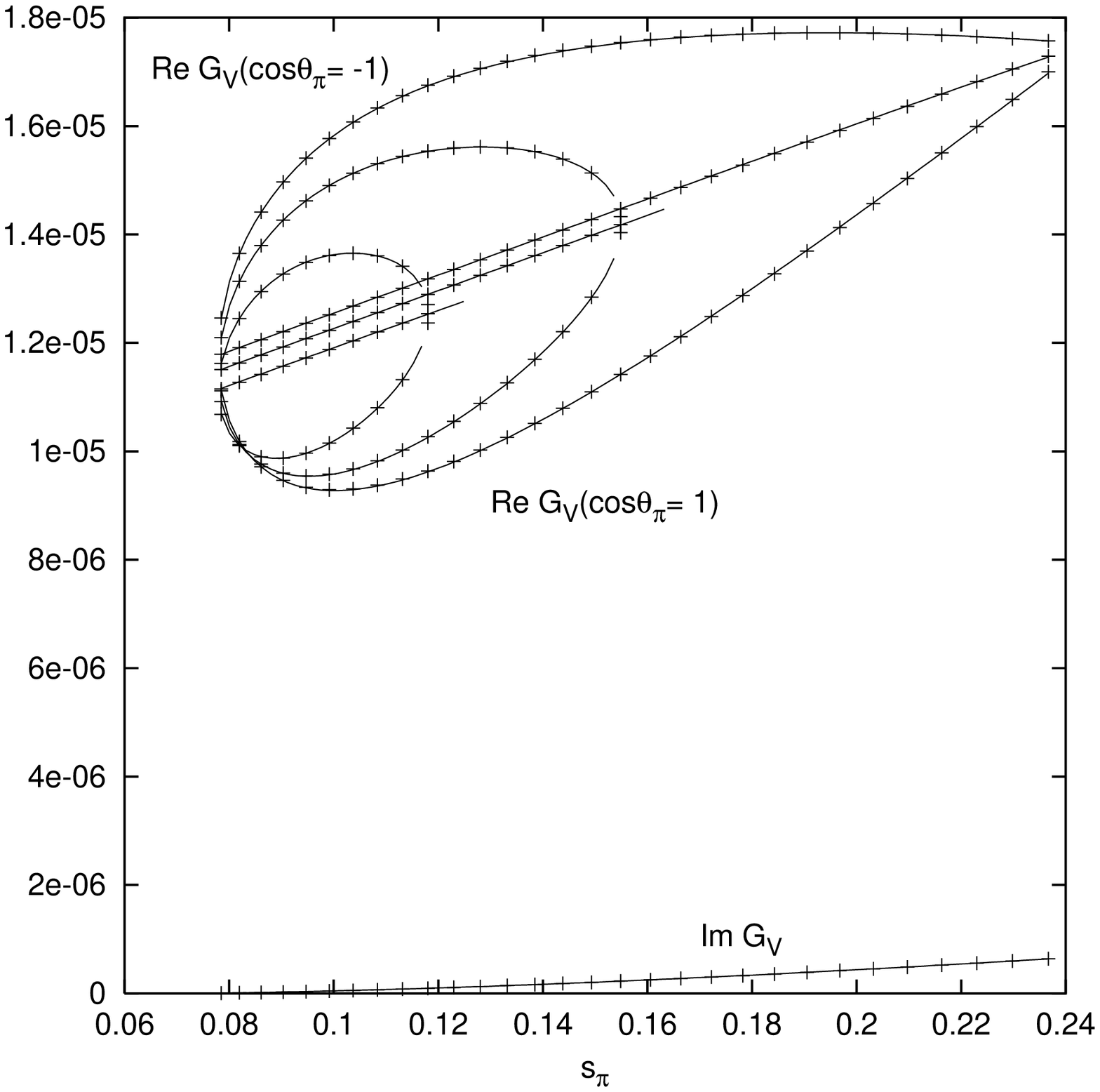,width=0.48\textwidth, angle=-90}
\end{center}
\caption{\label{figVVfitparam} 
 Fit to the $F_V$ and $G_V$ contribution. Plotted are the fit with
lines and the full calculation with {\bf +}. In all cases
we plotted $\cos\theta_\pi=0,\pm1$ and $s_\ell=0,0.01,0.0225$~GeV$^2$.}
\end{figure}

The parametrization is given by
\ba
\label{para}
F(G)&=&
(a_1 +\cos\theta_\pi a_2)(1+s_\ell a_{11})
+s_\pi (a_3 +\cos\theta_\pi a_4)
+s_\pi^2 (a_5 +\cos\theta_\pi a_6)
\nonumber\\&&
+s_\pi^3 (a_7 +\cos\theta_\pi a_8)
+\sigma_\pi X (a_9 +\cos\theta_\pi a_{10})\,.
\ea
where $X=1/2\,\sqrt{(m_K^2-s_\pi-s_\ell)^2-4 s_\pi s_\ell}$
and $\sigma_\pi=\sqrt{1-4 m_\pi^2/s_\pi}$ depend on $s_\pi$ and $s_\ell$.
The  values of the parameters are given in Table \ref{tabfitparams}. 
In Fig. \ref{figVVfitparam} we show how well this parametrization
fits the calculated expressions.

\begin{table}
\caption{\label{tabfitparams}The parameters in the fit to $F_V$ and $G_V$
for the three mass cases.}
\begin{center}
\begin{tabular}{c||c|c|c|c|c|c}

\hline
       &   $F_V$  &  $G_V$ &   $F_V$ &  $G_V$ &   $F_V$ &  $G_V$ \\
\hline
masses & \multicolumn{2}{c|}{ $m_{K^+},m_{\pi^+}$} 
& \multicolumn{2}{c|}{$m_{K^+},m_{\pi^0}$}
& \multicolumn{2}{c}{$m_{K^+},m_{\pi\mbox{\small av}}$}\\
\hline
$10^5 a_{1}$ &$1.14408$ &$0.92331$&
              $1.4238 $ &$0.90367$&
              $1.456$   &$0.91997$\\
             &$-i1.0993$ &$+i0.00198$&
              $-i0.88056$ &$+i0.000468$&
              $-i0.9431$   &$-i0.000496$\\
$10^6 a_{2}$ &$0.1061$          &$0.060766$&
              $0.09106$         &$-0.06771$&
              $0.09514$         &$-0.068565$\\
$10^5 a_{3}$ &$-1.5445$ &$3.0270$&
              $-1.7479$ &$3.5074$&
              $-1.6368$ &$3.5262$\\
             &$+i14.659$ &$-i0.21235$&
              $+i11.457$ &$-i0.17035$&
              $+i12.041$ &$-i0.15502$\\
$10^7 a_{4}$ &$-9.3974$         &$-5.5732$&
              $-7.8875$         &$-23.294$&
              $-8.5278$         &$22.837$\\
$10^5 a_{5}$ &$-30.562$ &$3.4592$&
              $-31.254$ &$0.037925$&
              $-30.653$ &$0.035941$\\
             &$-i0.7821$ &$+i2.6824$&
              $+i17.946$ &$+i2.5087$&
              $+i14.810$ &$+i2.4139$\\
$10^7 a_{6}$ &$-0.59204$        &$2.1912$&
              $-7.1682$         &$-215.51$&
              $-5.5226$         &$-206.17$\\
$10^4 a_{7}$ &$-3.6231$ &$-0.803725$&
              $-3.5113$ &$-0.078793$&
              $-3.6377$ &$-0.0812$\\
             &$+i7.6144$ &$-i0.28955$&
              $+i4.1122$ &$-i0.26763$&
              $+i4.4861$ &$-i0.24452$\\
$10^6 a_{8}$ &$7.519$           &$3.1053$&
              $8.3521$          &$54.23$&
              $8.4695$          &$50.816$\\
$10^6 a_{9}$ &$-0.54512$&$0.69997$&
              $-0.78072$&$0.29854$&
              $-0.46175$&$0.27801$\\
             &$+i5.7286$&$-i0.02308$&
              $+i6.901$ &$-i0.29305$&
              $+i6.9589$&$-i0.17864$\\
$10^5 a_{10}$&$-7.6007$         &$-9.7289$&
              $-7.5832$         &$-9.7041$&
              $-7.6143$         &$-9.7514$\\
$a_{11}$ &$-2.7610$             &$-3.0277$&
          $-2.8042$             &$-3.1412$&
          $-2.7268$             &$-3.0970$\\
\hline
\end{tabular}
\end{center}
\end{table}

\setcounter{equation}{0}
\section{Estimates of Some ${\cal O}(p^6)$ Constants}
\label{estimates}

In this section we estimate some of the ${\cal O}(p^6)$ 
constants that appear in the results. We assume saturation 
by the lightest vector, axial-vector and scalar mesons, extending the formalism
used in \cite{resonance} to the present case.

With higher orders in the chiral Lagrangian 
the number of constants to estimate increases.
This is due to the fact that 
the chiral Lagrangian is constructed using only
some  general conditions and can describe 
all the possible theories compatible with these.
For any specific theory, QCD in our case,
there is a large
correlation expected  
between
these effective constants.
Of course in principle  they are calculable from the underlying theory
but this is not possible at present. The data are also not sufficient to
determine all of them so we need to estimate using some other method. 
Their main contribution is expected to come from the exchange of higher
mass resonances. For these we use
a Lagrangian with chiral symmetry incorporating extra, higher mass, fields.
The precise input we use for this is described in this section.

For the spin-1 
mesons we use the realization where the vector contribution to the chiral 
Lagrangian starts at ${\cal O}(p^6)$. We will only discuss terms relevant
for $K_{\ell4}$, masses and decay constants.
Specifically we use for 
the vector matrix $V_\mu$
\ba
{\cal L}_V &=& -\frac{1}{4}\langle V_{\mu\nu}V^{\mu\nu}\rangle
+\frac{1}{2}m_V^2\langle V_\mu V^\mu\rangle
-\frac{f_V}{2\sqrt{2}}\langle V_{\mu\nu}f_+^{\mu\nu}\rangle
\nonumber\\
&&-\frac{ig_V}{2\sqrt{2}}\langle V_{\mu\nu}[u^\mu,u^\nu]\rangle
+f_\chi\langle V_\mu[u^\mu,\chi_-]\rangle
+ i \alpha_V \langle V_\mu[u_\nu,f_-^{\mu\nu}]\rangle
\label{vector}
\ea
and for the axial-vector $A_\mu$
\ba
{\cal L}_A &=& -\frac{1}{4}\langle A_{\mu\nu}A^{\mu\nu}\rangle
+\frac{1}{2}m_A^2\langle A_\mu A^\mu\rangle
-\frac{f_A}{2\sqrt{2}}\langle A_{\mu\nu}f_-^{\mu\nu}\rangle
+\gamma_A^{(1)} \langle A_\mu u_\nu u^\mu u^\nu \rangle
+\gamma_A^{(2)} \langle A_\mu \{u^\mu, u_\nu u^\nu\}\rangle\,.\nonumber\\
\label{axial}
\ea
with 
\ba
V_{\mu \nu}  =  \nabla_\mu V_\nu - \nabla_\nu V_\mu \,,& &
f^{\mu \nu}_{\pm}  =  
u (v^{\mu \nu}-a^{\mu \nu})  u^\dag 
\pm u^\dag (v^{\mu \nu}+a^{\mu \nu}) u \nonumber \,,
\ea
\noindent 
$V_\mu$ and $A_\mu$ are three-by-three 
matrices describing the full vector and axial-vector nonets.
The rest of the notation is as in Sect. \ref{intro}. 

There are other terms with the same order in momenta as in the previous 
Lagrangian, however they are related to the anomalous sector, which we do
not consider here.

The scalar mesons are considered through 
\ba
{\cal L}_S &=& \frac{1}{2} \langle \nabla^\mu S \nabla_\mu S 
 - M^2_S S^2 \rangle  
 + c_d \langle Su^\mu u_\mu \rangle + c_m \langle S \chi_+ \rangle 
 + \frac{d_m}{2} \langle S^2 \chi_+ \rangle 
\nonumber\\&&
+ c_\gamma \langle S f_{+\mu \nu} f_+^{\mu \nu} \rangle
+ c^\prime_\gamma \langle S f_{-\mu \nu} f_-^{\mu \nu} \rangle\, . 
\label{scalar}
\ea
The last two terms are needed for the calculation
of the vector and axial-vector two-point functions and 
not interesting for the present discussion.
We remark that these 
are not all the allowed contributions from the scalar mesons. We can 
construct other terms similar to the terms in the ${\cal O}(p^4)$ Lagrangian. 
However that will increases the number of unknown constants to determine 
without improvement for the estimates. It is however one 
source of error to keep in mind.

There are two ways to calculate the contribution from these Lagrangians: 
using the equations of motion to integrate out directly the heavy fields; or 
diagrammatically, considering all the diagrams with explicit resonance 
fields and expanding the result to the ${\cal O}(p^6)$. In this case it seems 
faster to use the equations of motion. For the spin-1 mesons it is immediate 
to derive the Lagrangian; for the scalars also shifts in
their vacuum expectation values have to be considered. 

 The result after the integration of resonances is for the vector fields
\begin{eqnarray}
{\cal L}_V & = & - \frac{i f_\chi g_V}{\sqrt{2} M^2_V} 
\langle \nabla_\lambda ([ u^\lambda, u^\nu ] ) [ u^\nu, \chi_- ] \rangle 
+ \frac{g_V \alpha_V}{\sqrt{2} M^2_V} 
\langle [ u_\lambda, f_-^{\nu \lambda} ] 
( \nabla^\mu [ u_\mu,u_\nu ] )\rangle \nonumber \\ 
& & - \frac{i g_V f_V}{2 M^2_V} \langle ( \nabla_\lambda f_+^{\lambda \nu} )
(\nabla^\mu [ u_\mu, u_\nu ] ) \rangle 
- \frac{i \alpha_V f_\chi}{M^2_V} \langle [ u_\nu, \chi_- ] 
[ u_\lambda,f_-^{\nu \lambda} ] \rangle \nonumber \\
& & - \frac{f_\chi f_V}{\sqrt{2} M^2_V} 
\langle ( \nabla_\lambda f_+^{\lambda \mu} ) 
[ u_\mu, \chi_- ] \rangle \; ,
\label{LagInt}
\end{eqnarray}
for the axial-vector fields 
\begin{eqnarray}
{\cal L}_A & = & - \frac{f_A \gamma^{(1)}_A}{\sqrt{2} M^2_A}
\langle ( \nabla_\mu f_-^{\mu \nu} ) u^\lambda u_\nu u_\lambda \rangle 
- \frac{f_A \gamma^{(1)}_A}{\sqrt{2} M^2_A}
\langle ( \nabla_\mu f_-^{\mu \nu} ) \{ u_\nu,u_\lambda u^\lambda \} 
\rangle \; ,
\end{eqnarray}
and for the scalar mesons
\begin{eqnarray}
\label{LagInt2}
{\cal L}_S & = & \frac{c_d^2}{2 M^4_S} \langle 
\nabla_\nu (u_\mu u^\mu) \nabla^\nu (u_\lambda u^\lambda) \rangle 
+ \frac{c^2_m}{2 M^4_S} \langle ( \nabla_\nu \chi_+) 
( \nabla^\nu \chi_+ ) \rangle 
+ \frac{c_d c_m}{ M^4_S} \langle \nabla_\nu (u_\mu u^\mu) 
( \nabla^\nu \chi_+ ) \rangle \, . \nonumber\\
\end{eqnarray}
The three Lagrangians are ${\cal O}(p^6)$. However we remark that the 
spin-1 Lagrangians are directly of this order after the integration. For 
the scalar case the expansion of the scalar propagator produces 
${\cal O}(p^4)$ as leading contribution but that part is already included
via the values of the $L_i^r$ constants.
For the work here we take the next term 
in this expansion to estimate the unknown ${\cal O}(p^6)$ constants.

Besides these ways to estimate the constants it is 
in some cases possible to obtain the ${\cal O}(p^6)$ couplings
from experiment.
This has been done in previous calculations  
\cite{kamborsumrules}, where 
the use of finite sum rules could provide the values for the combination 
of the unknown constants appearing in the considered process.
It was also done for the pion form-factors \cite{BCT}. In all cases
the obtained value is in reasonable agreement with the resonance estimate.

Some of the terms of Eqs. (\ref{LagInt}-\ref{LagInt2})
are already directly listed in the full list of the  
${\cal O}(p^6)$ Lagrangian terms of \cite{BCElag}.
The others can be rewritten in terms of them but the
resulting list is rather long.
The combination of terms we are interested in, 
can be more easily calculated directly from
Eqs. (\ref{LagInt}-\ref{LagInt2}).
The resulting resonance contribution is displayed 
in App. \ref{Appresonance}. Together with these we take 
the couplings
\begin{eqnarray}
&&f_V = 0.20,\quad  f_\chi = -0.025,\quad  g_V = 0.09, \nonumber \\
&&\alpha_V = -0.014,\quad  f_A = 0.1,\quad  \gamma_A^{(1)} = 0.1, \nonumber \\
&&\gamma_A^{(2)} = -0.01,\quad  c_m = 42 \mbox{ MeV},\quad
  c_d = 32 \mbox{ MeV}, 
\nonumber \\
&&c_\gamma = 19 \cdot 10^{-3} \mbox{ GeV}^{-1},\quad
  c^\prime_\gamma \sim c_\gamma, 
\end{eqnarray}
and the masses are the experimental ones \cite{PDG}.
\begin{eqnarray}
m_V = m_\rho = 0.77 \mbox{ GeV}, & m_A = m_{a_1} = 1.23 \mbox{ GeV}, &
m_S = 0.98 \mbox{ GeV}.
\end{eqnarray}
The values for $f_V$, $g_V$, $f_\chi$, $f_A$
and $c_\gamma$
are taken from experiment \cite{resonance,pipi2}.
$c_m$ and $c_d$ are from comparison with the resonance saturation
at ${\cal O}(p^4)$ \cite{resonance}. For the remainder we use the
ENJL model estimates of \cite{Prades}.
We test the influence of the scalar fields by also performing the fit
with vector meson contributions only as discussed below (fit 6).

Before continuing with the next section, it 
is worth it to summarize some ideas. There 
are $90+4$ ${\cal O}(p^6)$ constants in the most general three-flavour 
chiral Lagrangian, that 
are supposed to be strongly correlated if they give the low-energy behaviour 
of QCD. Considering the Lagrangian for heavier states, with the same 
symmetries as the pseudoscalar one, some correlations 
among the constants are obtained. The behaviour of the resonances 
is different from 
that for the Goldstone bosons: no perturbative theory
as for  the pseudoscalar mesons exists. However the above procedure
is arguably the correct picture in the large $N_c$ limit.

\subsection{The $\eta'$ Field}

Another possible contribution from heavier states is given through the $\eta'$ 
meson that at low-energies does not play as essential a r\^ole as
for instance the $\rho$, except for $\eta$ properties.
The $\eta$-$\eta'$ mixing which affects considerably the $\eta$ properties
is known to be fully contained at ${\cal O}(p^4)$
in the low-energy constant $L_7^r$. 
In order to include explicitly the $\eta'$ degrees of freedom  
we can enlarge the pseudoscalar multiplet including the singlet state, in such 
a way that the $\eta$ and $\eta'$ fields are a combination of the singlet and 
octet fields when $SU(3)$ symmetry is broken but $SU(2)$ is still respected. 
An alternative equivalent way is to introduce it as a separate
pseudoscalar singlet field $P_1$ described by the Lagrangian \cite{resonance}
\be
\label{lageta}
{\cal L}_{\eta^\prime} = \frac{1}{2}\partial_\mu P_1\partial^\mu P_1
-\frac{1}{2}M_{\eta^\prime}^2 P_1^2
+\frac{i\gamma F_0}{2\sqrt{6}}P_1\langle\chi_- \rangle\,.
\ee
At ${\cal O}(p^4)$ the $\eta'$ contribution is entirely encoded in $L_7^r$
and dominates its value.
The next order expansion in the $P_1$ propagator,
${\cal O}(p^6)$, gives  
\be
\label{etaprime}
{\cal L}_{\eta'} =  
- \frac{\tilde{d}_m^2}{2 M^4_{\eta'}} 
\partial_\mu \langle \chi_- \rangle \partial^\mu \langle \chi_- \rangle
~\mbox{with}~ \tilde{d}_m = 20 \, \, \mbox{MeV}. 
\ee
Similar to the scalar case we can add terms to Eq. (\ref{lageta}) that
will contribute as well to ${\cal O}(p^6)$.
The contribution is of relevance
to our fit only via the $\eta$ mass and the prediction for the $\eta$
decay constant, affecting mainly the value for $L_7^r$ with 
a small modification for $L_8^r$. 
Again, rewriting Eq. (\ref{etaprime}) into the terms of \cite{BCElag}
leads to a long expression so we do not present it.

We remark that although the two-point functions, 
used to calculate the decay constants and the masses, are obtained with the 
external currents from the octet, the pole in the propagator
of course corresponds to the real $\eta$ and the decay constant is the
coupling of the real $\eta$ to the octet current.

A more complete list of terms including their integrating out shows
only small modifications to the phenomenological implications \cite{etap}. 

\setcounter{equation}{0}
\section{Phenomenological Applications}
\label{applications}

There are two points to address with the two-loop result for $K_{\ell 4}$.
The first one is to obtain a prediction for the form-factors to be 
compared with experiment and to use our prediction to check
the validity of assumptions made in the data analysis.
This is presented below in Sect. \ref{kl4parametrization}.
The second one is to extract a fit 
for the low-energy constants $L_i^r$ which values will be used to
predict other quantities of interest. In particular, they are needed 
to obtain predictions for low-energy $\pi$-$\pi$ scattering,
one phase-shift combination can be obtained
independently from the same decay data.
The comparison with the scattering quantities obtained directly 
is a consistency check of CHPT.
We remark that in this two-loop 
calculation the only way to obtain an imaginary part for $K_{\ell 4}$ is 
through the intermediate states of two pions. However the $\pi$-$\pi$ 
amplitude appearing here is one-loop. It is then not possible to obtain 
indirectly a three-flavour two-loop result for the rescattering quantities.
We therefore use below the known two-flavour two-loop result
for the scattering \cite{pipi,pipi2}.

\subsection{Values of the Low-Energy Constants}
\label{Values}

The previous values of $L_i^r$ are obtained from one-loop or 
one-loop improved calculations \cite{GL1,BCG}. These values are obtained 
through partially independent fits, in particular $L_1^r,L_2^r,L_3^r$ 
are obtained from $K_{\ell 4}$ taking fixed the other $L_i^r$. In the 
present work we avoid to fix them and 
perform the first global fit of most of the ${\cal O}(p^4)$ 
low-energy parameters simultaneously. The total number of parameters up to
${\cal O}(p^4)$ is 13. They are $B_0 \hat m$, $B_0 m_s$, $F_0$
and $L_1^r$--$L_{10}^r$. $L_{10}^r$ does not contribute to any
of the quantities considered here and is thus fully independent.
$L_9^r$ is discussed below.
In practice we thus need to fit 11 free parameters.
The values are updated fitting 
the results obtained for the masses and decay constants to two-loops 
\cite{Amoros} and the result of the present work, see also \cite{LEC}. 

The other free quantities in the full result are the $C_i^r$, 
that are estimated in the previous section, Sect. \ref{estimates}. 
The values of $L_i^r$ can be reasonably well predicted from
resonance exchange, so a similar rough estimate is expected to
work for the $C_i^r$. We test the dependence by removing all the
resonance contributions except for the vector ones, whose couplings
are determined experimentally elsewhere, and the $\eta^\prime$, and see if
the results change. Changing the values of the $C_i^r$ by 50\%
changes the fitted values within the fit errors.
We also present a reasonable variation of the
other inputs.

Unfortunately, it is not possible to find at present 11 fully independent
inputs from direct experimental quantities.
Below we use as inputs $m_\pi^2$, $m_K^2$, $m_\eta^2$, $F_\pi$,
$F_K$ and three inputs from $K_{\ell 4}$, $f_s(0)$, $g_p(0)$ and $\lambda_f$,
defined below. The three additional inputs are
the quark-mass ratio $m_s/\hat m$, $L_4^r$ and $L_6^r$. In the future we hope
to be able to include additional experimental input to constrain these
as well.

The fit is performed by minimizing the $\chi^2$ function defined by
\be
\chi^2 = \sum_{i=1,6} \chi_i^2 = \sum_{i=1,6} 
\left(\displaystyle \frac{x_i - x_{i\mbox{\small input}}}{\Delta 
x_{i\mbox{\small input}}}
\right)^2\,,
\ee
with as weight $\Delta x_{\mbox{input}}$ the error in the input values 
$x_{i\mbox{\small input}}$. 

This allows us to fit everything since in all higher orders we have
replaced quark masses and $F_0$ by the relevant physical quantities. 
Then we can use the mass and decay constant expressions up to ${\cal O}(p^6)$
to determine
$B_0\hat m$ and $F_0$ from the physical input afterwards.
Let us now discuss the various inputs and $\chi_i^2$.

\begin{description}
\item[Physical masses and decay constants:]
For these quantities we use
the pion masses $m_{\pi^+} =139.56995 \, \mbox{MeV} $, 
$m_{\pi^0} =0.1349764 \, \mbox{MeV} $,
the Kaon masses $m_{K^+}=493.677 \, \mbox{MeV}$,
$m_{K^+}=497.672 \, \mbox{MeV}$  and $\eta$ mass 
$m_\eta =547.30 \, \mbox{MeV}$, the pion decay constant 
$F_\pi = 92.4 \, \mbox{MeV}$ and 
the Kaon decay constant $F_K/F_\pi = 1.22 \pm 0.01$ \cite{PDG}. 
\item[$\mathbf{L_4^r,L_6^r}$:]
The first main feature that we use is the 
behaviour under the large $N_c$ limit. The consequences for
the $L_i^r$ are discussed in \cite{GL1} and for the $C_i^r$ in \cite{BCElag}.
This limit tells us that some of the constants are expected to vanish at some
unknown scale.
This is the case for $L_4^r$ and $L_6^r$ that are both set zero for
the main fit (at $m_\rho$ scale). Fit 5 in Table \ref{Tableresults} 
assumes them to be zero at a scale of $m_\eta^2$
instead. For other fits relaxing this assumption
we refer to Section \ref{largeNc}.
\item[$\mathbf{L_9^r}$:]
Another point is the value for $L_9^r$. This constant is related to 
the charge radius of the pion. The value is completely saturated by 
the $\rho$ resonance, and for this reason the contribution from the 
logarithms is negligible. 
Moreover $L_9^r$ appears with the factor $s_\ell $ in $K_{\ell4}$, such that
it is not necessary for the main fit where we use $s_\ell  =0$. 
For the other 
fits we can safely use the value $L_9^r= 6.9~10^{-3}$.
\item[Other $\mathbf{L_i^r}$:]
The remaining constants will be taken as free parameters, at the end six 
LEC's to fit.
\item[The quark-mass ratio $m_s/\hat m$:] 
To ${\cal O}(p^2)$ it is possible to relate the ratio of the strange quark 
mass $m_s$ over the isospin doublet quark mass $\hat m =(m_u+m_d)/2$, 
with a combination of
the meson masses. If we assume that ${\cal O}(p^4)$ and ${\cal O}(p^6)$ 
do not appreciably modify this ratio we can use this as input or
alternatively the values from QCD sum rules and/or the lattice.
As standard input we use
 $m_s/\hat m  = 24$ and we check that $m_s/\hat m  = 26$ does not
lead to significantly different results (fit 2 in Table \ref{Tableresults}). 
We remark that there are two possible ways 
to calculate this ratio from the lowest-order meson masses, 
using or not the $\eta$ mass
\be
\left. \frac{m_s}{\hat m} \right|_1 
= \frac{2\, m^2_{0K} -m^2_{0\pi}}{m^2_{0\pi}}\,;\quad
\left. \frac{m_s}{\hat m} \right|_2 
= \frac{3\, m^2_{0 \eta} -m^2_{0\pi}}{2 \, m^2_{0\pi}}\,, 
\ee
where $m_0$ stands for the lowest-order mass that we calculate
in terms of the physical masses, $F_\pi$ and the $L_i^r$.
We included both relations in the fit. The expression of the bare masses 
in terms of the physical masses to two-loop is taken from \cite{Amoros}.
For the pion mass we use the neutral pion mass. For the 
Kaon mass we subtract the electromagnetic contribution through the formula
\be
m^2_{K\mbox{\small av}} =\frac{1}{2} (m^2_{K^+} + m^2_{K^0} 
- 1.8 (m^2_{\pi^+} -m^2_{\pi^0})) =  (494.53 \, \mbox{MeV})^2 \, ,
\ee
where the factor $1.8$ is a modification due to the corrections 
to Dashen's theorem \cite{Dashen}.
Whenever we fit or calculate masses we use $m^2_{K\mbox{\small av}}$
and $m_{\pi^0}^2$. The contribution to $\chi^2$ is
\be
\chi_1^2+\chi_2^2 = \sum_{i=1,2}\left(\frac{
\left.\frac{m_s}{\hat m}\right|_i-
\left.\frac{m_s}{\hat m}\right|_{\mbox{\small input}}}
{0.1\,\left.\frac{m_s}{\hat m}\right|_{\mbox{\small input}}}
\right)^2 \, .
\ee

\item[$\mathbf{F_K/F_\pi}$:]
This quantity is calculated using the formulas of \cite{Amoros}\footnote{
The published version has in App. A.2 the formulas in terms of
lowest-order masses and decay constant, contrary to what is stated in
the text there. Here we use the formulas in terms of physical masses.},
and enters in the fit as
\be
\chi_3^2 = \left(\frac{(F_K/F_\pi)-1.22}{0.01}\right)^2\,.
\ee

\item[$\mathbf{K_{\ell 4}}$ decay:]
The expression for the form-factors $F$ and $G$ obtained in 
the theoretical part 
is compared with the experimental analysis from \cite{Rosselet}. The results 
we are interested in are the extraction of the s-wave contribution from the 
F form-factor and the p-wave from the $G$ form-factor. The others contributing 
waves are not appreciable \cite{Rosselet}. This is also confirmed
in the expressions as discussed below.
In  \cite{Rosselet} the s-wave 
$f_s$ and p-wave $g_p$ are fitted with the expressions
\ba
\label{slope}
f_s(q^2) &=& f_s(0)\, (1+ \lambda_f \, q^2) \, ,\nonumber\\
g_p(q^2) &=& g_p(0)\, (1+ \lambda_g \, q^2) \, ,
\ea
where $f_s(0)$ and $g_p(0)$ are the 
form-factors at threshold of the dipion system.
The slopes are taken to be the same $\lambda_f = \lambda_g$ .

In order to keep the computer time needed in reasonable limits,
we use the following approximation
\be
f_s(0) = F(s_\pi,s_\ell =0, \cos \theta_\pi =0).
\ee
We use $s_\pi=(2m_\pi+1\,\mbox{MeV})^2$ to avoid numerical
problems at threshold. The $s_\ell$ dependence is very mild, we checked
this by performing an alternative fit with $\sqrt{s_\ell }=100$~MeV 
(fit 3 in Table \ref{Tableresults}). 
Only using $\cos\theta_\pi=0$ corresponds to assume $d$ and higher waves 
to be negligible.
Similarly 
\be
g_p(0) = G(s_\pi,s_\ell =0, \cos \theta_\pi =0).
\ee
We do not make use of the derivative at threshold as the slope.
This has both numerical problems and does not correspond to the
experimental situation. Instead we shift a bit from the threshold. 
The slope is then given by 
\be
\lambda_f = \Big( 
\frac{|F(s'_\pi,s_\ell,\cos \theta_\pi =0)|}{|F(s_\pi,s_\ell,\cos \theta_\pi =0)|}
- 1 \Big) \frac {4 m^2_\pi}{s'_\pi -s_\pi}
\ee
with the value $s_\pi = ( 2 \, m_\pi +1 \, \mbox{MeV}\,)^2$ and 
$s'_\pi =(336 \, \mbox{MeV}\,)^2$. As an alternative we also
present a fit with $s_\pi^\prime = 293$~MeV 
(fit 4 in Table \ref{Tableresults}).

These quantities enter in the fit via
\be
\label{chi5}
\chi_4^2 = \left(\frac{f_s(0)-5.59}{0.14}\right)^2\,,\quad
\chi_5^2 = \left(\frac{g_p(0)-4.77}{0.27}\right)^2\,,\quad
\chi_6^2 = \left(\frac{\lambda_f-0.08}{0.02}\right)^2\,.
\ee
We also present an alternative fit, fit 9, where we include additionally 
the $g_p(0)$ measurement in $K^0_L\to\pi^0\pi^-e^+\nu$ of
Ref. \cite{Makoff}. Combining errors using the PDG procedure \cite{PDG}, 
Eq.~(\ref{chi5}) is modified with
\be
\chi_5^2 = \left(\frac{g_p(0)-4.93}{0.31}\right)^2\,.
\ee
\end{description}

With the previous fit procedure we find for the values of the
low-energy constants
\ba
\label{newfit}
&&
10^3 L_1^r=   0.52 \pm  0.23\,, \quad
10^3 L_2^r=   0.72 \pm  0.24\,, \quad
10^3 L_3^r=  -2.70 \pm  0.99\,, \nonumber\\&&
10^3 L_5^r=   0.65 \pm  0.12\,, \quad
10^3 L_7^r=  -0.26 \pm  0.15\,, \quad
10^3 L_8^r=   0.47 \pm  0.18\,. \quad
\ea
In Table \ref{Tableresults} we present the alternative fits discussed above.
The difference with the results of \cite{LEC} is due to the
$\eta^\prime$ inclusion everywhere, this mainly affects the
values of $L_7^r$.

We also included two fits (fit $7$ and $8$)
with a different choice for the scale, 
$\mu =0.5$ and $\mu =1$ GeV. The rescaling of the $L_i^r$ is considered such 
that in the table we quote the shifted value at $\mu =0.77$ GeV. These 
choices imply different scales for the resonance saturation on the $C_i^r$. 
As can be noticed the changes due to the scale are quite considerable, but
almost covered by the fitting errors. This suggest that the errors
quoted in the main fit are unlikely to be reduced if there
is more control over the  ${\cal O}(p^6)$ constants.

The errors are those quoted by MINUIT and are those that
give $\Delta\chi^2=1$ assuming the quadratic approximation near the minimum. 

In addition we have performed a fit where $V_{us}$
was changed to 0.226 which showed no significant deviations
from our main fit. This value is from using unitarity in the CKM matrix
and $V_{ud}=0.9740$ \cite{PDG,Stern}.

\begin{table}
\caption{\label{Tableresults}
Results for $L_i^r(\mu)$ for the various
fits described in the main text. Notice that $L_4^r, L_6^r$ and
$L_9^r$ are input, $L_4^r=0, L_6^r=0$ and  $L_9^r=6.90\  10^{-3}$.
Errors are fitting errors described in the text.
All $L_i^r(\mu)$ values quoted have been brought to the scale
$\mu=0.77$~GeV.
The standard values are $m_s/\hat m = 24$, $\sqrt{s_\pi^\prime}= 0.336$~GeV
and $s_\ell = 0$.}
\vspace{0.25cm}
\begin{small}
\begin{tabular*}{\textwidth}{*{11}{c@{\hspace{1.7mm}}}c}
\hline
               & Main Fit & \cite{GL1,BCG} &${\cal O}(p^4)$&fit 2&fit 3&
 fit 4 & fit 5 & fit 6 & fit 7 & fit 8 & fit 9\\
\hline
$10^3\,L^r_1$&   0.52$\pm$0.23&   0.37$\pm$0.23&   0.46&   0.53&   0.50&
   0.49&   0.52&   0.45&   0.44&   0.63&   0.65\\
$10^3\,L^r_2$&   0.72$\pm$0.24&   1.35$\pm$0.23&   1.49&   0.73&   0.67&
   0.74&   0.80&   0.51&   1.04&   0.73&   0.85\\
$10^3\,L^r_3$&$-$2.70$\pm$0.99&$-$3.5 $\pm$0.85&$-$3.18&$-$2.71&$-$2.58&
$-$2.73&$-$2.75&$-$2.16&$-$2.95&$-$2.67&$-$3.27\\
$10^3\,L^r_5$&   0.65$\pm$0.12&   1.4 $\pm$0.5 &   1.46&   0.62&   0.65&
   0.64&   0.82&   0.67&   1.00&   0.51&   0.60\\ 
$10^3\,L^r_7$&$-$0.26$\pm$0.15&$-$0.4 $\pm$0.2 &$-$0.49&$-$0.20&$-$0.26&
$-$0.26&$-$0.30&$-$0.26&$-$0.23&$-$0.25&$-$0.26\\
$10^3\,L^r_8$&   0.47$\pm$0.18&   0.9 $\pm$0.3 &   1.08&   0.35&   0.47&
   0.47&   0.58&   0.46&   0.52&   0.44&   0.48\\
\hline	     
changed      & & &${\cal O}(p^4)$&$m_s/\hat m$&$\sqrt{s_\ell}$&
$\sqrt{s_\pi^\prime}$&$L_4^r;L_6^r$&$V_\mu,\eta^\prime$&$\mu$&$\mu$&$g(0)$\\ 
quantity     &          &  &       & 26 & 0.1             & 0.293 
&$-0.3;-0.2$&   only&0.5&1.0&4.93\\
Unit         &          &  &       &    & GeV             & GeV   &
 $10^{-3}$&     & GeV & GeV &\\
\hline
\end{tabular*}
\end{small}
\end{table}

\subsubsection{Errors and Correlations}
\label{errors}

The problem in determining errors on derived quantities is that
the $L_i^r$ values given above are very strongly correlated
and that with 6 free parameters the chance that all of them are
within their one sigma error, is negligibly small.
We have thus generated a distribution of sets of $L_i^r$ that
have a correlated Gaussian distribution according to the $\chi^2$ function 
defined above. We then keep the sets that have a $\chi^2$ less
than the value corresponding to 68\% confidence level.
Projecting this distribution on the relevant variable leads
to larger ranges than the errors quoted above. Those are the 
errors quoted in the remainder since they are more
relevant than those calculated from propagating the errors of
Table \ref{Tableresults}.

\begin{figure}
\begin{center}
\epsfig{file=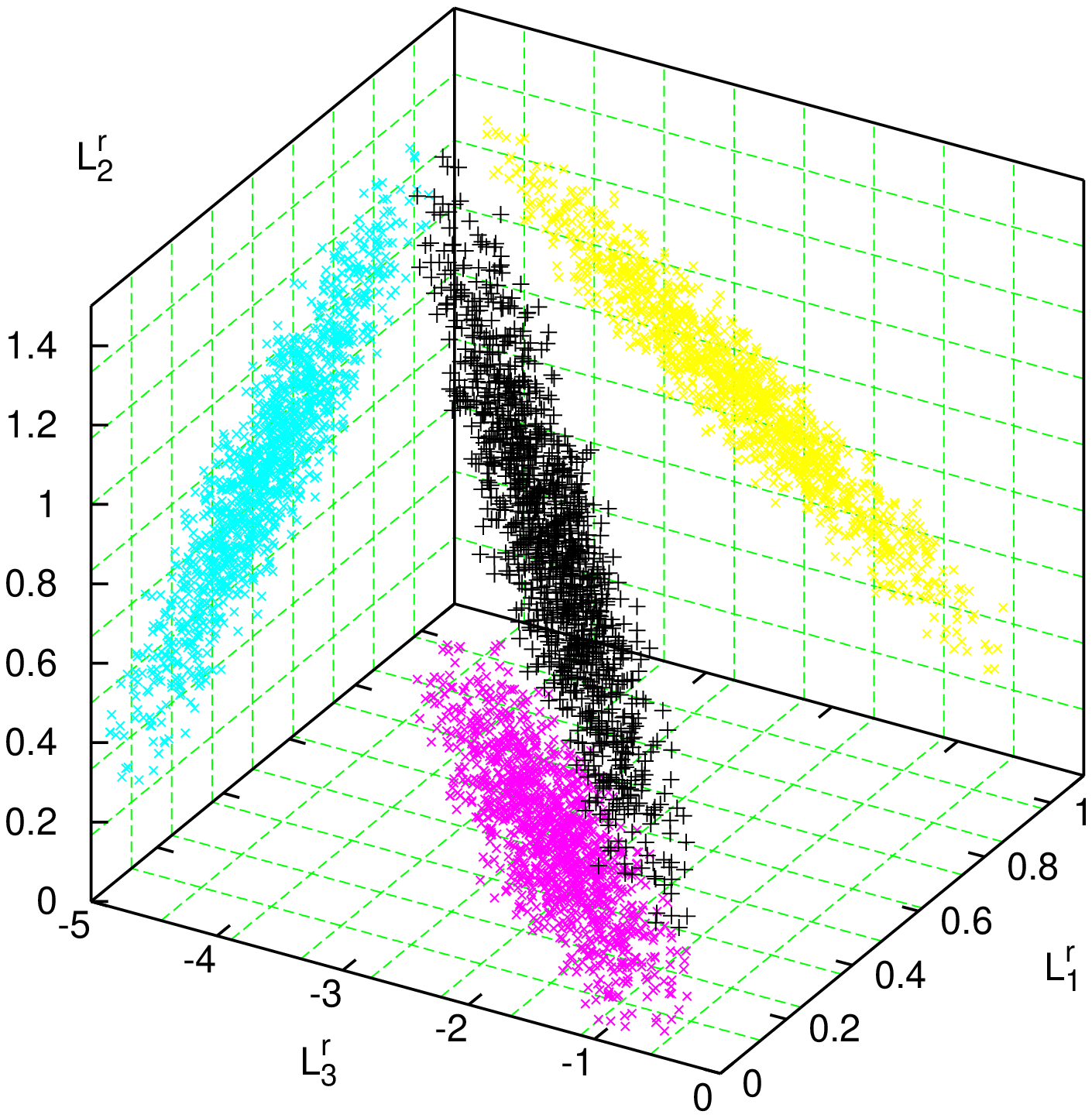,height=0.60\textwidth,angle=-90}
\raisebox{-1.5cm}{\epsfig{file=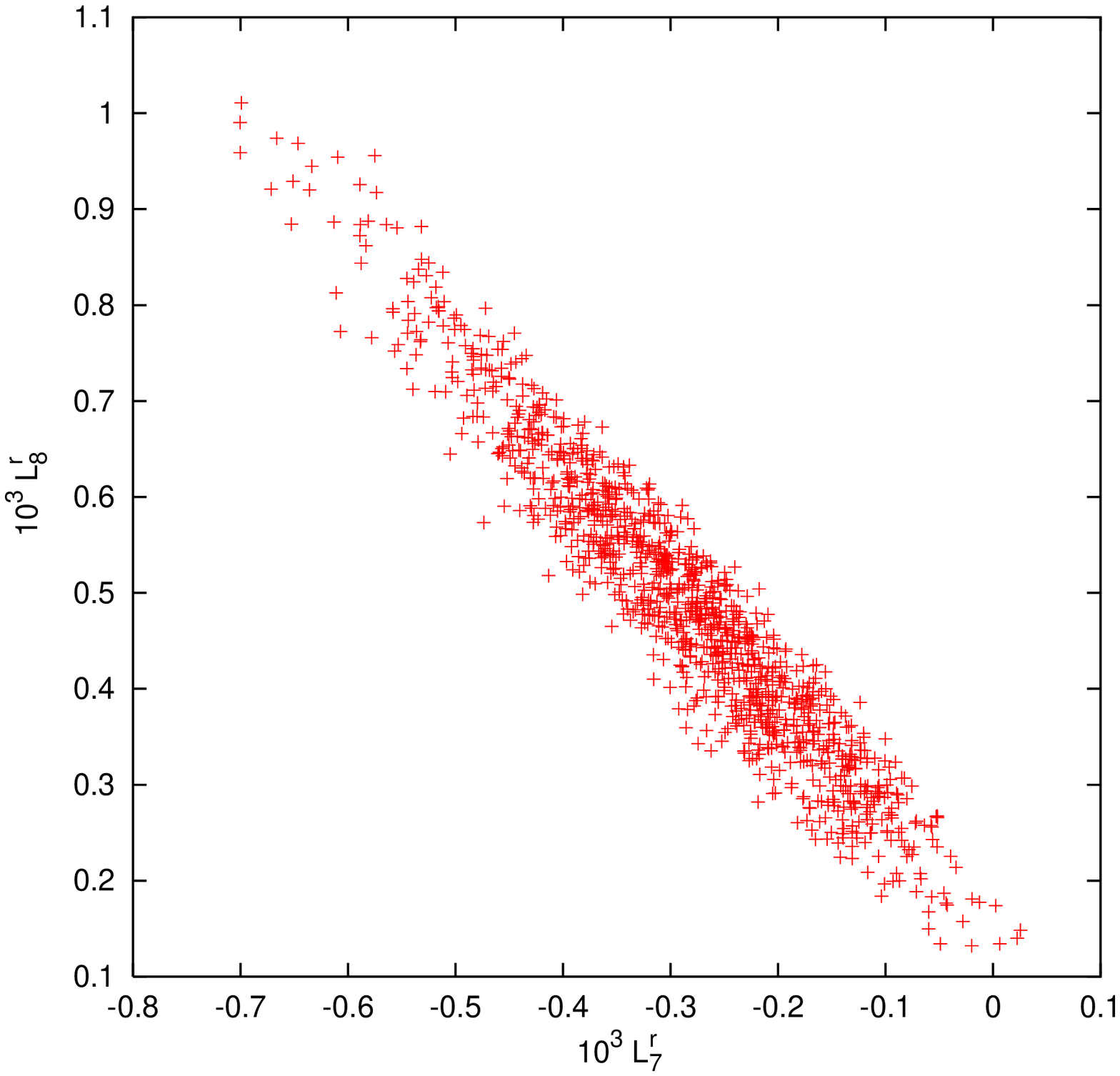,height=0.39\textwidth,angle=-90}}
\end{center}
\caption{\label{figcorrelations}The sets of $L_i^r$ within a 68\% confidence
level range of $\chi^2$. In the first plot we show {\bf +} the values
of the first three $L_i^r$ and {\bf $\times$} the projections
on the coordinate planes. The second plot shows the $L_7^r$--$L_8^r$
correlations.}
\end{figure}

As examples of the correlations we show the plot of $L_1^r$,
$L_2^r$ and $L_3^r$ together with their projections on the two-variable
planes and the plot of $L_7^r$ and $L_8^r$ in Figure \ref{figcorrelations}.
Each point correspond to one set of $L_i^r$ of the distribution described
above.
Similar plots can be made for all other combinations. 
Notice that this only shows the {\em experimental} type of error.
Not included are the errors due to the restriction to ${\cal O}(p^6)$
and other theory uncertainties.

\subsubsection{Comparison with Previous Results and Models}
\label{models}

\begin{figure}
\begin{center}
\epsfig{file=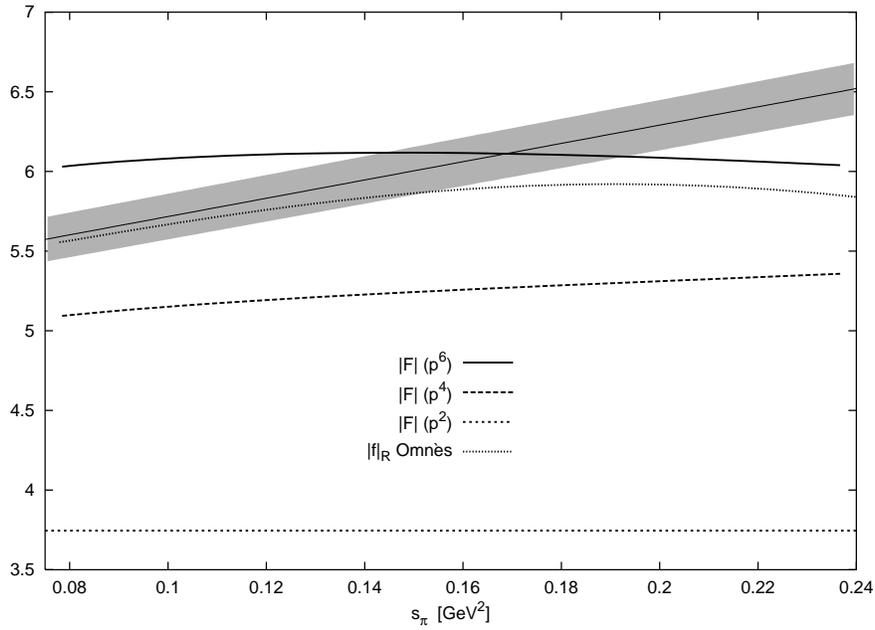,height=12cm,angle=-90}
\end{center}
\caption{\label{figBCG} Comparison of the Omn\`es improved estimate
with the full ${\cal O}(p^6)$ calculation using the same \emph{old}
set of values of $L_i^r$ as input together with $F_\pi=93.2$~MeV.
The shaded band is the experimental result.}
\end{figure}

In this section we want to give some insight about
the sizeable corrections of the LEC, Eq. (\ref{newfit}), 
with respect to the unitary
calculation, second column in Table \ref{Tableresults}, and to 
show the differences with some
existing models based on large $N_c$ \cite{domenech,michelle}. Due to the high
correlation between the various variables this
is not a trivial affair. To make the analysis more transparent we remove
first of  all
the constraints on the masses and decay constants, i.e we deal \emph{almost}
 with the same
assumptions as in previous calculations \cite{GL1,BCG}. Therefore we fix
$L_5^r, L_7^r$ and $L_8^r$  to their ${\cal O}(p^4)$ value. All in all this
 amounts to determine
$L_1^r, L_2^r$ and $L_3^r$ from the slope $(\lambda)$, the s--wave
$(f_s)$ and the p--wave $(g_p)$ --see Eq. \ref{slope}. The results obtained
after that are displayed in fit A of Table \ref{tabcomp}. 
A close look, shows that
the results are \emph{quite} similar to those quoted in Eq. (\ref{newfit}),
thus
the constraints imposed by the masses and decay constants do not drive at all
the differences. To disentangle the pieces from the full calculation
and compare
with \cite{BCG} is extremely delicate. For instance, using the dispersive
method only some pieces of diagrams (h), (j) in Fig. \ref{fig2loop1} and
of those depicted in Fig. \ref{fig2loop2} are kept. 
The rest has no new imaginary part
and is therefore modeled by a pure multiplicative constant only. 
Furthermore, the 
dispersive approach misses the full two-loop renormalization, only the one-loop
renormalization of $F_\pi$, masses  and WFR is introduced. Besides these
obvious remarks
a less straightforward point needs clarification: in \cite{BCG} the full
relativistic 
resonance propagator is kept, thus resumming all chiral orders. This has 
some influence on $G$.
The main point to stress is that already at threshold there is a significant
shift of the form-factors with respect to the dispersive calculation.
In Fig. \ref{figBCG} we show
this fact, comparing ${\cal O}(p^4)$, the Omn\`es improvement 
and the ${\cal O}(p^6)$
for the $F$ form-factor using the results of \cite{BCG} as input as well as
$F_\pi=93.2 MeV$.
Pieces
that can account for the difference  at threshold are
the $L_i \times L_j, L_i \times \log, \log\times
\log$ and the resonance saturation piece. All of them, except
the last one, are ${\cal O}(p^6)$ contributions not considered
in the dispersive approach \cite{BCG}. The Omn\`es improved calculation
for $F$
had the right sign but the wrong magnitude.
In Fig. \ref{figBCG} we have compared the Omn\`es improved calculation
with the full ${\cal O}(p^6)$. Notice the considerable shift 
already at threshold.
To take into account this difference the values of the LEC,
$L_1^r$, $L_2^r$ and $L_3^r$, have to
change substantially. In the figure we have used the values of the low-energy
constants obtained in \cite{BCG}, and as a consequence
the Omn\`es calculation matches the experimental result,
shadow area. With those parameters  the ${\cal O}(p^6)$ result is
outside the experimental errors.

\begin{table}
\caption{\label{tabcomp} Values of $L_i^r$ for comparison
with Eq.~(\ref{newfit}).
The first column is the fit 
without the constraints from the masses and decay constants. Second is the 
result from the Resonance Saturation model. Third and fourth 
columns are values from other large $N_c$ based models as described in the
text.}
\begin{center}
\vspace{.5cm}
\begin{tabular}{c||c|c|c|c} \hline
 & fit A  & Ref. \cite{resonance} & Ref. \cite{domenech}
 & Ref. \cite{michelle,peris} \\
\hline
$10^3 L_1^r$     & $0.53\pm 0.24$ & $0.6$   & $0.79$  & $0.81$  \\
$10^3 L_2^r$     & $0.67\pm 0.25$ & $1.2$   & $1.58$  & $1.62$  \\
$10^3 L_3^r$     & $-2.74\pm 1.09$& $-3.0$  & $-3.17$ & $-4.24$  \\
$10^3 L_5^r$     & fixed          & $1.4^{(a)}$   & $0.43$  & $1.21$  \\
$10^3 L_8^r$     & fixed          & $0.9^{(a)}$   & $0.46$  & $0.60$  \\
\hline
\end{tabular}
\end{center}
\vspace{-0.2cm} \hspace{3cm} \small $~^{(a)}$ Input
\end{table}

In our calculation we have assumed a resonance saturation for the
${\cal O}(p^6)$ coefficients.
To some extent this assumption was successful 
at ${\cal O}(p^4)$ \cite{resonance}.
In the second column of Table \ref{tabcomp}
we quote the values for this model-approach to some of the LECs. One can see
that $L_1^r$ and $L_3^r$ in Eq. (\ref{newfit}) are in good agreement
with the resonance saturation hypothesis. $L_2^r$ is off from this
hypothesis, and violates the relation
$2L_1^r = L_2^r$ by $20\%$, which is of the expected size of
deviations from the large $N_c$ limit.
As is discussed in \cite{descotes}
there is some indication that the resonance saturation assumption
fails in the scalar sector.
For this matter, one
should compare between different models.
It is clear that our result is within the range as exemplified
by the two examples discussed below.

We also compare in Table \ref{tabcomp} with two
model-predictions of CHPT
parameters, including also the values of $L_5^r$ and $L_8^r$. 
In Ref. \cite{domenech} the large $N_c$ and chiral limits lead
including the leading
gluonic contribution to\footnote{The predictions for $L_5^r$ and $L_8^r$
of Ref. \cite{domenech} are sensitive to the precise
regularization scheme chosen in the chiral quark model.}
\ba
\label{model1}
&&4L_1 = 2L_2 = \frac{N_c}{96 \pi^2} \Big[ 1 + 
{\cal O}(1/M_Q^6) \Big]\nonumber\,,\\&&
L_3 = -\frac{N_c}{96 \pi^2} \Big[ 1 + \frac{\pi^2}{5 N_c} 
\frac{\langle\frac{\alpha_S}{\pi} GG \rangle}{M_Q^4} 
+ {\cal O}(1/M_Q^6) \Big]\,,
\ea
where $M_Q$ can be interpreted as a constituent quark mass. Comparing
the second column in Table \ref{tabcomp} with Eq. (\ref{newfit}) one
can see an
improvement in the central value of $L_1^r, L_5^r$ and $L_8^r$ 
with respect to 
the one-loop result, while the values
for $L_2^r$ and $L_3^r$, which essentially agrees with the ${\cal O}(p^4)$
LEC's, went off
suggesting the relevance of higher gluonic or large $N_c$ corrections.

The model of \cite{domenech} was extended to the full ENJL model
with gluonic corrections in \cite{enjl}. The resulting relations
led to the proposal of a new model,
 LMD \cite{michelle}. 
It mimics QCD by taking into account the
first resonance plus the
continuum obtaining the relations
\be
\label{model2}
6 L_1 = 3L_2=-\frac{8}{7} L_3=4 L_5=8L_8=\frac{3}{8} \frac{F_\pi^2}{m_\rho^2}
=\frac{15}{8\sqrt{6}}\frac{1}{16\pi^2}\,,
\ee
where the last equality is obtained using higher orders in
 QCD sum rules \cite{peris}.
Its values are given in the third column of Table \ref{tabcomp} and the same 
comments as in the previous model apply for $L_1^r, L_2^r, L_3^r$
and $ L_8^r$, 
while the central value of $L_5^r$ turns out to be worse.

\subsection{$K_{\ell 4}$ Form-Factors and Partial Wave Expansions}
\label{FFandPWE}

\begin{figure}
\begin{center}
\parbox{0.49\textwidth}{
\epsfig{file=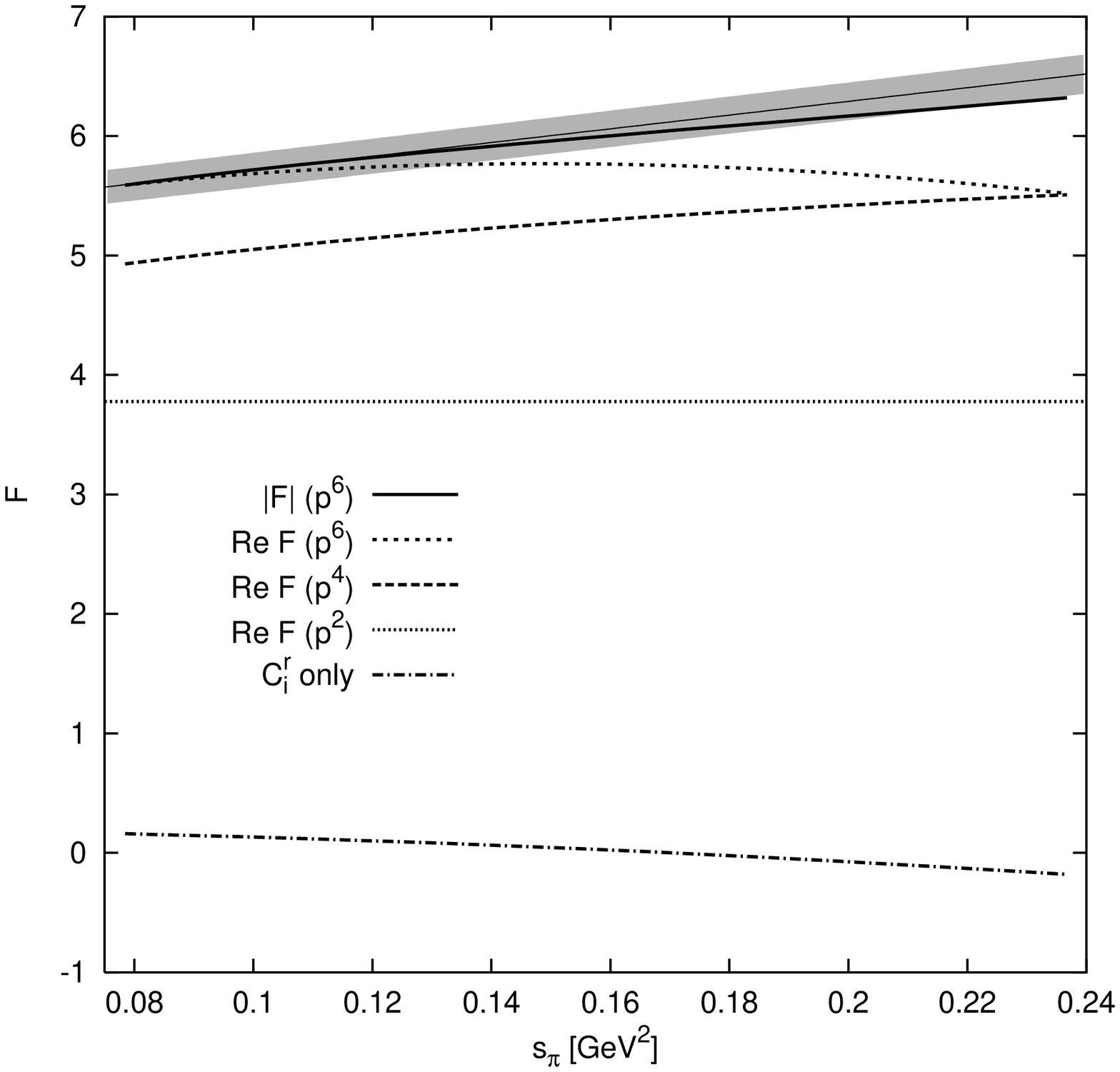,height=0.485\textwidth,angle=-90}\\
\begin{center}(a)\end{center}}
\parbox{0.49\textwidth}{
\epsfig{file=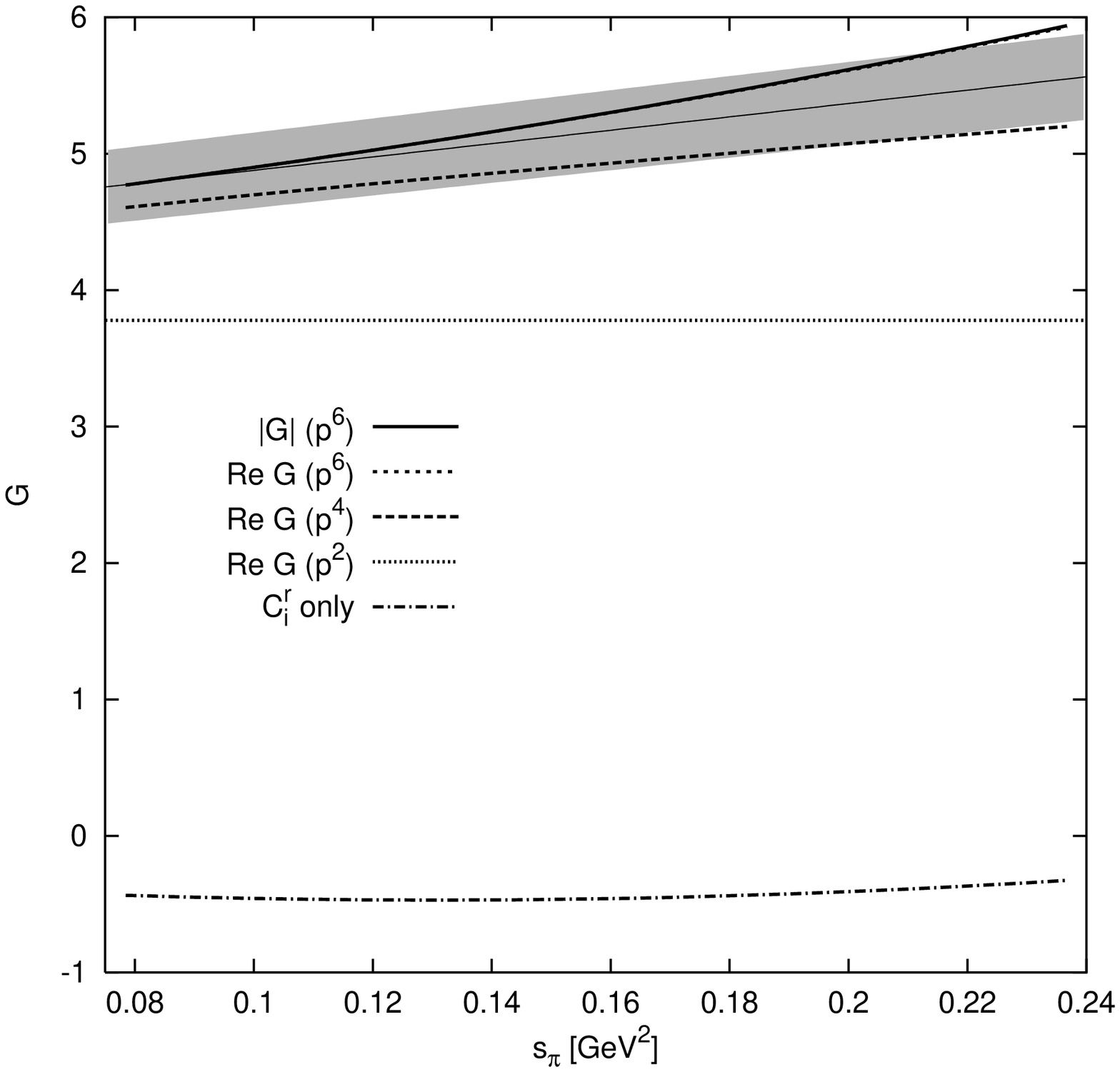,height=0.485\textwidth,angle=-90}
\begin{center}(b)\end{center}}
\end{center}
\caption{\label{figFmain}
The $F$ (a) and  $G$ (b) form-factor at $s_\ell =0$ and
$\cos\theta_\pi=0$.
The shaded band is the result of \cite{Rosselet}. Shown are
the full result for the absolute value, and the real part to lowest-order,
${\cal O}(p^4)$ and ${\cal O}(p^6)$. We also show the contribution from
the ${\cal O}(p^6)$ Lagrangian; this is dominated by the vector contribution.}
\end{figure}

We now plot the form-factors $F$ and $G$ at $\cos\theta_\pi=0$
for the main fit, Eq.~(\ref{newfit}), and the various separate contributions.
We show the full result for $F$ in Fig. \ref{figFmain}a
and for $G$ in Fig. \ref{figFmain}b.
Notice that the convergence
of the series is reasonable for both form-factors and that for 
$F$ the contribution from the
${\cal O}(p^6)$ Lagrangian is fairly small. For $G$ this is dominated by
vector exchange with experimentally determined parameters so the uncertainty
from this source is quite small.

\begin{figure}
\begin{center}
\epsfig{file=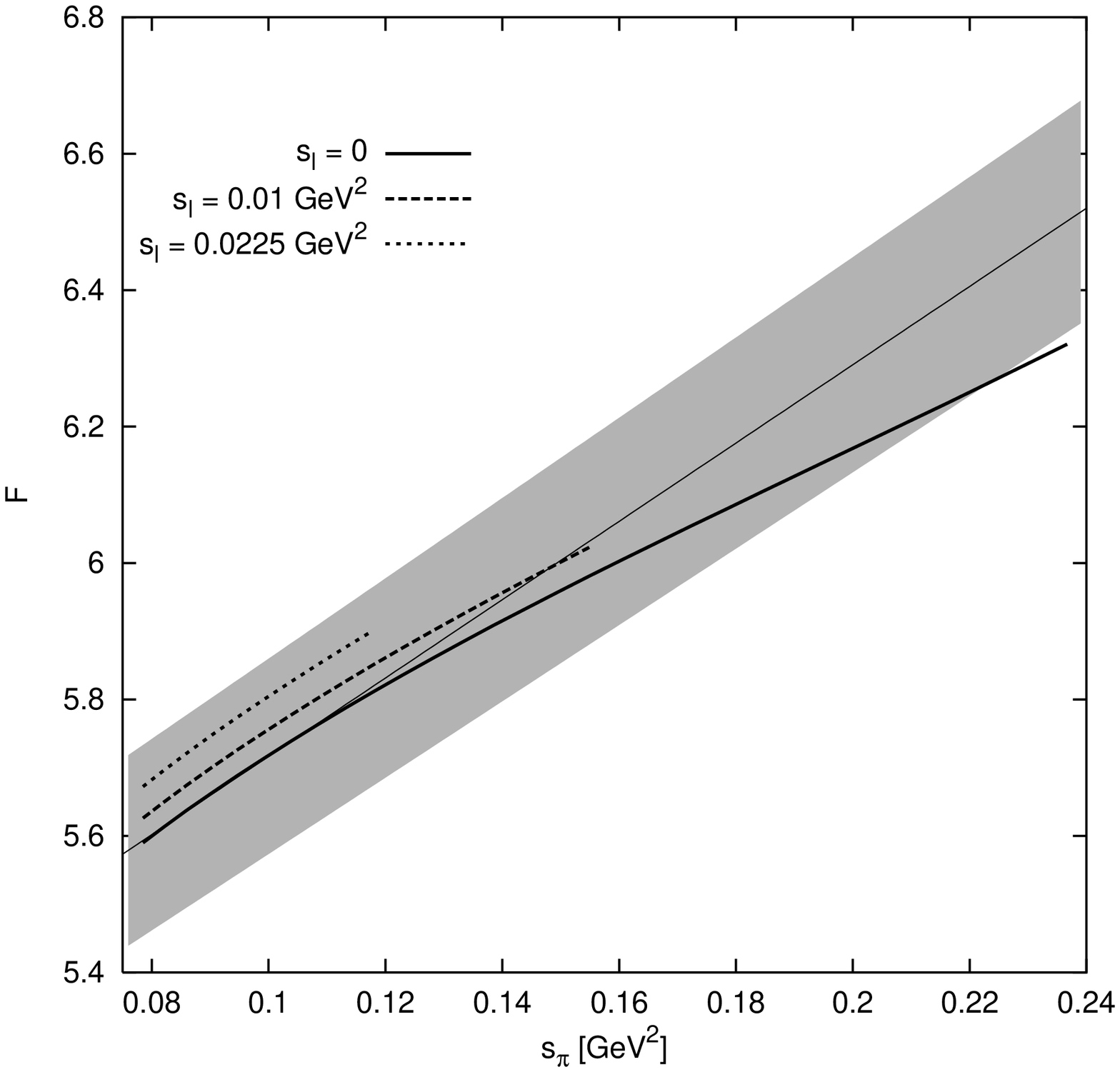,width=0.48\textwidth,angle=-90}
\epsfig{file=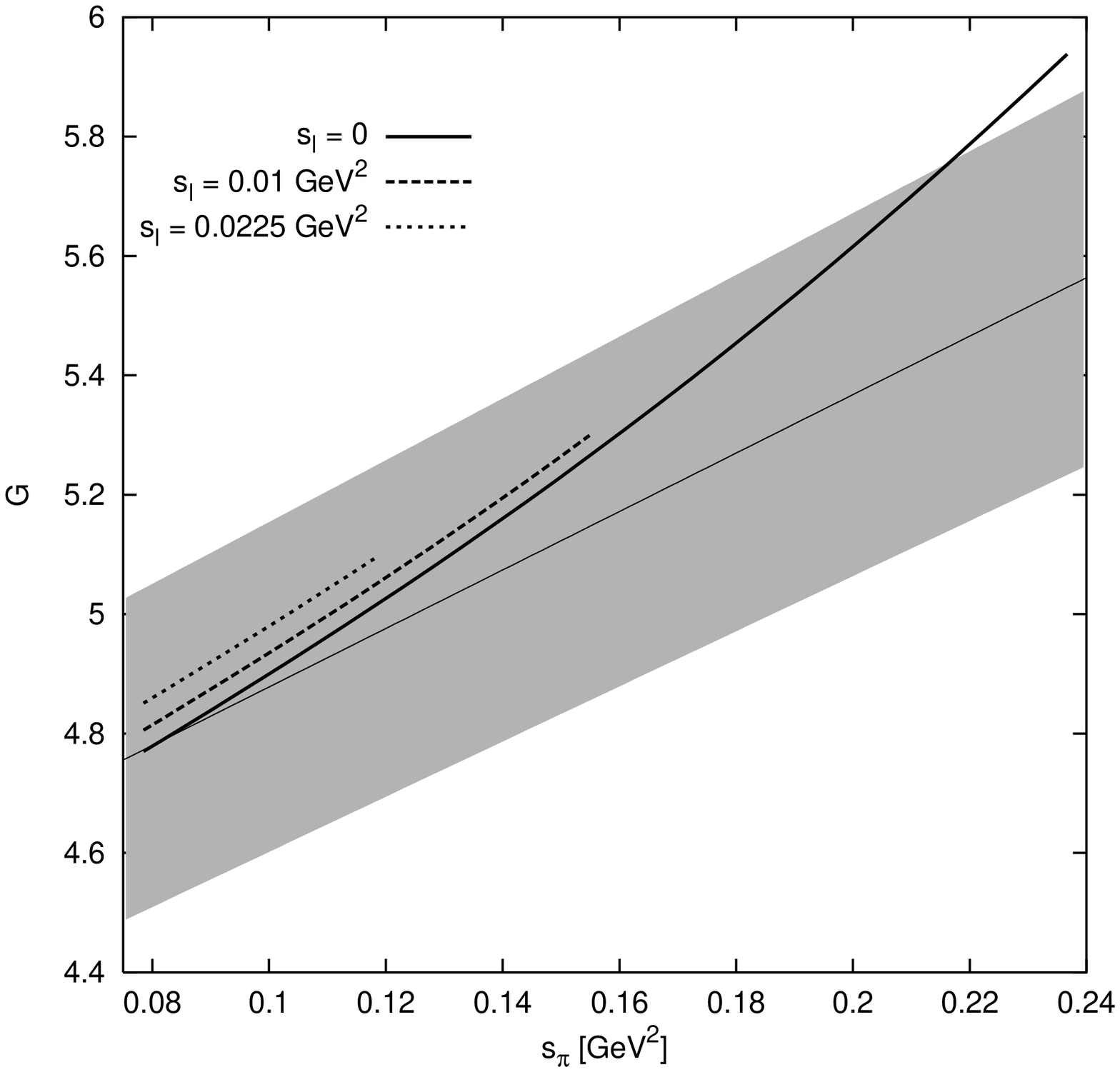,width=0.48\textwidth,angle=-90}
\end{center}
\caption{\label{figFGsl} The absolute value
of the $F$ and $G$ form-factor at $\sqrt{s_\ell }=0$, 100~MeV and 150~MeV. 
All cases with $\cos\theta_\pi=0$. The shaded band is 
the result of \cite{Rosselet}.}
\end{figure}

Let us now look at the dependence on $s_\ell $. In Fig. \ref{figFGsl} we show
the absolute value of $F$ and $G$ for $\sqrt{s_\ell }=0$, 100~MeV and 150~MeV,
together with the experimental error band. The lines for the 
higher $s_\ell $-values
stop at the kinematical limits. Notice the change in scale
with respect to Figures \ref{figFmain}a and \ref{figFmain}b. Given the
accuracy of the present data we can safely neglect the 
$s_\ell $ dependence completely.

\begin{figure}
\begin{center}
\parbox{0.49\textwidth}{
\epsfig{file=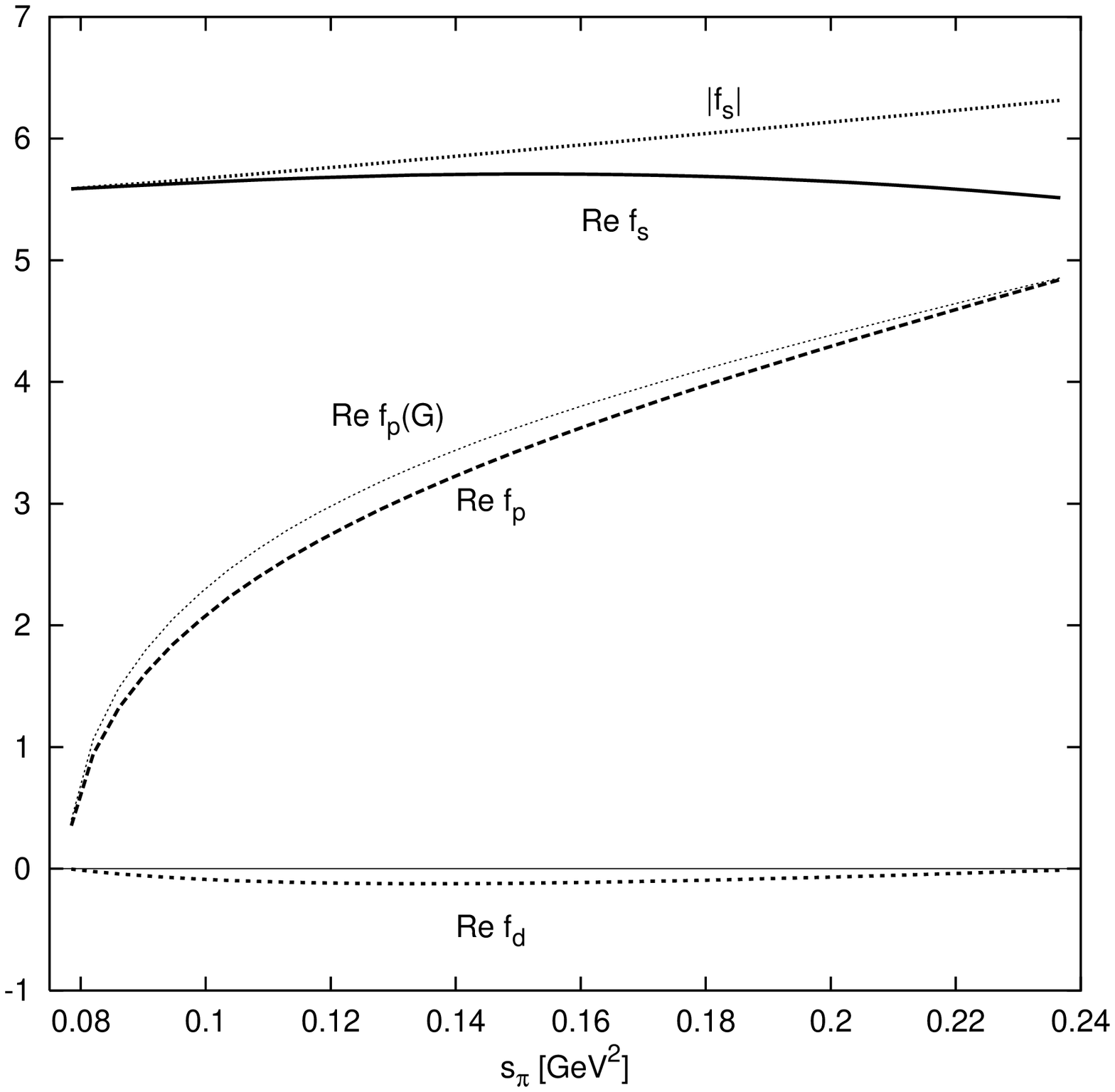,height=0.485\textwidth,angle=-90}
\begin{center}(a)\end{center}}
\parbox{0.49\textwidth}{
\epsfig{file=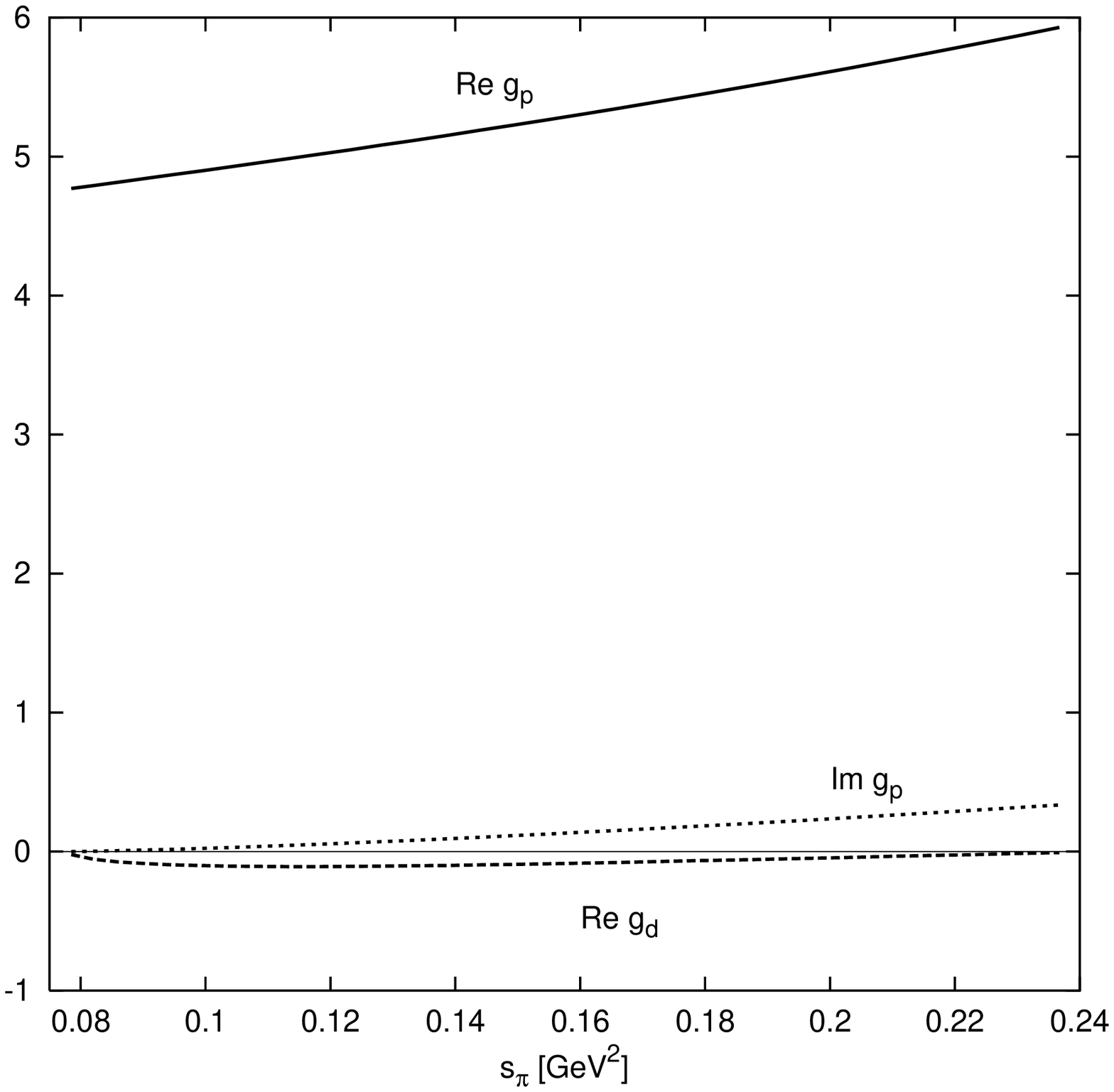,height=0.485\textwidth,angle=-90}
\begin{center}(b)\end{center}}
\end{center}
\caption{\label{figFpartial} (a) The partial wave expansion of
the combination $\bar F=F+\sigma_\pi PL \cos\theta_\pi G/X$.
Plotted are the absolute value of $f_s$ and the real part of
$f_s$, $f_p$ and $f_d$. We also show the $f_p$ contribution
from the second term in $\bar F$.
(b) The partial wave expansion of $G$.
Plotted are the real part of
$g_p$ and $g_d$ and the imaginary part of $g_p$.}
\end{figure}

The form-factors $F$ and $G$ can be decomposed in partial waves\footnote{The
partial wave expansion is not simply for $F$ and $G$ since the components
with a well defined $L_z=0,\pm1$ need to be expanded
\cite{cabibbo,PT}.}
 as \cite{cabibbo,PT}
\ba
\label{partial}
F(s_\pi,t_\pi,u_\pi,s_\ell ) &=& 
\sum_{l=0}^\infty P_l (\cos \theta_\pi) f_l(s_\pi,s_l) 
- \frac{\sigma_\pi P L}{X} \cos \theta_\pi G(s_\pi,t_\pi,u_\pi,s_\ell )
\nonumber\\
G(s_\pi,t_\pi,u_\pi,s_\ell )&=& \sum_{l=1}^\infty P_l'(\cos \theta_\pi)
 g_l(s_\pi,s_\ell )\,,
\ea
where
\be
P_l'(z) = \frac{d}{dz} P_l(z)\,.
\ee
$P_l(x)$ are the Legendre polynomials and the functions
$f_l(s_\pi,s_\ell )$, $g_l(s_\pi,s_\ell )$
have as phase the phase-shifts $\delta^I_l(s_\pi)$ from
$\pi$-$\pi$ scattering with isospin $I=0$ for even angular-momentum $l$, and 
$I=1$ for odd $l$. 
We perform this expansion numerically for the full $F$ and $G$ expressions.
The results are plotted in Figs. \ref{figFpartial}a
and \ref{figFpartial}b.

The $d$-waves are very small and the $f$-waves are not visible on
this scale. The contribution to $f_p$ coming from the
term with $G$, $f_p(G)$ is dominant. The difference between $f_p$ and $f_p(G)$
is negligible with the present data as was also found by \cite{Rosselet}.
$f_s$ has an appreciable imaginary part. The imaginary part of $f_p$
is also dominated by the part from $G$. The imaginary part for the other
waves is negligible.
Similarly we see that for $G$ the $d$-waves are very small. The $g_f$ and
higher
waves are not visible on this scale and only $g_p$ has
a visible imaginary part.

\begin{figure}
\begin{center}
\parbox{0.49\textwidth}{
\epsfig{file=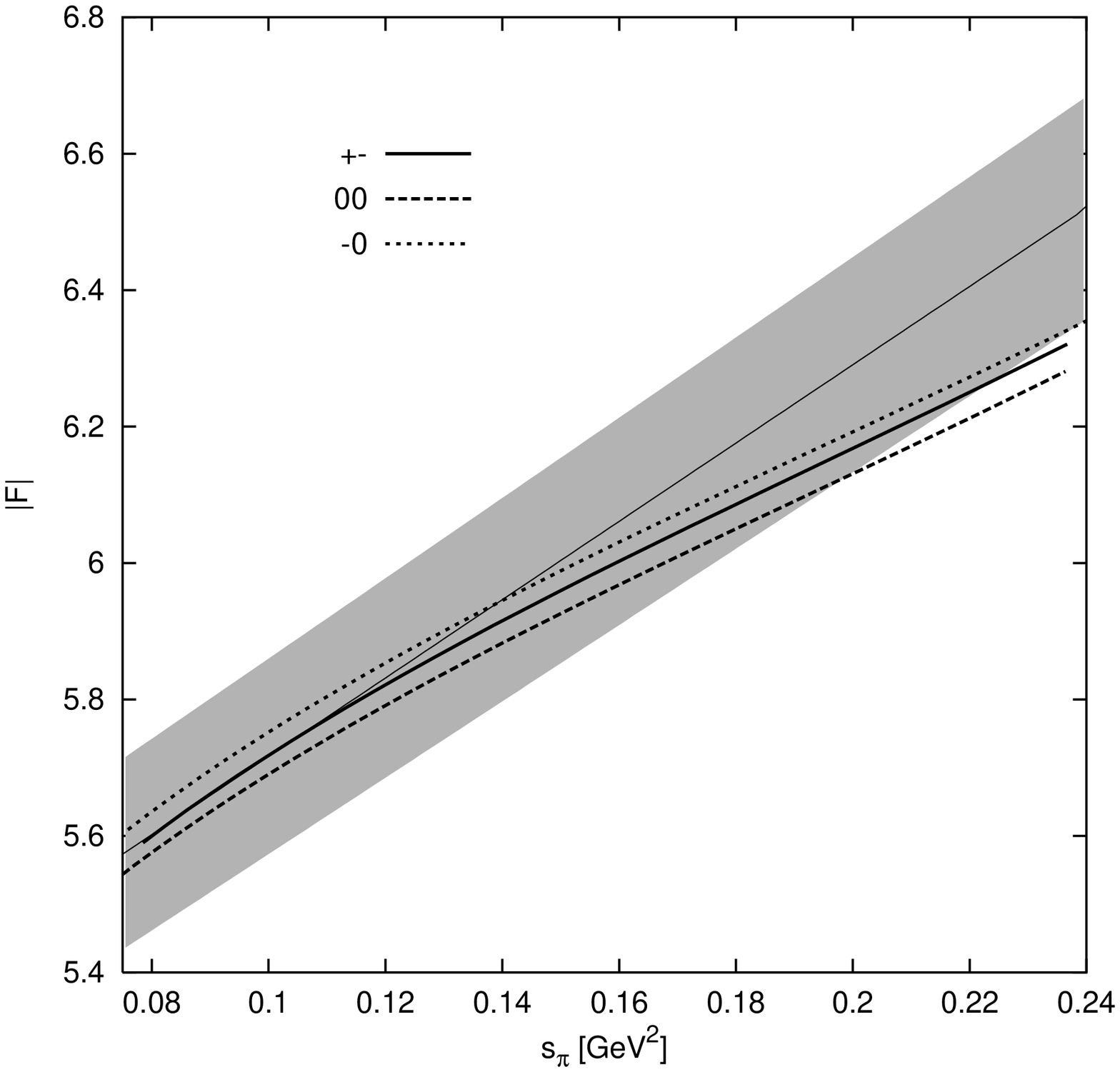,height=0.485\textwidth,angle=-90}
\begin{center}(a)\end{center}}
\parbox{0.49\textwidth}{
\epsfig{file=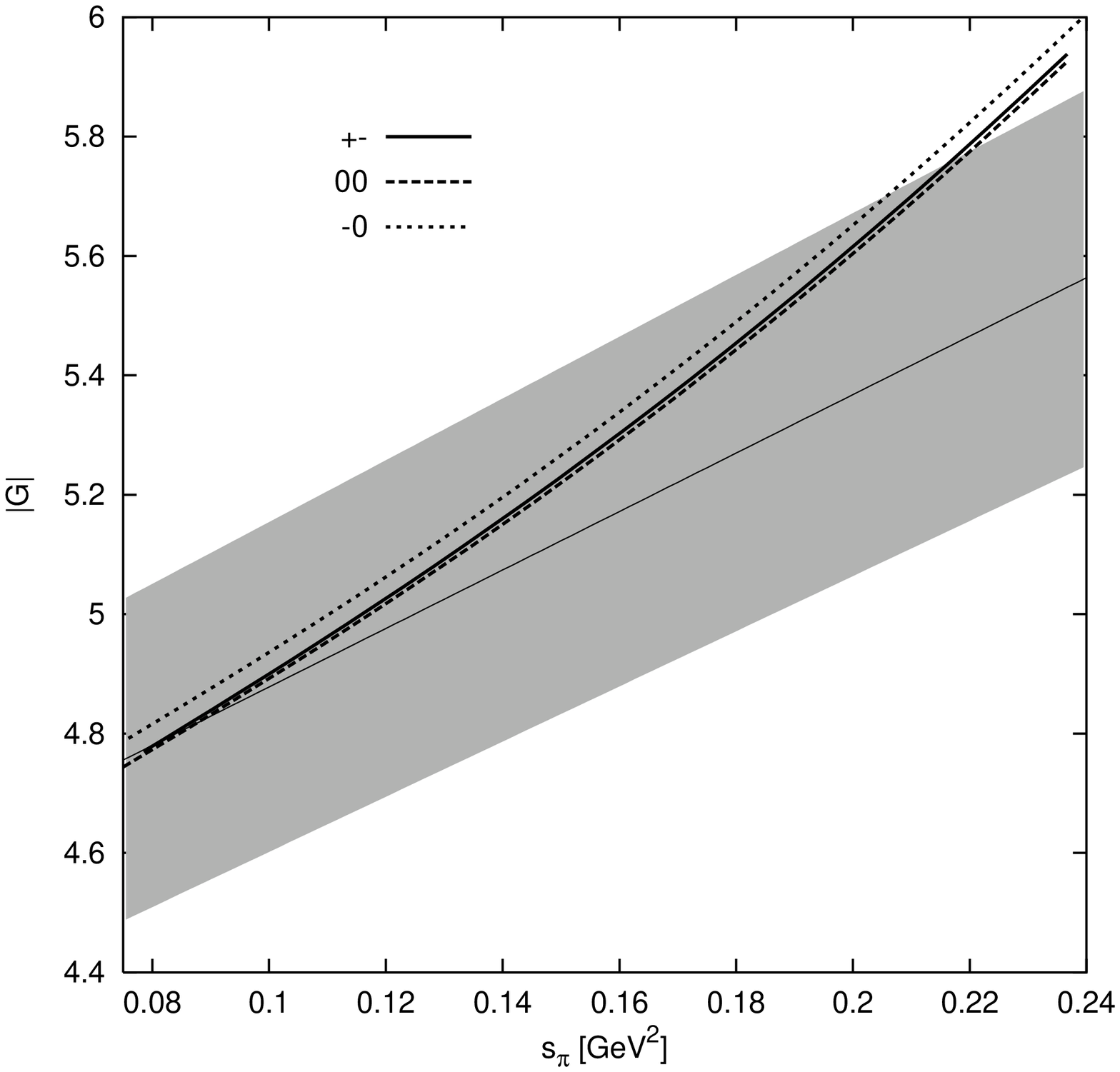,height=0.485\textwidth,angle=-90}
\begin{center}(b)\end{center}}
\end{center}
\caption{\label{figiso}
The absolute value form-factors calculated with the different masses
used for the three possible $K_{e4}$ decays.
The labels indicate the charges of the pion.
The curves shown are with $\cos\theta_\pi=s_\ell=0$.
 (a) $|F|$ (b) $|G|$. The experimental results are indicated with the grey
bands.}
\end{figure}

We can discuss one possible isospin breaking effect. As described
for the parametrization in Sect. \ref{tworesults}
we use slightly different masses for the three $K_{e4}$ decays.
The effect this has on $F$ and $G$ is plotted in Fig. \ref{figiso}.
It is obvious that with the present experimental accuracy this
source of isospin breaking is fully negligible.

\subsubsection{A Parametrization for $K^+ \to \pi^+ \pi^- e^+ \nu$ }
\label{kl4parametrization}

In \cite{AB} two of us proposed a parametrization to be used
to extract form-factors and phase-shifts from $K_{\ell 4}$ data.
Here we check how well the approximations inherent
in the parametrization are satisfied by our full two-loop
calculation.

The proposed parametrization was
\ba
\label{parametrization}
F &=& (\tilde f_s +\tilde f_s^\prime s_\pi +\tilde f_s^{\prime\prime}
+\tilde f_1 s_\ell )
       e^{i\delta_0^0(s_\pi)}
+\tilde f_p \sigma_\pi X \cos\theta_\pi e^{i\delta_1^1(s_\pi)}\,,
\nonumber\\
G &=& (\tilde g_p+\tilde g_p^\prime s_\pi+\tilde g_1 s_\ell )
 e^{i\delta_1^1(s_\pi)}
+\tilde g_d \sigma_\pi X \cos\theta_\pi e^{i\delta^0_2(s_\pi)}\,.
\ea
We can now fit these parametrizations to the partial waves obtained
in the previous section. The tilde notation is because these quantities
are related to the partial-wave expansion but not identical. 

\begin{figure}
\begin{center}
\epsfig{file=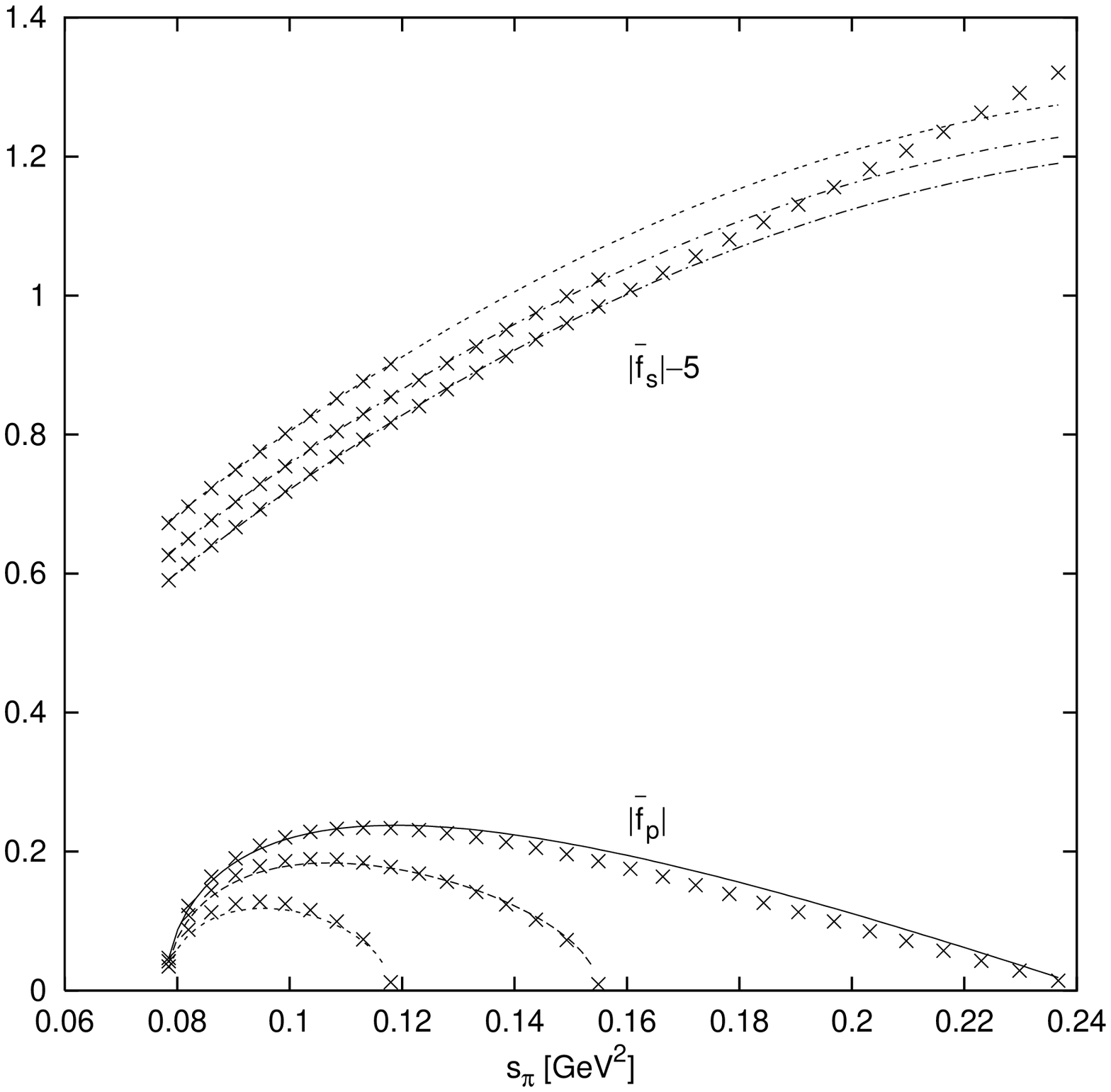,height=0.48\textwidth,angle=-90}
\epsfig{file=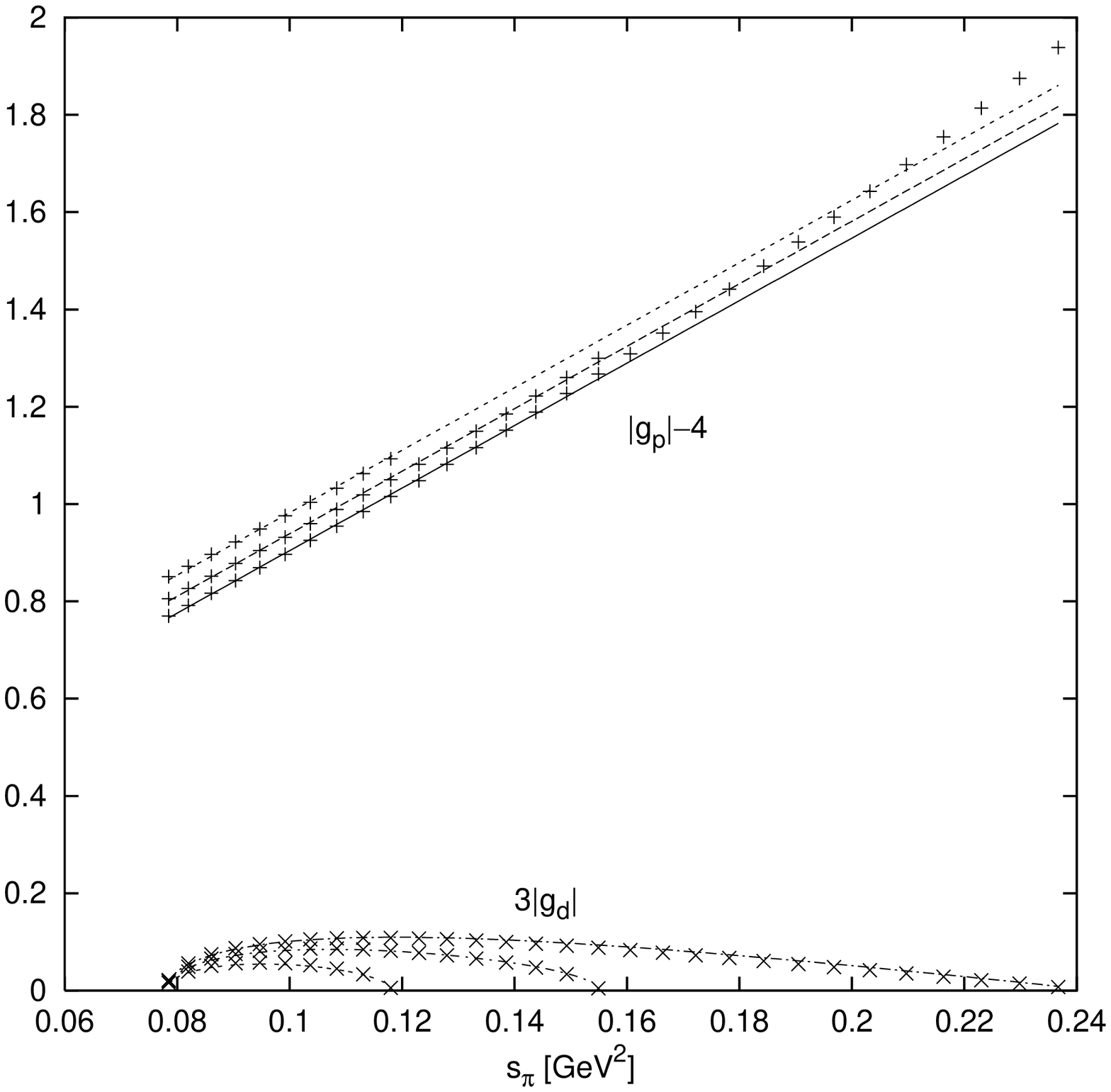,height=0.48\textwidth,angle=-90}
\end{center}
\caption{\label{figfitparam} The agreement of the parametrization with our full
two-loop result. The points are the full result.
The lines are the parametrization. We have shown $s_\ell=0,0.01,0.0225$
GeV$^2$.
Notice that for comparison purposes, the main parts have been 
strongly shifted downwards. The 
fits are performed for $s_\pi<0.16$~GeV$^2$ only.}
\end{figure}

As foreseen in \cite{AB} the parametrization works extremely well.
We have plotted in Fig. \ref{figfitparam} how well
the absolute value of the $s$ and $p$-waves of $F$, $\overline{f}_s$
and $\overline{f}_p$,
and the absolute value of the $p$ and $d$-waves of $G$ are parametrized
by Eq. (\ref{parametrization}). The higher partial waves are completely
negligible within the experimental precision expected in the near future.
The fits are performed for $s_\pi<0.16$~GeV$^2$ only, since that is the region
where
most events of the data will be.

\subsection{Partial Wave Expansion and Threshold Parameters 
for $\pi$-$\pi$ Scattering Amplitudes}
\label{pipi}

With the main fit result for the ${\cal O}(p^4)$ low-energy constants,
we can discuss how the threshold parameters and phase-shifts
in $\pi$-$\pi$ scattering are affected with
respect to the analysis performed in 
\cite{pipi2} and  \cite{Girlanda:1997ed}. For this purpose one expands the
$\pi$-$\pi$ amplitude as combinations of definite isospin amplitudes
in the s-channel
\ba
T^0(s,t)& =& 3 A(s,t,u)+A(t,u,s)+A(u,s,t)\,,\nonumber\\
T^1(s,t)& =& A(t,u,s)-A(u,s,t)\,,\nonumber\\
T^2(s,t)& =& A(t,u,s)+A(u,s,t)\,.
\ea
These isospin amplitudes are expanded into partial waves 
\be
T^I(s,t) = 32 \pi \sum_{I=0}^\infty (2 l +1)P_l(\cos\theta) t_l^I(s)\,,
\ee
with $s,t$ and $u$ the usual Mandelstam kinematical
variables, and $\theta$ is the scattering angle in the center of mass frame.
Unitary implies that
\be
\label{phaseshifts}
t^I_l(s) = \Big( \frac{s}{s-4m_\pi^2}\Big)^{1/2} 
\frac{1}{2i} \{e^{2 i \delta^I_l}-1\}
\ee
holds in the elastic region $4 m_\pi^2 \leq s \leq 16 m_\pi^2$
with real phase-shifts $\delta^I_l(s)$.
$s= 4(m_\pi^2+q^2)$ and $q$ is the center-of-mass three-momentum of
the pions. An expansion near threshold in Eq. 
(\ref{phaseshifts}) gives
\be
\label{ex}
{\cal R}e [t^I_l(s)] = q^{2l} \{a_l^I+q^2 b_l^I + {\cal O}(q^4)\}\,.
\ee
This last expansion defines the threshold parameters,
the scattering lengths $a_l^I$ and
the slopes $b_l^I$.

The $\pi$-$\pi$ scattering amplitude is known up to ${\cal O}(p^6)$
\cite{pipi2} --see also \cite{Knecht} for a dispersive calculation-- in 
two-flavour CHPT. Expanding this amplitude into partial
waves and in $q^2$, allows to obtain the threshold parameters  
in terms of the low-energy two-flavour ${\cal O}(p^4)$ constants, $l_i^r$. 
We can study the variations of these parameters with the $L_i^r$ of 
this work, Eq.~(\ref{newfit}). However the two-loop relations between
two- and three-flavour constants is unknown. 
A full matching would need the knowledge of processes involving the
relevant constants in both CHPT versions and to the same order, 
this is at present not the case. 
We use as a first estimate the relations obtained from a 
${\cal O}(p^4)$ matching~\cite{GL1} 
\ba
\label{relations2}
l_1^r& =& 4 L_1^r+2 L_3^r-\frac{1}{24} \nu_K +{\cal O}(p^6) \,, \nonumber\\
l_2^r& =& 4 L_2^r-\frac{1}{12} \nu_K+{\cal O}(p^6) \,, \nonumber\\
l_3^r& =& -8 L_4^r-4 L_5^r+16 L_6^r+8 L_8^r
- \frac{1}{16} \nu_\eta +{\cal O}(p^6)\,,\nonumber\\
l_4^r& =& 8 L_4^r+4 L_5^r-\frac{1}{2} \nu_K +{\cal O}(p^6)\,,
\ea
with
\be
l_i^r = \frac{\gamma_i}{32 \pi^2}\left\{\overline{l_i} 
+\ln (\frac{m_\pi^2}{\mu^2}) \right\}
\, \, \,  \mbox{and}  \, \, \,
\nu_P = \frac{1}{32 \pi^2} \left\{ \ln (\frac{m_P^2}{\mu^2}) +1 \right\}\,.
\ee
We want to emphasize that the relations
in Eq.~(\ref{relations2}) are affected by
higher --${\cal O}(p^6)$-- chiral corrections which might be
quite sizeable.
Bearing this last issue in mind, in Table \ref{parameters} 
we compare previous results and the experimental 
data \cite{Dumbrajs:1983jd} 
with the scattering lengths and slopes obtained with the
values
\be
\label{kl4ls}
\bar l_1= 0.40\pm2.40\,,\quad \bar l_2 = 4.90\pm1.0\,,\quad
\bar l_3 = 2.61^{+1.9}_{-2.4}\,,
\quad \bar l_4 = 4.05\pm0.18\,.
\ee
using the $L_i^r$ of Eq.~(\ref{newfit}) and Eq.~(\ref{relations2}). 
The errors are given by projections on
the relevant variable of the 68\% confidence level domain.

\begin{figure}
\begin{center}
\epsfig{file=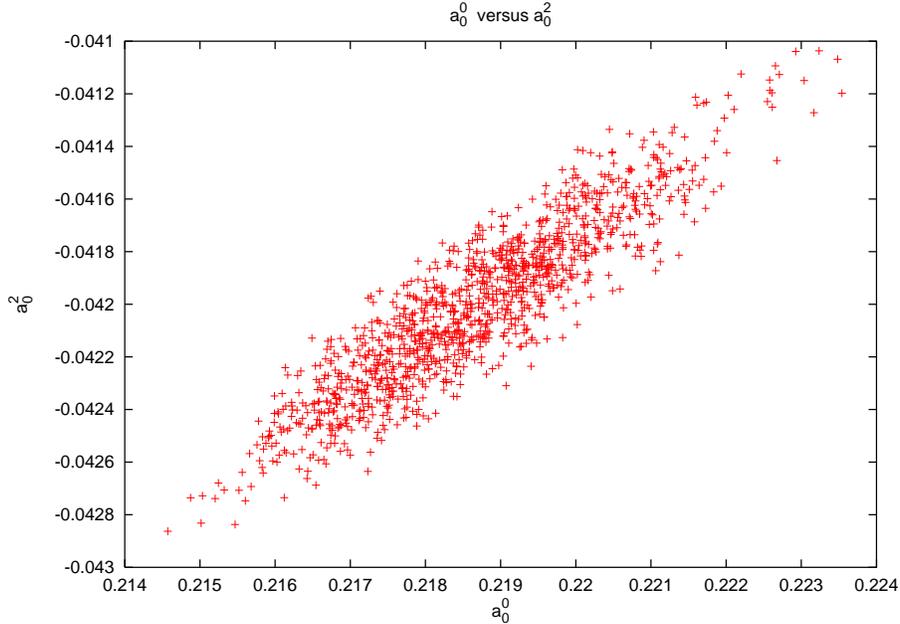,angle=-90,width=12cm}
\end{center}
\caption{\label{figa0a2} The prediction for the $a^0_0$ and $a^2_0$
scattering lengths for the set of 
$L_i^r$ distributed according to the correlations
of our main fit. Only points within the 68\% confidence level domain are
plotted.}
\end{figure}

In Figure \ref{figa0a2} we plot the predictions of $a^0_0$ and $a^2_0$ 
for the the set of
$L_i^r$ distributed according to the $\chi^2$ function as discussed
earlier. The projection
of the filled ellipse on the axis produces the 
errors in Table \ref{parameters}.

\begin{table}
\caption{ \label{parameters} Threshold parameters
in units of $m_{\pi^+}$. 
Set I is $\overline{l}_1=-1.7$ and $\overline{l}_2=6.1$ while 
Set II is $\overline{l}_1=-1.5$ and $\overline{l}_2=4.5$. Both sets
have $\overline{l}_3=2.9~\mbox{and}~\overline{l}_4=4.3$. Experimental values
are taken from \cite{Dumbrajs:1983jd}. Errors are from the
68\% confidence level ranges only.}
\begin{center}
\vspace{.5cm}
\begin{tabular}{c||c|c|c|c|c|c}\hline

       & Eq. (\ref{newfit}) & ${\cal O}(p^2)$ & ${\cal O}(p^4)$ &
        ${\cal O}(p^6)$ Set I & ${\cal O}(p^6)$ Set II & experiment\\ \hline

$a^0_0$     &$0.219\pm0.005$&$0.159$&$0.203$ &$0.222$&$0.212$&$0.26\pm0.05$\\
$b^0_0$     &$0.279\pm0.011$&$0.182$&$0.254$ &$0.282$&$0.256$ &$0.25\pm0.03$\\
$-10 a^2_0$ &$0.420\pm0.010$&$0.454$&$0.423$ &$0.420$&$0.450$ &$0.28\pm0.12$\\
$-10 b^2_0$ &$0.756\pm0.021$&$0.908$&$0.755$ &$0.729$&$0.81 $ &$0.82\pm0.08$\\
$10 a^1_1$  &$0.378\pm0.021$&$0.303$&$0.356$ &$0.404$&$0.38 $ &$0.38\pm0.02$\\
$10^2 b^1_1$&$0.59\pm0.12$  &$0$    &$0.31$  &$0.83 $&$0.56 $ &  --\\
$10^2 a^0_2$&$0.22\pm0.04$  &$0$    &$0.16$  &$0.28 $&input   & $0.17\pm0.03$\\
$-10^3 b^0_2$&$0.32\pm0.10$ &$0$    &$0.40$  &$0.24 $&$0.27 $ & --\\	     
$10^3 a^2_2$&$0.29\pm0.10$  &$0$    &$0.32$  &$0.24 $&input   & $0.13\pm0.30$\\
$-10^3 b^2_2$&$0.36\pm0.9$  &$0$    &$0.23$  &$0.33 $&$0.22 $ & --\\
$10^4 a^1_3$&$0.62\pm0.11$  &$0$    &$0.20$  &$0.65 $&$0.48 $ & --\\
$-10^4 b^1_3$&$0.36\pm0.02$  &$0$   &$0.15$  &$0.38 $&$0.37 $ & --\\
\hline
\end{tabular}
\end{center}
\end{table}

\begin{figure}
\begin{center}
\epsfig{file=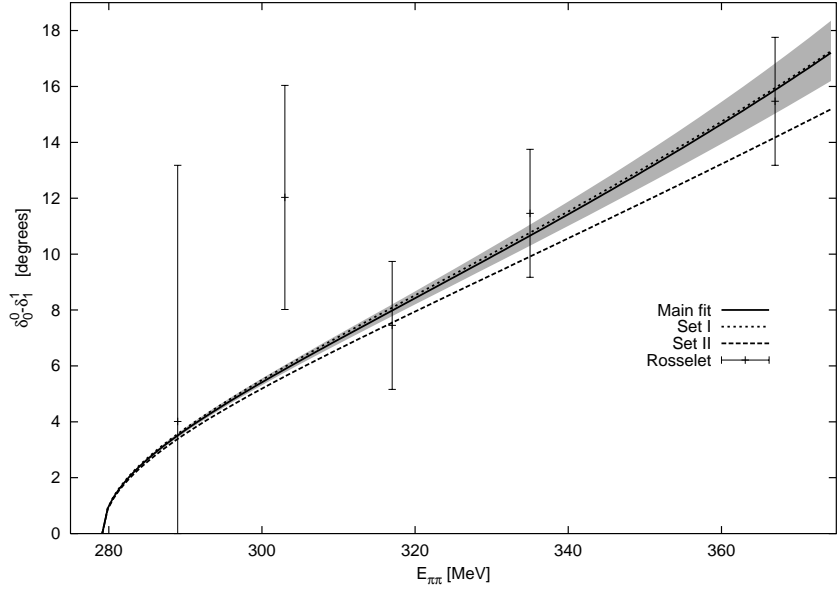,width=8cm,angle=-90}
\end{center}
\caption{\label{figd0d1} 
Difference of phase-shifts $\delta_0^0-\delta_1^1$ plotted as a function
of the center of mass energy $E_{\pi\pi}$. Data
points are taken from \cite{Rosselet}.
The shaded band is the 68\% confidence level range.}
\end{figure}

\begin{figure}
\begin{center}
\epsfig{file=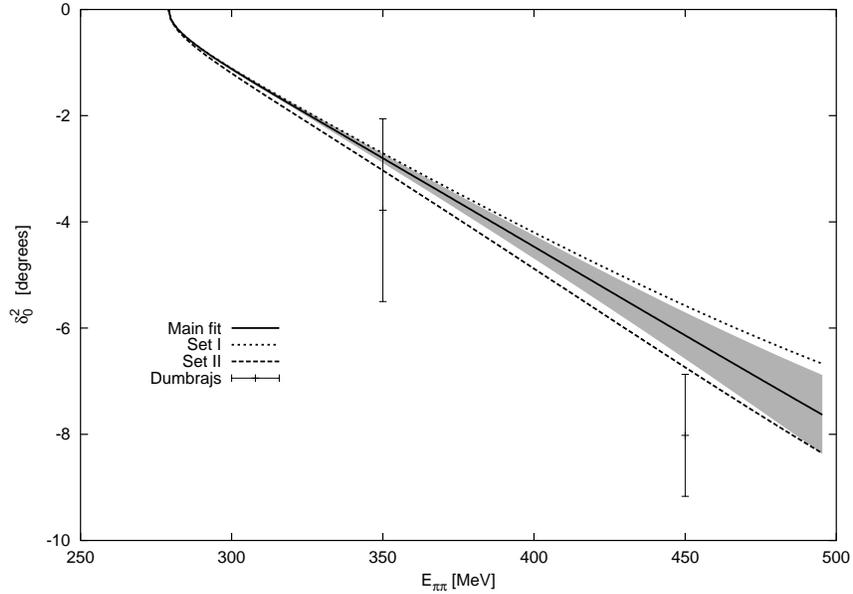,width=8cm,angle=-90}
\end{center}
\caption{\label{figd02} 
Phase-shift $\delta^2_0$ as function of the center of mass energy
$E_{\pi\pi}$. Data points are taken from \protect \cite{Hoogland}.
The shaded band indicates the 68\% confidence level range.}
\end{figure}

The interesting result that comes out from Table \ref{parameters}
is that with
the large difference in the input parameters, $\bar l_i$, 
the changes obtained with Eq. (\ref{newfit})
barely shift the results of set I \cite{pipi2}--with exception of $b^1_1$,
$a^0_2$ and $b^0_2$.
As was shown in \cite{pipi2},
that is due to the fact that the contribution from
the double logarithms almost saturates the result to ${\cal O}(p^6)$. 
However the new experiments 
are expected to have enough accuracy to distinguish small changes in the
scattering lengths and slopes.
The results of set I and II differ from 
\cite{pipi2} because
of the different value of $F_\pi$ and we have kept some ${\cal O}(p^8)$
parts which were dropped in the numerics there. Similar remarks
apply to Figures \ref{figd0d1} and \ref{figd02}.

The threshold parameters which were not listed in \cite{pipi2}
were calculated by doing the relevant expansions in the formulas
there to higher order, see App. \ref{scattering}.

For completeness we also display in Table \ref{bs} the value of the
variables $b_i$ defined in
\cite{pipi} as a finite combination of the two-flavour low-energy constants
and which determines the
polynomial piece in the $\pi$-$\pi$ scattering at ${\cal O}(p^6)$.

\begin{table}
\caption{ \label{bs} The $b_i$ quantities as defined in \cite{pipi}.
Set I and Set II are defined in the caption of Table \ref{parameters}.
Errors are the projections of the 68\% confidence level domain only.}
\begin{center}
\vspace{.5cm}
\begin{tabular}{c||c|c|c} \hline
               & Eq. (\ref{newfit}) & 
               ${\cal O}(p^6)$ Set I & ${\cal O}(p^6)$ Set II \\
\hline
$10^2 b_1$     & $-6.56\pm2.10$ & $-9.14$ & $-8.65$  \\
$10^2 b_2$     & $6.50\pm2.00$  & $8.88$  & $8.03$  \\
$10^3 b_3$     & $-1.27\pm3.20$ & $-4.33$ & $-2.64$  \\
$10^3 b_4$     & $5.44\pm1.25$  & $7.08$  & $4.83$  \\
$10^4 b_5$     & $2.32\pm1.20$  & $2.32$  & $-0.37$  \\
$10^4 b_6$     & $1.16\pm0.23$  & $1.49$  & $0.83$  \\
\hline
\end{tabular}
\end{center}
\end{table}

In Fig.~\ref{figd0d1} we show the difference of phase-shifts
$\delta_0^0-\delta_1^1$ as a function 
of the center of mass energy. This is the combination measured
in $K_{\ell 4}$ decays.
As can be appreciated the agreement between Eq. (\ref{kl4ls})
and set I is excellent and also with the
available data.

In Fig.~\ref{figd02} we plot
the $I=2$ $s$-wave phase-shift $\delta^2_0$ as function of the 
center of mass energy together with available 
experimental data. The phase-shift corresponding to
Eq. (\ref{kl4ls}) just interpolates between both sets, I and II.

All in all, the conclusions that can be drawn
from the $\pi$-$\pi$ scattering analysis is that
good agreement is obtained for all the threshold parameters and 
with $\delta_0^0-\delta_1^1$ and a substantial improvement 
with the data in the $\delta_0^2$ phase-shift case.

\subsection{Vacuum Expectation Values}
\label{vev}

The lowest-order effective Lagrangian of QCD is not determined by
the pion decay constant alone, but involves a second
low-energy constant, $B_0$, which is intimately related with the
vacuum expectation values of the scalar densities in the chiral limit
\be
\label{chirallimit}
\langle 0 \vert \overline{q}^i q^j \vert 0 \rangle = - F_0^2 B_0 \delta^{ij}\,.
\ee
In the absence of external scalar and pseudoscalar fields it is only
observable in combination with the quark masses. It is, however, a quantity
of theoretical interest and thus deserves study at the two-loop
level as well. The value of $B_0$ depends on the way the scalar quark bilinear
is defined in QCD. In particular it depends on the QCD renormalization
scale $\mu_{\mbox{\small QCD}}$.

The calculation of the one-point scalar Green function
without any external pseudoscalar field is rather
simple as it only involves one-loop$\times$one-loop diagrams at
${\cal O}(p^6)$. The results are displayed in App. \ref{vacuum} in terms
of the renormalized quantities, masses and decay constants.
The procedure is the same as described in Sect. \ref{technique}.
In fact, the contributing diagrams
form only a small subset
of those entering in the decay constant evaluation \cite{Amoros}.
This amounts to 
consider diagrams with one vertex involving the external field $s$
part of the matrix $\chi$ defined in Sect. \ref{CHPTsection}.

We write the full ${\cal O}(p^6)$ result as
\be
\langle 0 \vert \bar{q}q \vert 0 \rangle =
-F_0^2 B_0 \Big\{ 1 + \langle 0 \vert \bar{q}q \vert 0 \rangle^{(4)}
+ \langle 0 \vert \bar{q}q \vert 0 \rangle^{(6)} + \ldots \,, \Big\}\,.
\ee
Note that $\langle 0 \vert \bar{q}q \vert 0 \rangle^{(4)}$ and 
$\langle 0 \vert \bar{q}q \vert 0 \rangle^{(6)}$ are scale independent. 
The soft $SU(3)$ breaking, due to the quark masses,  introduces an additional 
dependence on the high-energy constant $H_2^r$, whose value also
depends on the precise definition of the quark bilinear in QCD.
This ambiguity is forced by the need of a
subtraction in the scalar two-point function at zero separation.
I.e. at the first order in perturbation theory
\be
\langle 0 \vert \bar{q}q \vert 0 \rangle = 
\langle 0 \vert \bar{q}q \vert 0 \rangle \Big\vert_{\mbox{\small chiral~limit}}
 - i \int d^4x \langle 0 \vert :(\bar{q}{\cal M} q)(x)(\bar{q}q)(0):
 \vert 0\rangle+{\cal O}({\cal M}^2)\,,
\ee
which blows up as $x \to 0$.
In contrast \cite{novikov} the non-analytic contribution is
unambiguous. As a last point 
we mention that as QCD predicts \cite{shifman} the vacuum condensate
increases once the quark masses are switched on.

In order to quote results we need to make certain assumptions
on the value of $H_2^r$. We use the scalar dominance assumption that
leads to
\be
H_2^r = 2 L_8^r\,,
\ee
with the value of $L_8^r$ that follows from Eq.~(\ref{newfit}).
The numerical results are
\ba
\label{vevvalues}
\langle0\vert\overline{u}u\vert 0\rangle &=& -B_0 F_0^2
\left[1+0.277+0.103\right]\,,
\nonumber\\
\langle0\vert\overline{s}s\vert 0\rangle &=& -B_0 F_0^2
\left[1+0.832+0.247\right]\,,
\ea
where the numbers correspond to lowest-order, ${\cal O}(p^4)$ and 
${\cal O}(p^6)$ respectively.

\begin{figure}
\begin{center}
\epsfig{file=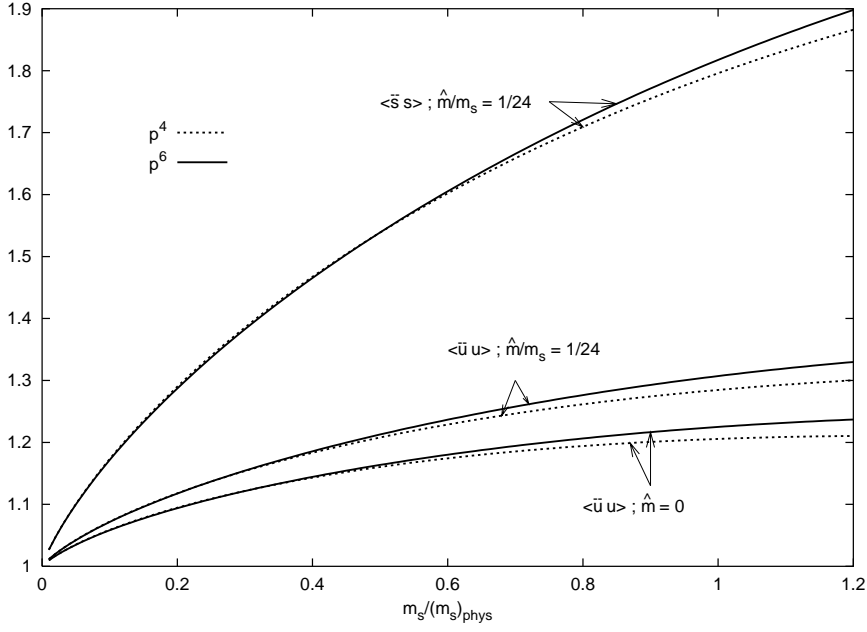,height=12cm,angle=-90}
\end{center}
\caption{\label{figvev} The vacuum expectation values (VEV) as
a function of $m_s/(m_s)_{\mbox{\small phys}}$. Plotted are the strange
and light quark VEV for the ratio $m_s/\hat m=24$ and the
light quark VEV for $\hat m=0$. The curves are normalized to the chiral limit.}
\end{figure}

We can also see how the vacuum expectation value changes as
a function of the quark masses. For this we can use for
$B_0\hat m$, $B_0m_s$ and $F_0$ the values derived using the main fit and the
higher order formulas expressed in terms of real pseudoscalar masses
and $F_\pi$. We then insert these values into the expressions for
the vacuum expectation values in terms of the unrenormalized masses
and decay constants and vary $B_0\hat m$ and $B_0 m_s$.
In Fig. \ref{figvev} we plotted the vacuum expectation values
as a function of $m_s/(m_s)_{\mbox{\small phys}}$. The difference
with the values in Eq. (\ref{vevvalues}) comes from higher orders.

\subsection{Masses and Decay Constants}
\label{massessect}

\begin{figure}
\parbox{0.49\textwidth}{
\epsfig{file=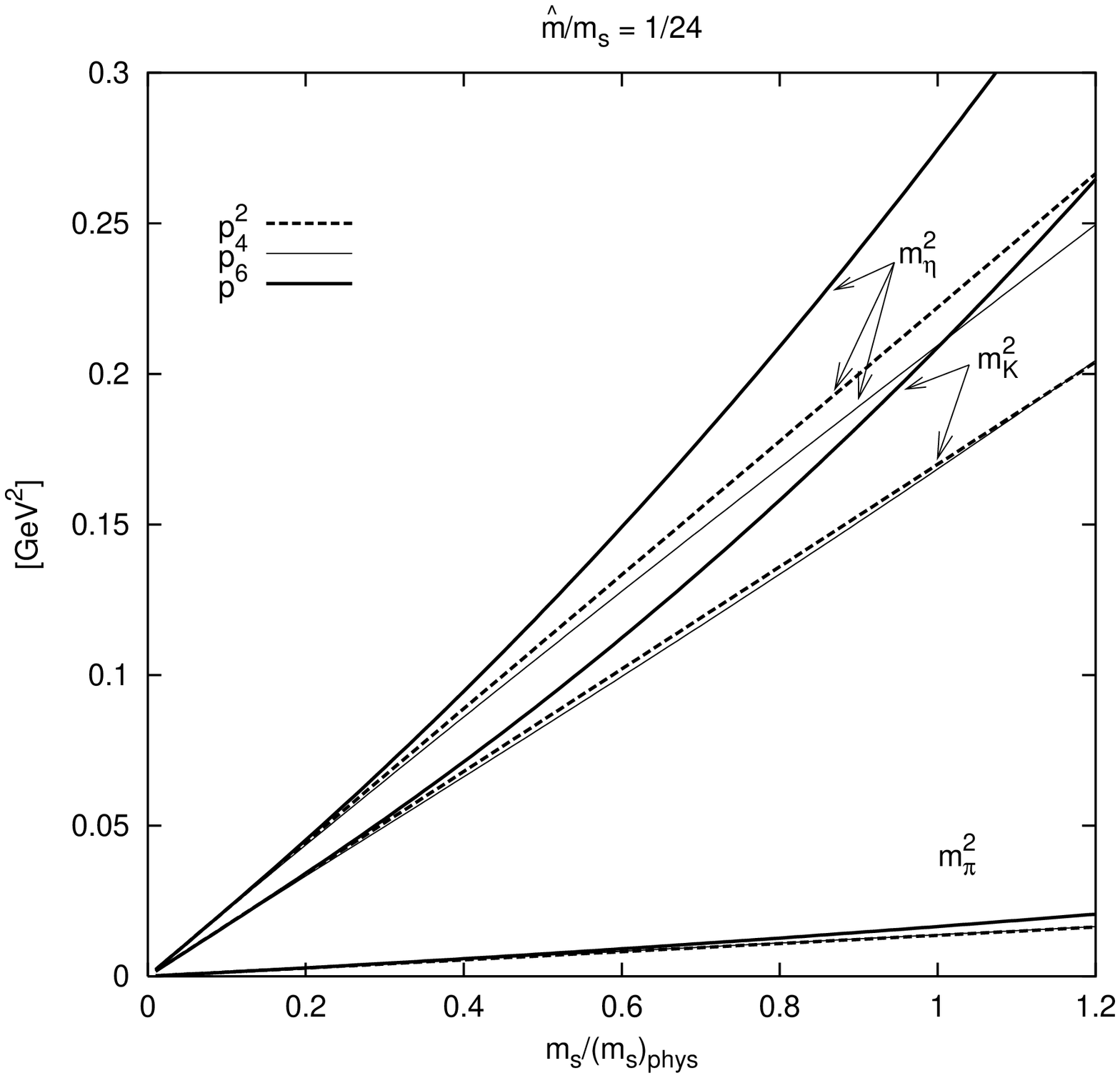,height=0.485\textwidth,angle=-90}
\begin{center}(a)\end{center}}
\parbox{0.49\textwidth}{
\epsfig{file=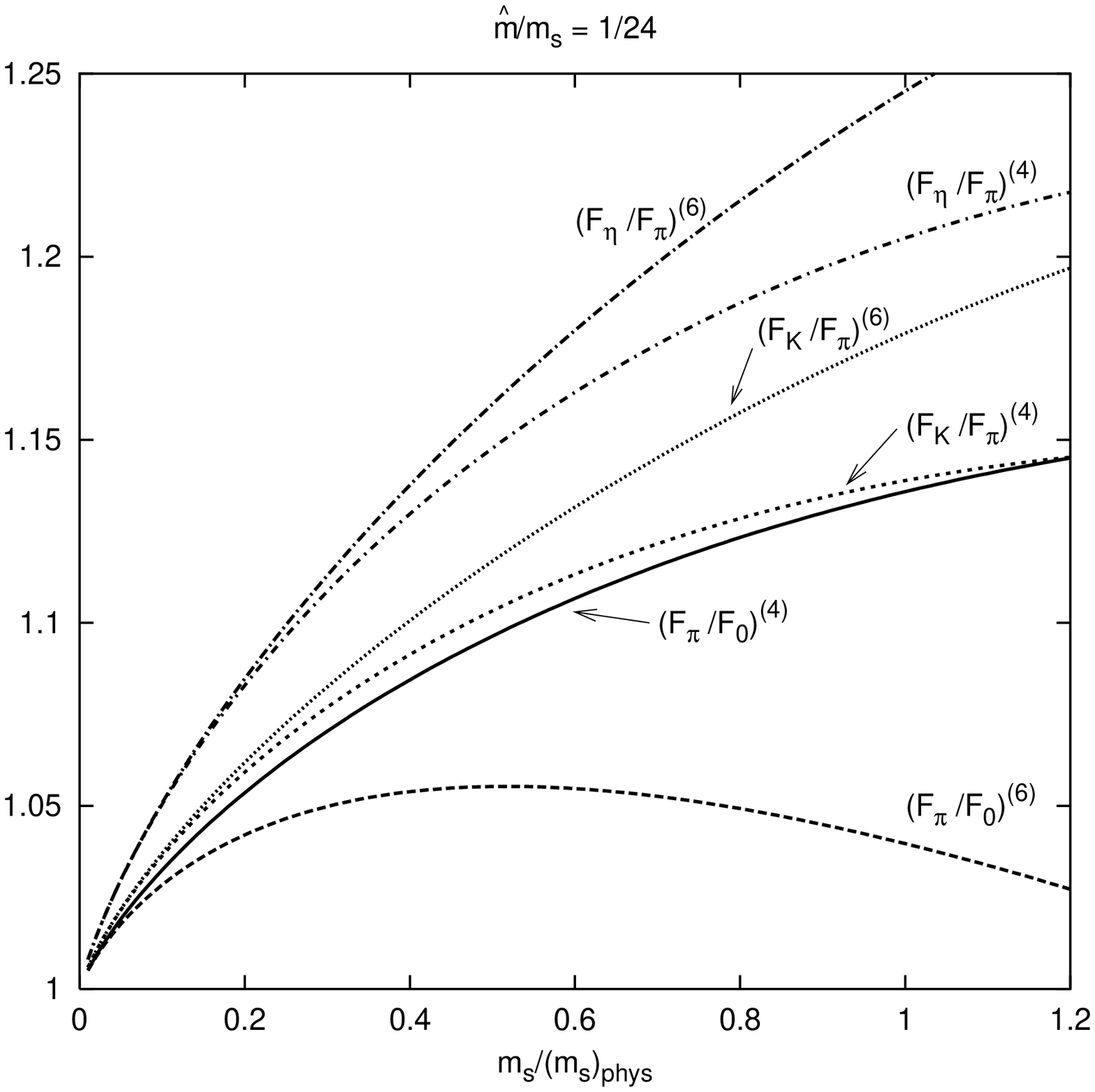,height=0.485\textwidth,angle=-90}
\begin{center}(b)\end{center}}
\caption{\label{figdecay}
(a) The pseudoscalar masses as a function of
the quark mass at fixed ratio $\hat m/m_s$. The results
up to ${\cal O}(p^2)$, ${\cal O}(p^4)$ and ${\cal O}(p^6)$ 
are shown for the three pseudoscalar
masses.
(b) The pseudoscalar decay constants as a function of
the quark mass at fixed ratio $\hat m/m_s$. The results
up to ${\cal O}(p^4)$ and ${\cal O}(p^6)$ are shown for the three pseudoscalar
decay constants. For the $\eta$ it is the coupling to the octet axial current.}
\end{figure}

In this section the pseudoscalar masses are studied using 
the new values for the LEC, Eq.~(\ref{newfit}). 

All masses are written as in Eq. (\ref{masses}) with the higher orders
expressed in terms of $F_\pi$ and the physical masses, \cite{Amoros}. 
The separate contributions for them are 
\ba
\label{resultmass}
m_{\pi}^2/(m_{\pi}^2)_{\mbox{\small phys}}&=&
0.746+0.007+0.247\,,
\nonumber\\
m_{K}^2/(m_{K}^2)_{\mbox{\small phys}}&=&
 0.695+0.019+0.286\,,
\nonumber\\
m_{\eta}^2/(m_{\eta}^2)_{\mbox{\small phys}}&=&
 0.742-0.040+0.298\,.
\ea
where the numbers are the lowest-order,  ${\cal O}(p^4)$ and
${\cal O}(p^6)$ contribution respectively.

We can now also check how the pseudoscalar masses vary with the quark masses.
For this purpose we rewrite the pseudoscalar masses as
\be
m^2_{phys} = m_0^2+ (m^2)^{(4)}+(m^2)^{(6)}    
\ee
where $(m^2)^{(i)}$
is a function of $B_0\hat{m}$, $B_0 m_s$ and $F_0$ is defined through 
\be
F_\pi = F_0 \left( 1 + F^{(4)} + F^{(6)} \right) \,.
\ee
Plugging in the values following from the lowest-order contribution
in Eq. (\ref{resultmass}) we can now vary them at a fixed ratio
$m_s/\hat m = 24$. The result is shown in Fig. \ref{figdecay}a.
The difference with the results in Eq. (\ref{resultmass}) is due
to higher orders.
We believe that the results using the physical decay constant and the
physical masses in the higher orders are numerically more reliable since
they incorporate part of the higher order corrections that we know
will appear.

As has been previously noticed \cite{Amoros,LEC}, Eq. (\ref{resultmass})
the light meson masses at 
${\cal O}(p^6)$ have a large contribution which seems
in conflict with the convergence of the chiral series.
The ${\cal O}(p^4)$ contribution is much smaller than the ${\cal O}(p^6)$ one.
This is due to two adding effects. First, there are cancellations that 
happen inside the ${\cal O}(p^4)$ contribution. E.g. for the Kaon mass the 
contribution
is given by 0.066($L_i^r$)$-$0.047(loops)=0.019(total $p^4$).
And second and mainly, the partial
contributions at ${\cal O}(p^6)$, 
$\log \times \log, L_i \times \log, L_i \times L_j$ and pure 
two-loop, all go in the same direction.
In other observables we typically had cancellations between these various
contributions.
The corrections to physically measurable quantities are in fact
not so large. Eq. (\ref{resultmass}) can be expressed as
\ba
\frac{m_K^2}{m_\pi^2} &=&  12.5  (p^2);~\;12.7 (p^4);~\;  13.4 (p^6)\,,
\nonumber\\
\frac{m_\eta^2}{m_\pi^2} &=&  16.3 (p^2);~\;15.3   (p^4);~\;16.4 (p^6)\,,
\ea
which shows that the corrections to the ratios are much smaller.

We have checked that relaxing some of the constraints used in the fits
can improve the above mentioned behaviour, one of these possibilities
is discussed in Sect. \ref{largeNc}. This does not qualitatively
change any of the predictions for decay constants and $K_{\ell4}$
form-factors made here. Other options are ad hoc choosing the values
of some of the $C_i^r$ to diminish the large ${\cal O}(p^6)$ corrections.
Some arguments for this can be found in
\cite{descotes} where it is argued that resonance saturation
seems to fail in the $0^{++}$ sector. 
Without new input we feel we cannot discuss further this question.

For the pseudoscalar decay constants we use the formulas
of \cite{Amoros} and obtain using our main fit
\ba
\label{numdecay}
F_\pi/F_0 &=&1+0.136-0.076\,,
\nonumber\\
F_K/F_\pi &=& 1+0.134+0.086\,,
\nonumber\\
F_\eta/F_\pi &=&1+0.202+0.069\,.
\ea
The $\eta$-decay constant here is the coupling of the physical $\eta$ to
the octet axial current.
The numerical result allows to obtain
\be
F_0 = 87~\mbox{MeV}\,.
\ee
Using this value of $F_0$ and the lowest-order masses obtained above we can
now plot the dependence of the decay constants on the quark masses.
We therefore rewrite the full formulas in terms of $F_0$ and the quark masses.
The results is shown in Fig. \ref{figdecay}b. Again, the difference
with Eq. (\ref{numdecay}) is higher order.

\subsubsection{Large $N_c$, Zweig Rule}
\label{largeNc}

In this subsection we relax some of our assumptions to see whether the unusual
relative size of the ${\cal O}(p^6)$ versus the  ${\cal O}(p^4)$
contribution in Eq. (\ref{resultmass}) can be avoided. 

We have performed fits with $m_s/\hat{m}=20$
and $m_s/\hat{m}=30$ but otherwise the same input as the main fit.
Neither of these changes the relative size of the contributions to the
masses qualitatively.

Another possibility is to relax
the accepted common wisdom about the large $N_c$ \cite{nc}
behaviour of the low-energy constants. In particular
we will focus on $L_4^r$ and $L_6^r$, 
recapitulating some of the reasoning
of \cite{GL1}.

In principle the main question is whether the corrections
of the relative order $m_s$ and $m_s^2$
to the ground state are small in comparison with the scale of the theory.
If the answer is no, clearly one can not trust large $N_c$ arguments in CHPT.
Large corrections would also cast doubt on CHPT
beyond one-loop calculations in the three-flavour sector. 

At ${\cal O}(p^4)$ the effects of $m_s$ on
$\langle 0 \vert \bar{u} u \vert 0\rangle$,
$F_\pi$ and $m^2_\pi$ are small if both $L_4^r$ and $L_6^r \sim 0$.
The main question is now
if higher chiral orders in those quantities  
are still protected against a  strong $m_s$ dependence.
As it is shown, there is a range
in $L_4^r$ and $L_6^r$ values
that allows large $N_c$ and CHPT at ${\cal O}(p^6)$ to coexist
without conflict.
Thus there is no reason to argue 
for large deviations from the Zweig rule.

Setting $U=\bf{1}$ in Eq. (\ref{l4}) the large $N_c$ counting comes
automatically, after using some trace identities
\cite{GL1}
\ba
{\cal O}(N_c) && \quad L_1, L_2, L_3, L_5, L_8, L_9, L_{10}, H_1
 ~\mbox{and}~ H_2\,, \nonumber\\
{\cal O}(1) && \quad 2 L_1-L_2, L_4, ~\mbox{and}~ L_6\,. \nonumber
\ea
The behaviour of the constant $L_7$ depends on the treatment
of the singlet $\eta_1$ \cite{GL1,eduard}.
In the constants of ${\cal O}(1)$ there are two different
behaviours: i) the
combination $2 L_1^r-L_2^r$ is scale independent, but 
ii) $L_4^r$ and $L_6^r$ are scale dependent. This last
dependence should cancel via the meson loops that explicitly
violate large $N_c$. In order to estimate possible sizes
we can just vary the scale
inside a reasonable range.
Using 
\be
\vert \Delta L_i^r \vert =
\vert L_i^r(\mu_2)-L_i^r(\mu_1)\vert=\left\vert\frac{\Gamma_i}{16 \pi^2} 
\ln\left(\frac{\mu_1}{\mu_2}\right)\right\vert
\ee
with the choice $\mu_1:\mu_2=2$, leads to
\be
\label{l4l6}
L_4^r = (0\pm 0.5) \cdot 10^{-3} 
~\mbox{and}~ L_6^r = (0\pm 0.3) \cdot 10^{-3}\,.
\ee
The central value correspond to the main
fit presented in Table \ref{Tableresults}, 
i.e. considering that
large $N_c$ is reached at the $\rho$ scale.
Fit 5 corresponds to use large $N_c$ at the
$\eta$ scale. Neither Eq. (\ref{newfit}), as shown in Eq. (\ref{resultmass}),
nor fit 5 have the expected ordering of ${\cal O}(p^4)$ and 
${\cal O}(p^6)$
contribution to the pseudoscalar masses.

Other constraints on $L_4^r$ and $L_6^r$ are
from a recent analysis \cite{descotes}
based on the positivity of the fermionic measure
giving a lower bound
$10^3 L_6^r(M_\rho) \geq 0.13$ and on
QCD sum-rules \cite{bachir} which allow a
band $0.2 \lapprox{10^3 L_6^r(M_\rho)} 
\lapprox{0.6}$. The latter reference also
quotes $10^3 L_4^r(M_\rho) \simeq 0.4$ with unknown errors.
The values and bounds from \cite{descotes,bachir} use ${\cal O}(p^4)$
expressions so these ranges 
 might well change when those calculations are performed
to ${\cal O}(p^6)$.

Let us now check for which ranges of $L_4^r$ and $L_6^r$ we find
results where {\em all} the quantities considered up to now
have an expansion of the form 
\be
\label{constraint}
X=X_2+X_4+X_6\quad\mbox{with}\quad |X_2|>|X_4|>|X_6|\,,
\ee
where the subscript refers to the chiral order.
This is for reasonable variations always the case for
$f_s(0)$, $g_p(0)$, $\lambda_f$, $\langle0\vert\bar{s}s\vert0\rangle$
and $F_\eta/F_\pi$. 
The ratios of the pseudoscalar masses always have small corrections even
if they do not always obey Eq. (\ref{constraint}).
We will refer to Eq. (\ref{constraint}) above as a convergent series.
For $\langle0\vert\bar{u}u\vert0\rangle$ $H_2^r$ has only a small
influence but it is very large for $\langle0\vert\bar{s}s\vert0\rangle$.
For this quantity we thus allow $|X_2|\lapprox|X_4|$ since this
can easily be changed by varying $H_2^r$.

The problematic quantities are $F_\pi/F_0$, $F_K/F_\pi$,
the pseudoscalar masses separately and the vacuum expectation value
$\langle0\vert\bar{u}u\vert0\rangle$.

\begin{figure}
\begin{center}
\epsfig{file=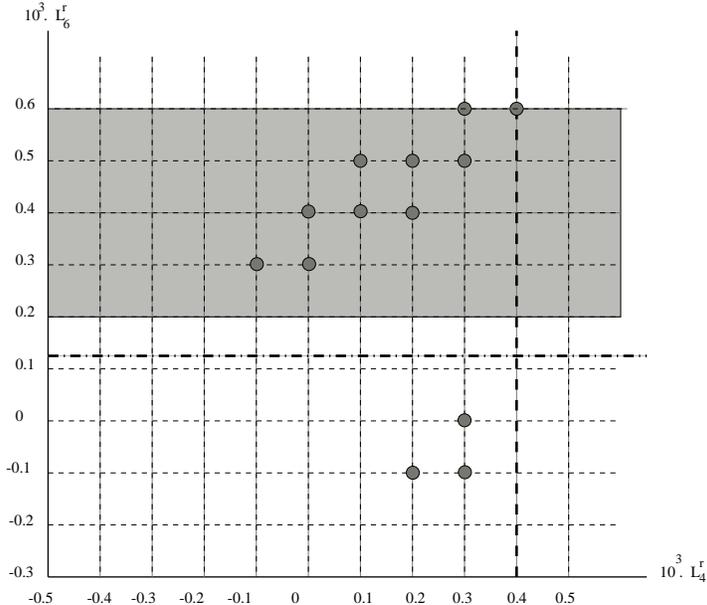,height=8cm,angle=0}
\end{center}
\caption{\label{l4l6plane} 
The $L_4^r$--$L_6^r$ plane. The grey circles are the points
where Eq. (\ref{constraint}) is satisfied for all quantities considered
in this manuscript. The grey band shows the bounds on $L_6^r$ from
QCD sum-rules.  The dash-dotted line is the lower bound that follows
from positivity constraints \cite{descotes}
and the thick dashed line the value of $L_4^r$ of Ref. \cite{bachir}.
Fits were only performed for points on the grid.}
\end{figure}

In Fig. \ref{l4l6plane} we 
show the allowed region in the plane $L_4^r-L_6^r$ consistent with
Eq. (\ref{constraint}) for all quantities\footnote{Quantities are
only calculated on the points of the grid.}.
The allowed bottom right region extends further to the right
but with $(F_K/F_\pi)_4 < (F_K/F_\pi)_6$ even though both are small.
As can be seen there is a region where 
Eq. (\ref{constraint}),
large $N_c$ constraints and the other
bounds are reasonably satisfied, taking into account possible higher
order corrections to the bounds.
Notice that a
large deviation from the Zweig rule is not needed thus the
bounds, Eq. (\ref{l4l6}), still hold.

Let us mention some general
features in the behaviour of these quantities as a function
of $L_4^r$ and $L_6^r$.
For instance, the decay constant has the broadest range (in
both parameters) of convergence, while the mass seems to be the most
restrictive. 
A low, negative value for both $L_4^r$ and $L_6^r$ is penalized
for the mass
and the vacuum expectation value.
If we increase $L_4^r$ fixing the value of $L_6^r$
the vacuum expectation value recover once more the convergent pattern. 
The mass behaviour is quite unexpected: large, absolute
values of $L_6^r$ reorder the
series making ${\cal O}(p^4) > {\cal O}(p^6)$,
but then ${\cal O}(p^4)$ starts
to compete with ${\cal O}(p^2)$ contribution.

\section{Summary}

Let us summarize our results.
We have described 
the calculation to ${\cal O}(p^6)$ in CHPT of $K_{\ell4}$ decays
and of the vacuum expectation values, giving some general lines
on the methods that we used
and all the various checks on the final results.
The explicit formulas have been presented as much as possible.
The long expressions involving the vertex- and sunset-type integrals
we simply parametrized, Eq. (\ref{para}). 
The rest of the expressions, involving the low-energy constants,
are fully listed in App. \ref{contributions}.

We have assumed resonance saturation to estimate the new
${\cal O}(p^6)$ parameters that appear. With all this at hand,
and large $N_c$ assumptions on $L_4^r$ and $L_6^r$ we updated most of the 
values of  the ${\cal O}(p^4)$ parameters, $L_i^r$, 
Eq. (\ref{newfit}), testing the validity of the large $N_c$
relation $(L_2^r-2L_1^r)/L_3^r \sim 0$.
To this end we made use of the existing $K_{\ell4}$ data, pseudoscalar
masses and decay constants. 
We varied the inputs over a fairly wide regime and discussed in
detail the underlying assumptions and their consequences once 
some of them are relaxed, see Table \ref{Tableresults}.
Among the assumptions it
is worth to mention the discussion on relaxing the large $N_c$ behaviour
of $L_4^r$ and $L_6^r$, Sect. \ref{largeNc}.
We show the allowed region in the $L_4^r$-$L_6^r$-plane, where 
CHPT, as a well defined perturbative series,
can exist without conflict with any of the known quantities
up to ${\cal O}(p^6)$.

We also present the comparison with previous results and model estimates,
pointing out the source of the difference.
As it becomes immediately clear the main source of the theoretical
uncertainties resides in the ${\cal O}(p^6)$ constants. 
We mention briefly how to treat the errors and show one way of
treating the correlations using a 68\% confidence level distribution
of the $L_i^r$.

Once the low-energy constants are found we use them to give the
phenomenological implications
on the forthcoming $K_{\ell4}$ experiments.
We review the experimental assumptions of 
\cite{Rosselet} finding some of them only borderline compatible
with the theoretical expression. We also
show the form-factors for different values of $s_\ell$ and present
the partial-waves predicted from the full result.
This  provides insight in the relevant assumptions
that the new experimental analysis
should make. 

The results were then used to obtain predictions for the $\pi$-$\pi$
threshold parameters and the $\pi$-$\pi$ phase-shifts. Good agreement
with the existing experimental results was obtained as is visible from
Table \ref{parameters} and Figs. \ref{figd0d1} and \ref{figd02}.
We also
warned the reader about the uncertainty in the relation
between the theoretical knowledge of 
$K_{\ell4}$ decays and $\pi$-$\pi$
scattering, reflected in the unknown ${\cal O}(p^6)$ corrections to the 
relation between the two- and three-flavour low-energy constants.
This is the reason we did not include
 $\pi$-$\pi$ scattering data in our fit as was done
in \cite{BCG}.
We do not expect large changes in the main results when this is done.

The variations of the vacuum expectation values, decay constants
and masses in terms of the lowest order parameters were shown 
in Figs. \ref{figvev} and \ref{figdecay} (at fixed ratio $\hat{m}:m_s$).
They show, up to
${\cal O}(p^6)$, no evidence of strong infrared singularities.

\vskip.4cm
\noindent {\bf Acknowledgements}\\
We thank G.~Colangelo, J.~Gasser, M.~Knecht, B.~Moussallam and J.~Stern
for discussions and comments.
Work supported in part by TMR, EC--Contract No. 
ERBFMRX--CT980169 (EURODAPHNE). 
P.T. was supported from the Swedish Natural Science Research 
Council (NFR).

\appendix
\setcounter{equation}{0}
\newcounter{zahler}
\addtocounter{zahler}{1}
\renewcommand{\thesection}{\Alph{zahler}}
\renewcommand{\theequation}{\Alph{zahler}.\arabic{equation}}

\section{Some ${\cal O}(p^6)$ Contributions}
\label{contributions}

In this appendix we give the explicit formulas for the ${\cal O}(p^6)$
part of the form-factors $F$ and $G$ of $K_{\ell4}$
split in parts as in Eq. (\ref{expansion2}).

\subsection{$L_i^r, L_i^r \times L_j^r$ and $L_i^r \times$ One-loop 
 Contributions}
\label{FGLL}
\newcommand{\newl}{\nonumber\\&&}

The $F_{LL}$ part of $F$ is
\begin{eqnarray}
F_{LL} &&\hspace{-0.7cm} = 
   m_\pi^4 /(16\pi^2)   \Big\{ 2/3 L_1^r + 2/9 L_2^r - 31/108 L_3^r \Big\}
\nonumber\\&&\hspace{-0.7cm}
  + m_\pi^4   \Big\{  - 256 L_1^r L_5^r + 64 L_2^r L_5^r - 48 L_3^r L_5^r + 256 (L_4^r)^2 + 416 
         L_4^r L_5^r - 512 L_4^r L_6^r - 512 L_4^r L_8^r 
\nonumber\\&&\hspace{-0.7cm}
+ 56 (L_5^r)^2 - 64 L_5^r L_6^r - 64 L_5^r 
         L_8^r - 16 L_5^r L_9^r \Big\}
\nonumber\\&&\hspace{-0.7cm}
  + m_\pi^2 m_K^2 /(16\pi^2)   \Big\{ 8/9 L_2^r + 8/27 L_3^r \Big\}
\nonumber\\&&\hspace{-0.7cm}
  + m_\pi^2 m_K^2   \Big\{ 
L_5^r \Big( 256 L_1^r + 64 L_3^r 
- 64 L_4^r- 32 L_5^r - 128 L_6^r + 8 L_9^r \Big)
+ 512 (L_4^r)^2  - 1024 
         L_4^r L_6^r  \Big\}
\nonumber\\&&\hspace{-0.7cm}
  + m_\pi^2 m_\eta^2 /(16\pi^2)   \Big\{  - 2 L_5^r \Big\}
\nonumber\\&&\hspace{-0.7cm}
  + m_\pi^2   \Big\{ L_5^r \Big(128 s_\pi L_1^r+ 32 s_\pi L_3^r  + 8 s_\pi L_9^r - 32 t_\pi L_2^r
- 16 t_\pi L_3^r + 8 t_\pi  L_9^r - 32 u_\pi L_2^r  
  + 8 u_\pi L_9^r \Big)
\nonumber\\&&\hspace{-0.7cm}
+ \overline{A}(m_\pi^2) \Big( - 404/3 L_1^r + 268/3 L_2^r - 13/3  L_3^r + 334/3
L_4^r + 31/2 L_5^r - 55/6 L_9^r \Big)
\nonumber\\&&\hspace{-0.7cm}
+\overline{A}(m_K^2) \Big(
- 48  L_1^r + 124/3 
L_2^r - 53/3 L_3^r + 32 L_4^r + 25 L_5^r - 11  
         L_9^r \Big) 
\nonumber\\&&\hspace{-0.7cm}
+ \overline{A}(m_\eta^2) \Big(
 16 L_1^r + 20/3 L_2^r - 73/18 L_3^r - 10 L_4^r
     + 73/6 L_5^r - 7/2 L_9^r \Big) \Big\}
\nonumber\\&&\hspace{-0.7cm}
  + m_K^4 /(16\pi^2)   \Big\{  - 2/3 L_1^r - 19/9 L_2^r - 55/108 L_3^r \Big\}
\nonumber\\&&\hspace{-0.7cm}
+m_K^4\Big\{-64 L_2^r L_5^r - 16 L_3^r L_5^r + 8 (L_5^r)^2 + 8 L_5^r L_9^r \Big\}
\nonumber\\&&\hspace{-0.7cm}
  + m_K^2 m_\eta^2 /(16\pi^2)   \Big\{ 8 L_5^r \Big\}
\nonumber\\&&\hspace{-0.7cm}
+m_K^2\Big\{ L_5^r \Big( -128 s_\pi L_1^r - 32 s_\pi L_3^r - 8 s_\pi L_9^r + 32 t_\pi 
         L_2^r 
+ 16 t_\pi L_3^r  - 8 t_\pi L_9^r + 32 u_\pi L_2^r  
- 8 u_\pi 
          L_9^r \Big)
\nonumber\\&&\hspace{-0.7cm}
+ \overline{A}(m_\pi^2) \Big( 54 L_2^r + 27/2 L_3^r + 4/3 L_4^r - 14/3 L_5^r
          - 55/12 L_9^r \Big)
\nonumber\\&&\hspace{-0.7cm}
+\overline{A}(m_K^2) \Big( 
- 4/3 L_1^r + 130/3 L_2^r + 1/2 L_3^r
     + 24 L_4^r + 8 L_5^r - 11/2 L_9^r \Big)
\nonumber\\&&\hspace{-0.7cm}
+ \overline{A}(m_\eta^2)\Big(
- 32/3  L_1^r + 10/
         3 L_2^r - 49/18 L_3^r + 4 L_4^r - 8/3 L_5^r - 7/4 
         L_9^r \Big) \Big\}
\nonumber\\&&\hspace{-0.7cm}
  + m_\eta^4 /(16\pi^2)   \Big\{ L_2^r + 1/2 L_3^r - 6 L_5^r \Big\}
\nonumber\\&&\hspace{-0.7cm}
  + m_\eta^2 \overline{A}(m_\eta^2)  \Big\{ 8 L_1^r + 2 L_2^r + L_3^r + 3 L_5^r \Big\}
\nonumber\\&&\hspace{-0.7cm}
  + \overline{B}(m_\pi^2,m_\pi^2,s_\pi) m_\pi^4   \Big\{ 80 L_1^r + 32 L_3^r - 80 L_4^r - 46 L_5^r
     + 48 L_6^r + 24 L_8^r + 6 L_9^r \Big\}
\nonumber\\&&\hspace{-0.7cm}
  + \overline{B}(m_\pi^2,m_\pi^2,s_\pi) m_\pi^2 m_K^2   \Big\{  - 32 L_4^r + 3 L_9^r \Big\}
\nonumber\\&&\hspace{-0.7cm}
  + \overline{B}(m_\pi^2,m_\pi^2,s_\pi) m_\pi^2/3   \Big\{  s_\pi \Big(- 200 L_1^r - 68 L_3^r + 116
L_4^r + 46 L_5^r - 21 L_9^r \Big) - 9 t_\pi L_9^r - 9 u_\pi L_9^r \Big\}
\nonumber\\&&\hspace{-0.7cm}
  + \overline{B}(m_\pi^2,m_\pi^2,s_\pi) \Big\{
m_K^2   \Big( 32/3 s_\pi L_4^r - 2 s_\pi L_9^r \Big)
  +  2 s_\pi^2 L_9^r + 2 s_\pi t_\pi L_9^r + 2 s_\pi u_\pi 
         L_9^r \Big\}
\nonumber\\&&\hspace{-0.7cm}
  + \overline{B}(m_\pi^2,m_K^2,t_\pi) m_\pi^4   \Big\{ 8/3 L_2^r - 8/3 L_3^r + 14/3 L_4^r + 3 
         L_5^r - L_9^r \Big\}
\nonumber\\&&\hspace{-0.7cm}
  + \overline{B}(m_\pi^2,m_K^2,t_\pi) m_\pi^2 m_K^2   \Big\{  - 8/3 L_2^r - 16/3 L_3^r + 8/3 L_4^r
     + 5 L_5^r - 20/3 L_6^r - 3/2 L_9^r \Big\}
\nonumber\\&&\hspace{-0.7cm}
  + \overline{B}(m_\pi^2,m_K^2,t_\pi) m_\pi^2   \Big\{ s_\pi/2 L_9^r + 
 t_\pi \Big( 8/3 L_2^r + 16/3  
L_3^r + 14/3 L_4^r - 3 L_5^r + 3/2 L_9^r \Big) +u_\pi/2 L_9^r \Big\}
\nonumber\\&&\hspace{-0.7cm}
  + \overline{B}(m_\pi^2,m_K^2,t_\pi) m_K^4   \Big\{ 8/3 L_4^r + 4/3 L_5^r - 40/3 L_6^r - 20/3
          L_8^r - 1/2 L_9^r \Big\}
\nonumber\\&&\hspace{-0.7cm}
  + \overline{B}(m_\pi^2,m_K^2,t_\pi) m_K^2   \Big\{ 1/2 s_\pi L_9^r + 4 t_\pi L_4^r + 2 t_\pi L_5^r
     + t_\pi L_9^r + 1/2 u_\pi L_9^r \Big\}
\nonumber\\&&\hspace{-0.7cm}
  + \overline{B}(m_\pi^2,m_K^2,t_\pi)   \Big\{  - 1/2 s_\pi t_\pi L_9^r - 1/2 t_\pi^2 L_9^r - 1/2 
         t_\pi u_\pi L_9^r \Big\}
\nonumber\\&&\hspace{-0.7cm}
  + \overline{B}(m_\pi^2,m_K^2,u_\pi) m_\pi^4   \Big\{ 56/3 L_2^r + 64/3 L_4^r + 20/3 L_5^r - 8/
         3 L_9^r \Big\}
\nonumber\\&&\hspace{-0.7cm}
  + \overline{B}(m_\pi^2,m_K^2,u_\pi) m_\pi^2 m_K^2   \Big\{ 40/3 L_2^r + 32 L_6^r + 16 L_8^r - 8/3 
         L_9^r \Big\}
\nonumber\\&&\hspace{-0.7cm}
  + \overline{B}(m_\pi^2,m_K^2,u_\pi) 4/3 m_\pi^2   \Big\{ s_\pi L_9^r + t_\pi L_9^r - 10 u_\pi 
         L_2^r - 8 u_\pi L_4^r - 4 u_\pi L_5^r + 2 u_\pi L_9^r \Big\}
\nonumber\\&&\hspace{-0.7cm}
  + \overline{B}(m_\pi^2,m_K^2,u_\pi) m_K^4   \Big\{  - 16/3 L_4^r - 8/3 L_5^r - 2/3 L_9^r \Big\}
\nonumber\\&&\hspace{-0.7cm}
  + \overline{B}(m_\pi^2,m_K^2,u_\pi) m_K^2   \Big\{ 2/3 s_\pi L_9^r + 2/3 t_\pi L_9^r + 16/3 u_\pi 
         L_4^r + 8/3 u_\pi L_5^r + 4/3 u_\pi L_9^r \Big\}
\nonumber\\&&\hspace{-0.7cm}
  + \overline{B}(m_\pi^2,m_K^2,u_\pi)   \Big\{  - 2/3 s_\pi u_\pi L_9^r - 2/3 t_\pi u_\pi L_9^r - 2/
         3 u_\pi^2 L_9^r \Big\}
\nonumber\\&&\hspace{-0.7cm}
  + \overline{B}(m_K^2,m_K^2,s_\pi) m_\pi^2 m_K^2   \Big\{ 96 L_1^r + 24 L_3^r - 96 L_4^r - 24 L_5^r
     + 96 L_6^r + 48 L_8^r \Big\}
\nonumber\\&&\hspace{-0.7cm}
  + \overline{B}(m_K^2,m_K^2,s_\pi) m_\pi^2 s_\pi  \Big\{  - 256/3 L_1^r + 80/3  L_2^r - 16 
         L_3^r + 42 L_4^r + 10 L_5^r - 4 L_9^r \Big\}
\nonumber\\&&\hspace{-0.7cm}
  + \overline{B}(m_K^2,m_K^2,s_\pi) m_K^2 s_\pi  \Big\{  - 80 L_1^r + 16 L_2^r - 20 
         L_3^r + 28 L_4^r - 2 L_9^r \Big\}
\nonumber\\&&\hspace{-0.7cm}
 + \overline{B}(m_K^2,m_K^2,s_\pi)   \Big\{ 48 s_\pi^2 L_1^r + 48 s_\pi^2 L_2^r + 24 s_\pi^2 
L_3^r + 2 s_\pi^2 L_9^r - 16/3 s_\pi t_\pi L_1^r - 40/3 s_\pi t_\pi L_2^r 
\nonumber\\&&\hspace{-0.7cm}
- 4 s_\pi t_\pi 
L_3^r 
+ 2 s_\pi t_\pi L_9^r - 16/3 s_\pi u_\pi L_1^r - 40/3 s_\pi u_\pi L_2^r - 4 s_\pi u_\pi
L_3^r + 2 s_\pi u_\pi L_9^r \Big\}
\nonumber\\&&\hspace{-0.7cm}
+\overline{B}(m_\eta^2,m_K^2,t_\pi) m_\pi^4 \Big\{  - 4 L_3^r - 28/3 L_4^r - 16/9 L_5^r + 8/
         3 L_6^r + 32/3 L_7^r + 8 L_8^r - 8/3 L_9^r \Big\}
\nonumber\\&&\hspace{-0.7cm}
+\overline{B}(m_\eta^2,m_K^2,t_\pi) m_\pi^2 m_K^2/3 \Big\{ 2/3 L_3^r-46/3 L_4^r - 22/3 L_5^r + 
         20 L_6^r + 32 L_7^r + 16 L_8^r - 2 L_9^r \Big\}
\nonumber\\&&\hspace{-0.7cm}
+\overline{B}(m_\eta^2,m_K^2,t_\pi) m_\pi^2/3   \Big\{ 
\Big(3 L_3^r + 4  L_9^r\Big) \Big( s_\pi + u_\pi\Big)
 + 43/3 t_\pi L_3^r
 + 12 t_\pi L_4^r - 2 t_\pi L_5^r + 10 t_\pi L_9^r \Big\}
\nonumber\\&&\hspace{-0.7cm}
  + \overline{B}(m_\eta^2,m_K^2,t_\pi) m_K^4/3 \Big\{  - 14/3 L_3^r + 20 L_4^r + 38/3 L_5^r
     + 8 L_6^r - 64 L_7^r - 28 L_8^r + L_9^r \Big\}
\nonumber\\&&\hspace{-0.7cm}
  + \overline{B}(m_\eta^2,m_K^2,t_\pi) m_K^2/3\Big\{
\Big( -3 L_3^r - L_9^r \Big) \Big( s_\pi +u_\pi\Big) + 47/3 t_\pi 
L_3^r + 24 t_\pi L_4^r - 4 t_\pi L_5^r + 2 t_\pi L_9^r \Big\}
\nonumber\\&&\hspace{-0.7cm}
+\overline{B}(m_\eta^2,m_K^2,t_\pi) \Big\{ 
- 4 L_4^r m_K^2 m_\eta^2   
-2 L_4^r m_\pi^2 m_\eta^2 
- 2 L_5^r m_\eta^4    \Big\}
\nonumber\\&&\hspace{-0.7cm}
+\overline{B}(m_\eta^2,m_K^2,t_\pi)\Big\{  - 1/3 s_\pi t_\pi L_3^r - s_\pi t_\pi L_9^r - 7/3 
 t_\pi^2 L_3^r - t_\pi^2 L_9^r - 1/3 t_\pi u_\pi L_3^r - t_\pi u_\pi L_9^r \Big\}
\nonumber\\&&\hspace{-0.7cm}
+\overline{B}(m_\eta^2,m_\eta^2,s_\pi) m_\pi^4\Big\{ 56/9 L_4^r + 10/3 L_5^r - 16 L_6^r + 64 
         L_7^r + 24 L_8^r - 2 L_9^r \Big\}
\nonumber\\&&\hspace{-0.7cm}
+ \overline{B}(m_\eta^2,m_\eta^2,s_\pi) m_\pi^2 m_K^2 \Big\{-224/9 L_4^r-16/3 L_5^r+64 L_6^r
          - 64 L_7^r - L_9^r \Big\}
\nonumber\\&&\hspace{-0.7cm}
+\overline{B}(m_\eta^2,m_\eta^2,s_\pi) m_\pi^2 m_\eta^2 \Big\{112/3 L_1^r+32/9 L_3^r-24 L_4^r - 4
          L_5^r \Big\}
\nonumber\\&&\hspace{-0.7cm}
  + \overline{B}(m_\eta^2,m_\eta^2,s_\pi) m_\pi^2\Big\{  - 4 s_\pi L_4^r - 2 s_\pi L_5^r + s_\pi L_9^r + 
         t_\pi L_9^r + u_\pi L_9^r \Big\}
\nonumber\\&&\hspace{-0.7cm}
  + \overline{B}(m_\eta^2,m_\eta^2,s_\pi) \Big\{ 16 s_\pi L_4^r m_K^2 + 
  m_\eta^2   \Big(  - 24 s_\pi L_1^r - 4 s_\pi L_3^r \Big) \Big\}
\nonumber\\&&\hspace{-0.7cm}
  + \overline{B}_1(m_\pi^2,m_\pi^2,s_\pi) m_\pi^4   \Big\{  - 8 L_3^r + 32 L_4^r + 40 L_5^r + 64 L_6^r
     + 32 L_8^r - 8 L_9^r \Big\}
\nonumber\\&&\hspace{-0.7cm}
  + \overline{B}_1(m_\pi^2,m_\pi^2,s_\pi) m_\pi^2 m_K^2   \Big\{  - 32 L_2^r - 8 L_3^r + 64 L_4^r - 4 
         L_9^r \Big\}
\nonumber\\&&\hspace{-0.7cm}
  + \overline{B}_1(m_\pi^2,m_\pi^2,s_\pi) m_\pi^2   \Big\{  - 368/3 s_\pi L_1^r + 80/3 s_\pi L_2^r - 32 
         s_\pi L_3^r + 24 s_\pi L_4^r + 8/3 s_\pi L_9^r + 16 t_\pi L_2^r 
\nonumber\\&&\hspace{-0.7cm}
+ 4 t_\pi L_3^r + 4 t_\pi 
         L_9^r + 16 u_\pi L_2^r + 4 u_\pi L_3^r + 4 u_\pi L_9^r \Big\}
\nonumber\\&&\hspace{-0.7cm}
  + \overline{B}_1(m_\pi^2,m_\pi^2,s_\pi) m_K^2   \Big\{ 128/3 s_\pi L_2^r + 32/3 s_\pi L_3^r - 32/3 
         s_\pi L_4^r - 2/3 s_\pi L_9^r \Big\}
\nonumber\\&&\hspace{-0.7cm}
  + \overline{B}_1(m_\pi^2,m_\pi^2,s_\pi)   \Big\{ 200/3 s_\pi^2 L_1^r + 12 s_\pi^2 L_2^r + 68/3 
         s_\pi^2 L_3^r + 2/3 s_\pi^2 L_9^r - 64/3 s_\pi t_\pi L_2^r 
\nonumber\\&&\hspace{-0.7cm}
- 16/3 s_\pi t_\pi L_3^r 
+ 2/3
s_\pi t_\pi L_9^r - 64/3 s_\pi u_\pi L_2^r - 16/3 s_\pi u_\pi L_3^r + 2/3 s_\pi u_\pi L_9^r
          \Big\}
\nonumber\\&&\hspace{-0.7cm}
  + \overline{B}_1(m_\pi^2,m_K^2,t_\pi) m_\pi^4   \Big\{  - 8 L_1^r - 20 L_2^r - 18 L_3^r - 46/3 
         L_4^r + 16/3 L_5^r - 4/3 L_9^r \Big\}
\nonumber\\&&\hspace{-0.7cm}
  + \overline{B}_1(m_\pi^2,m_K^2,t_\pi) m_\pi^2 m_K^2/3   \Big\{ 16 L_1^r + 8 L_2^r - 2 L_3^r + 56
L_4^r + 36 L_5^r - 76 L_6^r - 48 L_8^r - 4 L_9^r \Big\}
\nonumber\\&&\hspace{-0.7cm}
  + \overline{B}_1(m_\pi^2,m_K^2,t_\pi) m_\pi^2/3   \Big\{ 8 s_\pi L_1^r + 4 s_\pi L_2^r + 5 s_\pi 
  L_3^r + 2 s_\pi L_9^r - 24 t_\pi L_1^r + 12 t_\pi L_2^r + 39 t_\pi L_3^r 
\nonumber\\&&\hspace{-0.7cm}
+ 22 t_\pi L_4^r
-20 t_\pi L_5^r+10 t_\pi L_9^r+8 u_\pi L_1^r+4 u_\pi L_2^r + 5 u_\pi 
         L_3^r + 2 u_\pi L_9^r \Big\}
\nonumber\\&&\hspace{-0.7cm}
  + \overline{B}_1(m_\pi^2,m_K^2,t_\pi) m_K^4/3   \Big\{ 8 L_1^r + 4 L_2^r - 4 L_3^r - 40 
         L_4^r - 20 L_5^r + 40 L_6^r + 20 L_8^r - L_9^r \Big\}
\nonumber\\&&\hspace{-0.7cm}
+\overline{B}_1(m_\pi^2,m_K^2,t_\pi) m_K^2/3 \Big\{-8 s_\pi L_1^r
-4 s_\pi L_2^r +  
s_\pi L_3^r+s_\pi L_9^r - 8 t_\pi L_1^r + 4 t_\pi L_2^r 
\nonumber\\&&\hspace{-0.7cm}
+ 17 t_\pi L_3^r - 4
t_\pi L_4^r-2 t_\pi L_5^r+5 t_\pi L_9^r -8 u_\pi L_1^r-4 u_\pi L_2^r 
         + u_\pi L_3^r + u_\pi L_9^r \Big\}
\nonumber\\&&\hspace{-0.7cm}
  + \overline{B}_1(m_\pi^2,m_K^2,t_\pi) \Big\{ 8 s_\pi t_\pi L_1^r + 4 s_\pi t_\pi L_2^r -  
         s_\pi t_\pi L_3^r 
- 4 s_\pi t_\pi L_9^r - 8 t_\pi^2 L_2^r - 13 t_\pi^2 L_3^r 
\nonumber\\&&\hspace{-0.7cm}
- 4
t_\pi^2 L_9^r + 8 t_\pi u_\pi L_1^r + 4 t_\pi u_\pi L_2^r -t_\pi u_\pi L_3^r - 4
t_\pi u_\pi L_9^r \Big\}/3
\nonumber\\&&\hspace{-0.7cm}
  + \overline{B}_1(m_\pi^2,m_K^2,u_\pi) m_\pi^4   \Big\{  - 40 L_1^r - 20 L_2^r - 20 L_3^r - 56/3 
         L_4^r \Big\}
\nonumber\\&&\hspace{-0.7cm}
  + \overline{B}_1(m_\pi^2,m_K^2,u_\pi) m_\pi^2 m_K^2   \Big\{  - 32/3 L_1^r - 16/3 L_2^r - 16/3 L_3^r
          - 16/3 L_4^r \Big\}
\nonumber\\&&\hspace{-0.7cm}
  + \overline{B}_1(m_\pi^2,m_K^2,u_\pi) m_\pi^2/3   \Big\{ 32 s_\pi L_1^r + 16 s_\pi L_2^r + 16
         s_\pi L_3^r + 32 t_\pi L_1^r + 16 t_\pi L_2^r + 16 t_\pi L_3^r 
\nonumber\\&&\hspace{-0.7cm}
+ 48 u_\pi L_1^r - 
48 u_\pi L_2^r + 24 u_\pi L_3^r + 8 u_\pi L_4^r - 4 u_\pi L_5^r + 4 u_\pi L_9^r \Big\}
\nonumber\\&&\hspace{-0.7cm}
  + \overline{B}_1(m_\pi^2,m_K^2,u_\pi) m_K^4/3   \Big\{  - 40 L_1^r - 20 L_2^r - 20 L_3^r
     + 24 L_4^r \Big\}
\nonumber\\&&\hspace{-0.7cm}
  + \overline{B}_1(m_\pi^2,m_K^2,u_\pi) m_K^2   \Big\{ 16/3 s_\pi L_1^r + 8/3 s_\pi L_2^r + 8/3 s_\pi 
L_3^r + 16/3 t_\pi L_1^r + 8/3 t_\pi L_2^r 
\nonumber\\&&\hspace{-0.7cm}
+ 8/3 t_\pi L_3^r 
+ 16/3 u_\pi L_1^r - 32/3 
u_\pi L_2^r + 8/3 u_\pi L_3^r - 56/3 u_\pi L_4^r - 16/3 u_\pi L_5^r + 2/3 u_\pi L_9^r \Big\}
\nonumber\\&&\hspace{-0.7cm}
  + \overline{B}_1(m_\pi^2,m_K^2,u_\pi) 1/3  \Big\{  - 16 s_\pi u_\pi L_1^r - 8 s_\pi u_\pi L_2^r - 8
s_\pi u_\pi L_3^r - 2 s_\pi u_\pi L_9^r - 16 t_\pi u_\pi L_1^r 
\nonumber\\&&\hspace{-0.7cm}
- 8 t_\pi u_\pi L_2^r
- 8 t_\pi u_\pi L_3^r - 2 t_\pi u_\pi L_9^r + 24 u_\pi^2 L_1^r + 52 u_\pi^2 L_2^r
+ 12 u_\pi^2 L_3^r - 2 u_\pi^2 L_9^r \Big\}
\nonumber\\&&\hspace{-0.7cm}
  + \overline{B}_1(m_K^2,m_\pi^2,t_\pi) m_\pi^2 m_K^2   \Big\{  - 10/3 L_4^r + 20/3 L_6^r \Big\}
\nonumber\\&&\hspace{-0.7cm}
  + \overline{B}_1(m_K^2,m_\pi^2,t_\pi) m_K^4   \Big\{  - 20/3 L_4^r - 10/3 L_5^r + 40/3 L_6^r
     + 20/3 L_8^r \Big\}
\nonumber\\&&\hspace{-0.7cm}
  + \overline{B}_1(m_K^2,m_K^2,s_\pi) m_\pi^2   \Big\{ 256/3 s_\pi L_1^r - 80/3 s_\pi L_2^r + 16 s_\pi
          L_3^r - 36 s_\pi L_4^r - 2 s_\pi L_5^r + 2 s_\pi L_9^r \Big\}
\nonumber\\&&\hspace{-0.7cm}
  + \overline{B}_1(m_K^2,m_K^2,s_\pi) m_K^2   \Big\{ 32/3 s_\pi L_1^r - 64/3 s_\pi L_2^r + 16 s_\pi 
         L_4^r + s_\pi L_9^r \Big\}
\nonumber\\&&\hspace{-0.7cm}
  + \overline{B}_1(m_K^2,m_K^2,s_\pi)   \Big\{  - 80/3 s_\pi^2 L_1^r - 200/3 s_\pi^2 L_2^r - 22 
s_\pi^2 L_3^r - s_\pi^2 L_9^r + 16/3 s_\pi t_\pi L_1^r 
\nonumber\\&&\hspace{-0.7cm}
+ 40/3 s_\pi t_\pi L_2^r
 + 4 s_\pi 
         t_\pi L_3^r - s_\pi t_\pi L_9^r + 16/3 s_\pi u_\pi L_1^r + 40/3 s_\pi u_\pi L_2^r + 4 s_\pi 
         u_\pi L_3^r - s_\pi u_\pi L_9^r \Big\}
\nonumber\\&&\hspace{-0.7cm}
  + \overline{B}_1(m_K^2,m_\eta^2,t_\pi) m_\pi^4   \Big\{ 1/3 L_4^r + 1/3 L_5^r - 8/3 L_6^r - 8/3 
         L_8^r \Big\}
\nonumber\\&&\hspace{-0.7cm}
 + \overline{B}_1(m_K^2,m_\eta^2,t_\pi) \Big\{ 
m_\pi^2 m_K^2   \Big( 25/3 L_4^r + 7 L_5^r - 20/3 L_6^r \Big)
 + m_\pi^2   \Big(  - 3 t_\pi L_4^r - 3 t_\pi L_5^r \Big) \Big\}
\nonumber\\&&\hspace{-0.7cm}
  + \overline{B}_1(m_K^2,m_\eta^2,t_\pi) \Big\{ m_K^4 \Big( 46/3 L_4^r + 2/3 L_5^r - 8/3 L_6^r - 4/3 
         L_8^r \Big)
  - 6 t_\pi L_4^r   m_K^2   \Big\}
\nonumber\\&&\hspace{-0.7cm}
  + \overline{B}_1(m_\eta^2,m_K^2,t_\pi) m_\pi^4   \Big\{ 25/3 L_3^r + 55/3 L_4^r + 16/9 L_5^r - 8/
         3 L_6^r - 32/3 L_7^r - 8 L_8^r + 17/3 L_9^r \Big\}
\nonumber\\&&\hspace{-0.7cm}
  + \overline{B}_1(m_\eta^2,m_K^2,t_\pi) m_\pi^2 m_K^2/3   \Big\{  - 56/3 L_3^r + 73 L_4^r + 22/3 L_5^r
          - 20 L_6^r - 32 L_7^r - 16 L_8^r - 5/2 L_9^r \Big\}
\nonumber\\&&\hspace{-0.7cm}
  + \overline{B}_1(m_\eta^2,m_K^2,t_\pi) m_\pi^2   \Big\{  - 19/9 s_\pi L_3^r - 17/6 s_\pi L_9^r - 79/9
t_\pi L_3^r - 3 t_\pi L_4^r + 2 t_\pi L_5^r - 35/6 t_\pi L_9^r 
\nonumber\\&&\hspace{-0.7cm}
- 19/9 u_\pi L_3^r - 17/6
          u_\pi L_9^r \Big\}
\nonumber\\&&\hspace{-0.7cm}
  + \overline{B}_1(m_\eta^2,m_K^2,t_\pi) m_K^4/3   \Big\{  - 7/3 L_3^r - 74 L_4^r - 38/3 L_5^r - 
         8 L_6^r + 64 L_7^r + 28 L_8^r - 11/2 L_9^r \Big\}
\nonumber\\&&\hspace{-0.7cm}
+ \overline{B}_1(m_\eta^2,m_K^2,t_\pi) \Big\{ 2 L_4^r m_\pi^2 m_\eta^2  
+ 4 L_4^r  m_K^2 m_\eta^2   
+ 2 L_5^r m_\eta^4   \Big\}
\nonumber\\&&\hspace{-0.7cm}
  + \overline{B}_1(m_\eta^2,m_K^2,t_\pi) m_K^2   \Big\{ 25/9 s_\pi L_3^r + 11/6 s_\pi L_9^r - 41/9 
         t_\pi L_3^r - 6 t_\pi L_4^r + 4 t_\pi L_5^r + 1/3 t_\pi L_9^r 
\nonumber\\&&\hspace{-0.7cm}
+ 25/9 u_\pi L_3^r + 11/6 
         u_\pi L_9^r \Big\}
\nonumber\\&&\hspace{-0.7cm}
  + \overline{B}_1(m_\eta^2,m_K^2,t_\pi)   \Big\{ 
\Big(1/3 L_3^r + 3/2  L_9^r\Big) \Big(s_\pi+u_\pi\Big) t_\pi + 4 
         t_\pi^2 L_3^r + 3/2 t_\pi^2 L_9^r \Big\}
\nonumber\\&&\hspace{-0.7cm}
  + \overline{B}_1(m_\eta^2,m_\eta^2,s_\pi) \Big\{ \Big( m_\pi^4+m_\pi^2 m_K^2\Big)
   \Big( 32/3 L_2^r + 8/9 L_3^r \Big)
  \Big\}
\nonumber\\&&\hspace{-0.7cm}
  + \overline{B}_1(m_\eta^2,m_\eta^2,s_\pi) m_\pi^2   \Big\{  - 112/3 s_\pi L_1^r - 32/9 s_\pi L_3^r + 24 
         s_\pi L_4^r + 4 s_\pi L_5^r - 16/3 t_\pi L_2^r 
\nonumber\\&&\hspace{-0.7cm}
- 4/9 t_\pi L_3^r 
- 16/3 u_\pi L_2^r - 4/9
          u_\pi L_3^r \Big\}
\nonumber\\&&\hspace{-0.7cm}
  + \overline{B}_1(m_\eta^2,m_\eta^2,s_\pi)   \Big\{ 24 s_\pi^2 L_1^r + 12 s_\pi^2 L_2^r + 8 s_\pi^2 L_3^r
          \Big\}
\nonumber\\&&\hspace{-0.7cm}
  + \overline{B}_{21}(m_\pi^2,m_\pi^2,s_\pi) \Big\{
\Big(m_\pi^4 +m_\pi^2 m_K^2\Big)  \Big( 32 L_2^r + 8 L_3^r \Big) \Big\}
\nonumber\\&&\hspace{-0.7cm}
  + \overline{B}_{21}(m_\pi^2,m_\pi^2,s_\pi) \Big\{
m_\pi^2  \Big(  - 32 s_\pi L_2^r - 32/3 s_\pi L_4^r - 40/3 
         s_\pi L_5^r + 8/3 s_\pi L_9^r - 16 t_\pi L_2^r 
\nonumber\\&&\hspace{-0.7cm}
- 4 t_\pi L_3^r 
- 16 u_\pi L_2^r - 4 u_\pi 
L_3^r \Big)
  + m_K^2   \Big(  - 64/3 s_\pi L_4^r + 4/3 s_\pi L_9^r \Big) \Big\}
\nonumber\\&&\hspace{-0.7cm}
  + \overline{B}_{21}(m_\pi^2,m_\pi^2,s_\pi) 4/3  \Big\{ 44 s_\pi^2 L_1^r + 9 s_\pi^2 L_2^r + 16 
s_\pi^2 L_3^r - s_\pi^2 L_9^r - s_\pi t_\pi L_9^r - s_\pi u_\pi L_9^r \Big\}
\nonumber\\&&\hspace{-0.7cm}
  + \overline{B}_{21}(m_\pi^2,m_K^2,t_\pi) m_\pi^4   \Big\{ 40 L_1^r + 20 L_2^r + 10 L_3^r + 16 L_4^r
     + 9 L_5^r - 3 L_9^r \Big\}
\nonumber\\&&\hspace{-0.7cm}
  + \overline{B}_{21}(m_\pi^2,m_K^2,t_\pi) m_\pi^2 m_K^2   \Big\{  - 112/3 L_1^r - 56/3 L_2^r - 10/3 
         L_3^r - 8 L_4^r - 9 L_5^r + 3/2 L_9^r \Big\}
\nonumber\\&&\hspace{-0.7cm}
  + \overline{B}_{21}(m_\pi^2,m_K^2,t_\pi) m_\pi^2/3   \Big\{  - 32 s_\pi L_1^r - 16 s_\pi L_2^r - 8
          s_\pi L_3^r + s_\pi/2 L_9^r + 48 t_\pi L_2^r + 36 t_\pi L_3^r 
\nonumber\\&&\hspace{-0.7cm}
- 52 t_\pi L_4^r - 23
t_\pi L_5^r + 19/2 t_\pi L_9^r - 32 u_\pi L_1^r - 16 u_\pi L_2^r - 8 u_\pi L_3^r
     + u_\pi/2 L_9^r \Big\}
\nonumber\\&&\hspace{-0.7cm}
  + \overline{B}_{21}(m_\pi^2,m_K^2,t_\pi) m_K^4   \Big\{  - 8/3 L_1^r - 4/3 L_2^r + 4/3 L_3^r - 8
          L_4^r + 3/2 L_9^r \Big\}
\nonumber\\&&\hspace{-0.7cm}
  + \overline{B}_{21}(m_\pi^2,m_K^2,t_\pi) m_K^2   \Big\{ 32/3 s_\pi L_1^r + 16/3 s_\pi L_2^r + 2/3 
         s_\pi L_3^r - 3/2 s_\pi L_9^r - 16/3 t_\pi L_4^r 
\nonumber\\&&\hspace{-0.7cm}
- 20/3 t_\pi L_5^r 
- 2/3 t_\pi L_9^r + 
         32/3 u_\pi L_1^r + 16/3 u_\pi L_2^r + 2/3 u_\pi L_3^r - 3/2 u_\pi L_9^r \Big\}
\nonumber\\&&\hspace{-0.7cm}
 + \overline{B}_{21}(m_\pi^2,m_K^2,t_\pi)   \Big\{  - 5/6 s_\pi t_\pi L_9^r - 8 t_\pi^2 L_1^r - 44/3 
         t_\pi^2 L_2^r - 34/3 t_\pi^2 L_3^r - 5/6 t_\pi^2 L_9^r - 5/6 t_\pi u_\pi L_9^r \Big\}
\nonumber\\&&\hspace{-0.7cm}
  + \overline{B}_{21}(m_\pi^2,m_K^2,u_\pi) \Big\{ \Big( 20 m_\pi^4  +16/3 m_\pi^2 m_K^2\Big) 
\Big( 2 L_1^r + L_2^r + L_3^r \Big) \Big\}
\nonumber\\&&\hspace{-0.7cm}
  + \overline{B}_{21}(m_\pi^2,m_K^2,u_\pi) \Big\{ 
\Big(16/3 m_\pi^2 +8/3 m_K^2\Big)  \Big(s_\pi+t_\pi\Big) \Big(  - 2 L_1^r -  L_2^r -
          L_3^r \Big)
          \Big\}
\nonumber\\&&\hspace{-0.7cm}
  + \overline{B}_{21}(m_\pi^2,m_K^2,u_\pi) \Big\{ m_K^4 
 \Big( 40/3 L_1^r + 20/3 L_2^r + 20/3 L_3^r \Big)
   - 8 u_\pi^2 L_1^r + 4/3 u_\pi^2 L_2^r - 4 u_\pi^2
          L_3^r \Big\}
\nonumber\\&&\hspace{-0.7cm}
  + \overline{B}_{21}(m_K^2,m_K^2,s_\pi)   \Big\{ 64/3 s_\pi^2 L_1^r - 56/3 s_\pi^2 L_2^r + 2 
         s_\pi^2 L_3^r \Big\}
\nonumber\\&&\hspace{-0.7cm}
  + \overline{B}_{21}(m_K^2,m_\eta^2,t_\pi) \Big\{
m_\pi^4   \Big( 9 L_4^r + 9 L_5^r \Big)
  +  m_\pi^2 m_K^2   \Big( 9 L_4^r - 9 L_5^r \Big)
  + m_\pi^2   \Big(  - 3 t_\pi L_4^r - 3 t_\pi L_5^r \Big) \Big\}
\nonumber\\&&\hspace{-0.7cm}
  + \overline{B}_{21}(m_K^2,m_\eta^2,t_\pi) \Big\{
m_K^4   \Big(  - 18 L_4^r \Big)
  +  m_K^2   \Big(  - 6 t_\pi L_4^r \Big) \Big\}
\nonumber\\&&\hspace{-0.7cm}
  + \overline{B}_{21}(m_\eta^2,m_K^2,t_\pi) \Big\{
m_\pi^4   \Big(  - 13/3 L_3^r - 9 L_4^r - 3 L_9^r \Big)
  + m_\pi^2 m_K^2   \Big( 56/9 L_3^r - 9 L_4^r + 3/2 L_9^r \Big) \Big\}
\nonumber\\&&\hspace{-0.7cm}
  + \overline{B}_{21}(m_\eta^2,m_K^2,t_\pi)  \Big\{ m_\pi^2   \Big( 
\Big(10/9 s_\pi + 4 t_\pi+ 10/9 u_\pi\Big) L_3^r   
- t_\pi L_4^r - 4/3 t_\pi L_5^r 
\nonumber\\&&\hspace{-0.7cm}
+ \Big(3 s_\pi + 5 t_\pi + 3 u_\pi \Big)
      L_9^r/2 \Big)
+ m_K^4   \Big( 7/9 L_3^r + 18 L_4^r + 3/2 L_9^r \Big)
  + m_K^2   \Big(  - 16/9 s_\pi L_3^r 
\nonumber\\&&\hspace{-0.7cm}
- 3/2 s_\pi L_9^r - 2 
         t_\pi L_4^r - 8/3 t_\pi L_5^r - t_\pi L_9^r - 16/9 u_\pi L_3^r - 3/2 u_\pi L_9^r \Big) \Big\}
\nonumber\\&&\hspace{-0.7cm}
  + \overline{B}_{21}(m_\eta^2,m_K^2,t_\pi)   \Big\{  - 1/2 s_\pi t_\pi L_9^r - 5/3 t_\pi^2 L_3^r - 1/2
          t_\pi^2 L_9^r - 1/2 t_\pi u_\pi L_9^r \Big\}
\nonumber\\&&\hspace{-0.7cm}
- \overline{B}_{21}(m_\eta^2,m_\eta^2,s_\pi) \Big\{ 
\Big(m_\pi^4+m_\pi^2 m_K^2\Big)   \Big( 32/3 L_2^r + 8/9 L_3^r \Big) \Big\}
\nonumber\\&&\hspace{-0.7cm}
  + \overline{B}_{21}(m_\eta^2,m_\eta^2,s_\pi) m_\pi^2   \Big\{ 16/3 t_\pi L_2^r + 4/9 t_\pi L_3^r + 16/3 
         u_\pi L_2^r + 4/9 u_\pi L_3^r \Big\}
\nonumber\\&&\hspace{-0.7cm}
  + \overline{B}_{21}(m_\eta^2,m_\eta^2,s_\pi)   \Big\{  - 12 s_\pi^2 L_2^r - 4 s_\pi^2 L_3^r \Big\}
\nonumber\\&&\hspace{-0.7cm}
  + \overline{B}_{22}(m_\pi^2,m_\pi^2,s_\pi) m_\pi^2   \Big\{ 272/3 L_2^r + 56/3 L_3^r - 32/3 L_4^r - 40/
         3 L_5^r + 8/3 L_9^r \Big\}
\nonumber\\&&\hspace{-0.7cm}
  + \overline{B}_{22}(m_\pi^2,m_\pi^2,s_\pi) m_K^2   \Big\{ 128/3 L_2^r + 32/3 L_3^r - 64/3 L_4^r + 4/
         3 L_9^r \Big\}
\nonumber\\&&\hspace{-0.7cm}
  + \overline{B}_{22}(m_\pi^2,m_\pi^2,s_\pi)   \Big\{ 176 s_\pi L_1^r - 128 s_\pi L_2^r + 24 s_\pi 
  L_3^r - 4 s_\pi L_9^r - 64 t_\pi L_2^r - 4 t_\pi L_3^r - 4 t_\pi L_9^r 
\nonumber\\&&\hspace{-0.7cm}
- 64
         u_\pi L_2^r - 28 u_\pi L_3^r - 4 u_\pi L_9^r \Big\}/3
\nonumber\\&&\hspace{-0.7cm}
  + \overline{B}_{22}(m_\pi^2,m_K^2,t_\pi) m_\pi^2/3   \Big\{  - 32 L_1^r - 112 L_2^r + 88 L_3^r
   - 196 L_4^r - 104 L_5^r + 32 L_9^r \Big\}
\nonumber\\&&\hspace{-0.7cm}
  + \overline{B}_{22}(m_\pi^2,m_K^2,t_\pi) m_K^2   \Big\{ 4 L_3^r - 88/3 L_4^r - 20/3 L_5^r + 16/3 
         L_9^r \Big\}
\nonumber\\&&\hspace{-0.7cm}
  + \overline{B}_{22}(m_\pi^2,m_K^2,t_\pi)   \Big\{  - 16/3 s_\pi L_1^r + 40/3 s_\pi L_2^r - 16/3 
         s_\pi L_3^r - 16/3 s_\pi L_9^r + 64/3 t_\pi L_1^r 
\nonumber\\&&\hspace{-0.7cm}
- 24 t_\pi L_2^r 
- 4 t_\pi L_3^r - 16/3
          t_\pi L_9^r + 32/3 u_\pi L_1^r + 40/3 u_\pi L_2^r - 4/3 u_\pi L_3^r - 16/3 u_\pi L_9^r \Big\}
\nonumber\\&&\hspace{-0.7cm}
  + \overline{B}_{22}(m_\pi^2,m_K^2,u_\pi) \Big\{
m_\pi^2/3   \Big( 160 L_1^r + 80 L_2^r + 80 L_3^r + 16
         L_4^r \Big)
  + 16 m_K^2   \Big(2 L_1^r + L_2^r + L_3^r \Big) \Big\}
\nonumber\\&&\hspace{-0.7cm}
  - \overline{B}_{22}(m_\pi^2,m_K^2,u_\pi) 8/3  \Big\{
\Big( 2 L_1^r + L_2^r + 
         L_3^r \Big)\Big(s_\pi  + t_\pi\Big)
+ 10 u_\pi L_1^r + 3 u_\pi
          L_2^r + 5 u_\pi L_3^r \Big\}
\nonumber\\&&\hspace{-0.7cm}
  + \overline{B}_{22}(m_K^2,m_K^2,s_\pi) \Big\{
m_\pi^2   \Big( 32/3 L_1^r + 368/3 L_2^r + 32 L_3^r \Big)
+ m_K^2   \Big( 32/3 L_1^r + 80/3 L_2^r + 8 L_3^r \Big) \Big\}
\nonumber\\&&\hspace{-0.7cm}
  + \overline{B}_{22}(m_K^2,m_K^2,s_\pi)   \Big\{ 32/3 s_\pi L_1^r - 208/3 s_\pi L_2^r - 12 s_\pi 
         L_3^r - 16/3 t_\pi L_1^r - 64/3 t_\pi L_2^r - 6 t_\pi L_3^r 
\nonumber\\&&\hspace{-0.7cm}
- 16/3 u_\pi L_1^r - 16/3 
         u_\pi L_2^r - 2 u_\pi L_3^r \Big\}
\nonumber\\&&\hspace{-0.7cm}
  + \overline{B}_{22}(m_K^2,m_\eta^2,t_\pi) \Big\{ m_\pi^2   
\Big(  - 30 L_4^r - 30 L_5^r \Big)
 - 60   m_K^2  L_4^r \Big\}
\nonumber\\&&\hspace{-0.7cm}
+ \overline{B}_{22}(m_\eta^2,m_K^2,t_\pi) m_\pi^2   \Big\{ 268/9 L_3^r + 26 L_4^r - 4/3 L_5^r + 10 
         L_9^r \Big\}
\nonumber\\&&\hspace{-0.7cm}
  + \overline{B}_{22}(m_\eta^2,m_K^2,t_\pi) m_K^2   \Big\{ 8/9 L_3^r + 52 L_4^r - 8/3 L_5^r + 5 L_9^r \Big\}
\nonumber\\&&\hspace{-0.7cm}
  + \overline{B}_{22}(m_\eta^2,m_K^2,t_\pi)   \Big\{  - 22/3 s_\pi L_3^r - 5 s_\pi L_9^r + 20/3 t_\pi 
         L_3^r - 5 t_\pi L_9^r - 16/3 u_\pi L_3^r - 5 u_\pi L_9^r \Big\}
\nonumber\\&&\hspace{-0.7cm}
  + \overline{B}_{22}(m_\eta^2,m_\eta^2,s_\pi) \Big\{ 
m_\pi^2   \Big( 112/3 L_2^r + 136/9 L_3^r \Big)
  - 24 s_\pi L_2^r - 8 s_\pi L_3^r \Big\}
\nonumber\\&&\hspace{-0.7cm}
  + \overline{B}_{31}(m_\pi^2,m_\pi^2,s_\pi) \Big(m_\pi^2+m_K^2\Big)
   \Big\{  - 128/3 s_\pi L_2^r - 32/3 s_\pi L_3^r \Big\}
\nonumber\\&&\hspace{-0.7cm}
  + \overline{B}_{31}(m_\pi^2,m_\pi^2,s_\pi) 8/3  \Big\{  - 6 s_\pi^2 L_1^r - 3 s_\pi^2 L_2^r - 3 s_\pi^2 
L_3^r + \Big( 8 L_2^r + 2 L_3^r\Big) \Big(u_\pi+t_\pi\Big) s_\pi
        \Big\}
\nonumber\\&&\hspace{-0.7cm}
  + \overline{B}_{31}(m_\pi^2,m_K^2,t_\pi) \Big\{
m_\pi^4   \Big(  - 32 L_1^r - 16 L_2^r - 4 L_3^r \Big)
+ m_\pi^2 m_K^2 \Big( 32 L_1^r + 16 L_2^r + 4 L_3^r \Big) \Big\}
\nonumber\\&&\hspace{-0.7cm}
  + \overline{B}_{31}(m_\pi^2,m_K^2,t_\pi) m_\pi^2   \Big\{ 
\Big( 8 L_1^r + 4 L_2^r + L_3^r \Big) \Big( s_\pi+ u_\pi\Big)+ 8 
         t_\pi L_1^r + 4 t_\pi L_2^r - t_\pi L_3^r \Big\}
\nonumber\\&&\hspace{-0.7cm}
  + \overline{B}_{31}(m_\pi^2,m_K^2,t_\pi) m_K^2   \Big\{ 
-4  \Big( s_\pi 
- t_\pi/3+ u_\pi \Big) \Big(2 L_1^r +L_2^r \Big)
-\Big( s_\pi + t_\pi/3+u_\pi\Big) L_3^r
          \Big\}
\nonumber\\&&\hspace{-0.7cm}
  + \overline{B}_{31}(m_\pi^2,m_K^2,t_\pi)   \Big\{  \Big( - 8 L_1^r 
- 4 L_2^r + 
 L_3^r\Big) \Big(s_\pi + u_\pi \Big) t_\pi/3 + \Big(8 L_1^r + 4 L_2^r + L_3^r\Big)t_\pi^2
          \Big\}
\nonumber\\&&\hspace{-0.7cm}
  + \overline{B}_{31}(m_\pi^2,m_K^2,u_\pi) \Big(m_\pi^2+ m_K^2/3\Big)
   \Big\{  - 16 u_\pi L_1^r - 8 u_\pi L_2^r - 8 u_\pi 
         L_3^r \Big\}
\nonumber\\&&\hspace{-0.7cm}
  + \overline{B}_{31}(m_\pi^2,m_K^2,u_\pi) 8/3  \Big\{ 2 s_\pi u_\pi L_1^r +  s_\pi u_\pi L_2^r + 
          s_\pi u_\pi L_3^r + 2 t_\pi u_\pi L_1^r +  t_\pi u_\pi L_2^r +  t_\pi u_\pi L_3^r \Big\}
\nonumber\\&&\hspace{-0.7cm}
  + \overline{B}_{31}(m_K^2,m_K^2,s_\pi) \Big(m_\pi^2+m_K^2\Big)   \Big\{  - 32/3 s_\pi L_1^r - 80/3 s_\pi L_2^r - 8 
         s_\pi L_3^r \Big\}
\nonumber\\&&\hspace{-0.7cm}
  + \overline{B}_{31}(m_K^2,m_K^2,s_\pi) 4/3   \Big\{ 
\Big(4 L_1^r + 10 L_2^r + 3 
         L_3^r \Big) \Big(t_\pi  +u_\pi\Big) s_\pi\Big\}
\nonumber\\&&\hspace{-0.7cm}
  + \overline{B}_{31}(m_K^2,m_\eta^2,t_\pi) \Big\{ 
m_\pi^2 \Big(  - s_\pi  - t_\pi  - u_\pi  \Big) L_3^r
+ m_K^2 \Big( s_\pi - 1/3 t_\pi + u_\pi \Big) L_3^r \Big\}
\nonumber\\&&\hspace{-0.7cm}
  + \overline{B}_{31}(m_K^2,m_\eta^2,t_\pi) \Big\{ 4 L_3^r 
\Big(m_\pi^4-m_\pi^2 m_K^2\Big)
  + \Big( 1/3 s_\pi t_\pi - t_\pi^2 + 1/3 t_\pi 
         u_\pi \Big) L_3^r \Big\}
\nonumber\\&&\hspace{-0.7cm}
  + \overline{B}_{32}(m_\pi^2,m_\pi^2,s_\pi) \Big\{ m_\pi^2 
  \Big( 64 L_1^r - 160/3 L_2^r + 32/3 L_3^r \Big)
  -  m_K^2  \Big( 256/3 L_2^r + 64/3 L_3^r \Big) \Big\}
\nonumber\\&&\hspace{-0.7cm}
  - \overline{B}_{32}(m_\pi^2,m_\pi^2,s_\pi) 32/3  \Big\{ 6 s_\pi L_1^r + 7 s_\pi L_2^r + 4 
         s_\pi L_3^r - 4 t_\pi L_2^r - t_\pi L_3^r - 4 u_\pi L_2^r - u_\pi L_3^r
          \Big\}
\nonumber\\&&\hspace{-0.7cm}
  + \overline{B}_{32}(m_\pi^2,m_K^2,t_\pi) \Big\{ 
m_\pi^2   \Big( 96 L_1^r + 48 L_2^r + 8 L_3^r \Big)
  - m_K^2   \Big(  32/3 L_1^r + 16/3 L_2^r + 8/3 L_3^r \Big) \Big\}
\nonumber\\&&\hspace{-0.7cm}
  + \overline{B}_{32}(m_\pi^2,m_K^2,t_\pi)   \Big\{  - 64/3 s_\pi L_1^r - 32/3 s_\pi L_2^r - 4/3 s_\pi
          L_3^r + 112/3 t_\pi L_1^r + 56/3 t_\pi L_2^r 
\nonumber\\&&\hspace{-0.7cm}
 + 10/3 t_\pi L_3^r - 112/3 u_\pi L_1^r - 
         56/3 u_\pi L_2^r - 10/3 u_\pi L_3^r \Big\}
\nonumber\\&&\hspace{-0.7cm}
  - \overline{B}_{32}(m_\pi^2,m_K^2,u_\pi) \Big\{
16/3 \Big(3 m_\pi^2 + m_K^2 +s_\pi+  t_\pi -  u_\pi \Big)
  \Big( 2 L_1^r + L_2^r + L_3^r\Big) \Big\}
\nonumber\\&&\hspace{-0.7cm}
  + \overline{B}_{32}(m_K^2,m_K^2,s_\pi) \Big(m_\pi^2+m_K^2\Big)   \Big\{  - 64/3 L_1^r - 160/3 L_2^r - 16 L_3^r \Big\}
\nonumber\\&&\hspace{-0.7cm}
  + \overline{B}_{32}(m_K^2,m_K^2,s_\pi)   \Big\{  - 32 s_\pi L_1^r - 80 s_\pi L_2^r - 24 s_\pi 
L_3^r + 80 t_\pi L_1^r + 104 t_\pi L_2^r + 48 t_\pi L_3^r 
\nonumber\\&&\hspace{-0.7cm}
- 16 u_\pi L_1^r + 56
          u_\pi L_2^r \Big\}/3
\nonumber\\&&\hspace{-0.7cm}
  + \overline{B}_{32}(m_K^2,m_\eta^2,t_\pi) L_3^r 
\Big\{  -12 m_\pi^2 +4/3m_K^2
  + 8/3 s_\pi  - 14/3 t_\pi + 14/3 u_\pi 
         \Big\}
\nonumber\\&&\hspace{-0.7cm}
+\overline{A}(m_\pi^2)
\Big\{ - 4/3 s_\pi L_1^r - 16 s_\pi L_2^r - 19/3 s_\pi L_3^r + 55/12 s_\pi L_9^r
- 27 t_\pi  L_2^r - 27/2 t_\pi L_3^r
\nonumber\\&&\hspace{-0.7cm}
+ 55/12 t_\pi L_9^r
- 27 u_\pi
         L_2^r 
+ 55/12 u_\pi L_9^r\Big\} 
\nonumber\\&&\hspace{-0.7cm}
+\overline{A}(m_K^2)
\Big\{ - 56/3 s_\pi L_1^r - 32/3 s_\pi  L_2^r - 10 s_\pi  
         L_3^r + 11/2 s_\pi L_9^r 
- 14/3 t_\pi L_2^r + 17/3 t_\pi L_3^r 
\nonumber\\&&\hspace{-0.7cm}
+ 11/2 t_\pi L_9^r
- 110/3 u_\pi L_2^r + 11/2 u_\pi L_9^r\Big\}
\nonumber\\&&\hspace{-0.7cm}
+\overline{A}(m_\eta^2)
\Big\{ - 12 s_\pi L_1^r - s_\pi L_3^r 
 - 
   3 t_\pi L_2^r + 11/6 t_\pi L_3^r 
- 3 u_\pi L_2^r + 7/4 \Big(s_\pi+t_\pi+u_\pi\Big) L_9^r\Big\}
\end{eqnarray}
\noindent
For $G$ form-factor the expression reads
\noindent
\begin{eqnarray}
G_{LL} &&\hspace{-0.7cm} = 
   m_\pi^4 /(16\pi^2)   \Big\{  - 2 L_1^r - 10/9 L_2^r - 49/108 L_3^r \Big\}
\nonumber\\&&\hspace{-0.7cm}
  + m_\pi^4   \Big\{  - 16 L_3^r L_5^r + 32 L_4^r L_5^r + 56 (L_5^r)^2 - 64 L_5^r L_6^r - 64 L_5^r
          L_8^r - 16 L_5^r L_9^r \Big\}
\nonumber\\&&\hspace{-0.7cm}
  + m_\pi^2 m_K^2 /(16\pi^2)   \Big\{ 8/9 L_2^r + 8/27 L_3^r \Big\}
\nonumber\\&&\hspace{-0.7cm}
  + m_\pi^2 m_K^2   \Big\{ 64 L_4^r L_5^r - 32 (L_5^r)^2 - 128 L_5^r L_6^r + 8 L_5^r L_9^r \Big\}
\nonumber\\&&\hspace{-0.7cm}
  + m_\pi^2 m_\eta^2 /(16\pi^2)   \Big\{  - 2/3 L_5^r \Big\}
\nonumber\\&&\hspace{-0.7cm}
  + m_\pi^2   \Big\{ 8 s_\pi L_5^r L_9^r + 32 t_\pi L_2^r L_5^r + 16 t_\pi L_3^r L_5^r + 8 t_\pi L_5^r 
         L_9^r - 32 u_\pi L_2^r L_5^r + 8 u_\pi L_5^r L_9^r 
\nonumber\\&&\hspace{-0.7cm}
+ \overline{A}(m_\pi^2) \Big( - 4 L_1^r - 10  L_2^r - 
19 L_3^r + 46/3 L_4^r + 161/6 L_5^r - 83/6 L_9^r\Big) 
\nonumber\\&&\hspace{-0.7cm}
+ \overline{A}(m_K^2) \Big( 32
L_2^r + 17/3 L_3^r - 8/3 L_4^r + 25/3 L_5^r - 19/3 
L_9^r \Big)
\nonumber\\&&\hspace{-0.7cm}
+ \overline{A}(m_\eta^2) \Big(8/3 L_1^r + 2/3 L_2^r + 53/18 L_3^r - 2 
          L_4^r - 11/6 L_5^r + 1/2  L_9^r \Big)\Big\}
\nonumber\\&&\hspace{-0.7cm}
  + m_K^4 /(16\pi^2)   \Big\{ 2 L_1^r - 7/9 L_2^r - 1/108 L_3^r \Big\}
\nonumber\\&&\hspace{-0.7cm}
  + m_K^4   \Big\{ 16 L_3^r L_5^r + 8 (L_5^r)^2 + 8 L_5^r L_9^r \Big\}
\nonumber\\&&\hspace{-0.7cm}
  + m_K^2 m_\eta^2 /(16\pi^2)   \Big\{ 8/3 L_5^r \Big\}
\nonumber\\&&\hspace{-0.7cm}
  + m_K^2   \Big\{  - 8 s_\pi L_5^r L_9^r - 32 t_\pi L_2^r L_5^r - 16 t_\pi L_3^r L_5^r - 8 t_\pi 
         L_5^r L_9^r + 32 u_\pi L_2^r L_5^r - 8 u_\pi L_5^r L_9^r 
\nonumber\\&&\hspace{-0.7cm}
+ \overline{A}(m_\pi^2) \Big( - 27/2 L_3^r - 4/3 
L_4^r - 2/3 L_5^r - 83/12 L_9^r\Big) 
\nonumber\\&&\hspace{-0.7cm}
+\overline{A}(m_K^2)\Big( 4 L_1^r + 26 L_2^r
     + 5/6 L_3^r + 32/3 L_4^r - 8/3 L_5^r - 19/6 L_9^r \Big)
\nonumber\\&&\hspace{-0.7cm}
+ \overline{A}(m_\eta^2) \Big(- 
         32/3 L_1^r - 8/3 L_2^r - 47/18 L_3^r - 4 L_4^r - 8/3 
L_5^r + 1/4 L_9^r \Big) \Big\}
\nonumber\\&&\hspace{-0.7cm}
  + m_\eta^4 /(16\pi^2)   \Big\{ L_2^r + 1/6 L_3^r - 2 L_5^r \Big\}
\nonumber\\&&\hspace{-0.7cm}
  + m_\eta^2   \Big\{ 8 \overline{A}(m_\eta^2) L_1^r + 2 \overline{A}(m_\eta^2) L_2^r + 3 \overline{A}(m_\eta^2) L_3^r + \overline{A}(m_\eta^2) L_5^r \Big\}
\nonumber\\&&\hspace{-0.7cm}
  + \overline{B}(m_\pi^2,m_K^2,t_\pi) m_\pi^4   \Big\{  - 8/3 L_2^r + 8/3 L_3^r - 14/3 L_4^r - 3
          L_5^r + L_9^r \Big\}
\nonumber\\&&\hspace{-0.7cm}
  + \overline{B}(m_\pi^2,m_K^2,t_\pi) m_\pi^2 m_K^2   \Big\{ 8/3 L_2^r + 16/3 L_3^r - 8/3 L_4^r - 5 
         L_5^r + 20/3 L_6^r + 3/2 L_9^r \Big\}
\nonumber\\&&\hspace{-0.7cm}
  - \overline{B}(m_\pi^2,m_K^2,t_\pi) m_\pi^2/2   \Big\{ s_\pi L_9^r 
+2 t_\pi \Big(  8/3 L_2^r + 16/3 
L_3^r + 14/3 L_4^r - 3 L_5^r + 3/2 L_9^r\Big) + u_\pi L_9^r \Big\}
\nonumber\\&&\hspace{-0.7cm}
  + \overline{B}(m_\pi^2,m_K^2,t_\pi) m_K^4   \Big\{  - 8/3 L_4^r - 4/3 L_5^r + 40/3 L_6^r + 
         20/3 L_8^r + 1/2 L_9^r \Big\}
\nonumber\\&&\hspace{-0.7cm}
  + \overline{B}(m_\pi^2,m_K^2,t_\pi) m_K^2   \Big\{  - 1/2 s_\pi L_9^r - 4 t_\pi L_4^r - 2 t_\pi 
         L_5^r - t_\pi L_9^r - 1/2 u_\pi L_9^r \Big\}
\nonumber\\&&\hspace{-0.7cm}
  + \overline{B}(m_\pi^2,m_K^2,t_\pi)   \Big\{ 1/2 s_\pi t_\pi L_9^r + 1/2 t_\pi^2 L_9^r + 1/2 t_\pi
          u_\pi L_9^r \Big\}
\nonumber\\&&\hspace{-0.7cm}
  + \overline{B}(m_\pi^2,m_K^2,u_\pi) m_\pi^4   \Big\{ 56/3 L_2^r + 64/3 L_4^r + 20/3 L_5^r - 8/
         3 L_9^r \Big\}
\nonumber\\&&\hspace{-0.7cm}
  + \overline{B}(m_\pi^2,m_K^2,u_\pi) m_\pi^2 m_K^2   \Big\{ 40/3 L_2^r + 32 L_6^r + 16 L_8^r - 8/3 
         L_9^r \Big\}
\nonumber\\&&\hspace{-0.7cm}
  + \overline{B}(m_\pi^2,m_K^2,u_\pi) 4/3 m_\pi^2
   \Big\{ s_\pi L_9^r + t_\pi L_9^r - 10 u_\pi 
         L_2^r - 8 u_\pi L_4^r - 4 u_\pi L_5^r + 2 u_\pi L_9^r \Big\}
\nonumber\\&&\hspace{-0.7cm}
  + \overline{B}(m_\pi^2,m_K^2,u_\pi) m_K^4   \Big\{  - 16/3 L_4^r - 8/3 L_5^r - 2/3 L_9^r \Big\}
\nonumber\\&&\hspace{-0.7cm}
  + \overline{B}(m_\pi^2,m_K^2,u_\pi) m_K^2   \Big\{ 2/3 s_\pi L_9^r + 2/3 t_\pi L_9^r + 16/3 u_\pi 
         L_4^r + 8/3 u_\pi L_5^r + 4/3 u_\pi L_9^r \Big\}
\nonumber\\&&\hspace{-0.7cm}
  + \overline{B}(m_\pi^2,m_K^2,u_\pi)   \Big\{  - 2/3 s_\pi u_\pi L_9^r - 2/3 t_\pi u_\pi L_9^r - 2/
         3 u_\pi^2 L_9^r \Big\}
\nonumber\\&&\hspace{-0.7cm}
  + \overline{B}(m_\eta^2,m_K^2,t_\pi) m_\pi^4   \Big\{ 4 L_3^r + 28/3 L_4^r + 16/9 L_5^r - 8/3 
         L_6^r - 32/3 L_7^r - 8 L_8^r + 8/3 L_9^r \Big\}
\nonumber\\&&\hspace{-0.7cm}
  + \overline{B}(m_\eta^2,m_K^2,t_\pi) m_\pi^2 m_K^2/3   \Big\{  - 2/3 L_3^r + 46 L_4^r + 22/3 L_5^r
          - 20 L_6^r - 32 L_7^r - 16 L_8^r + 2 L_9^r \Big\}
\nonumber\\&&\hspace{-0.7cm}
  + \overline{B}(m_\eta^2,m_K^2,t_\pi) m_\pi^2 m_\eta^2   \Big\{ 2 L_4^r - 2 L_5^r \Big\}
\nonumber\\&&\hspace{-0.7cm}
  - \overline{B}(m_\eta^2,m_K^2,t_\pi) m_\pi^2/3   \Big\{  
\Big(s_\pi+u_\pi\Big) \Big(3 L_3^r - 4 L_9^r \Big) +t_\pi\Big( - 43/3  
L_3^r - 12 L_4^r + 2 L_5^r - 10 L_9^r \Big) 
          \Big\}
\nonumber\\&&\hspace{-0.7cm}
  + \overline{B}(m_\eta^2,m_K^2,t_\pi) m_K^4   \Big\{ 14/9 L_3^r - 20/3 L_4^r - 38/9 L_5^r - 8/
         3 L_6^r + 64/3 L_7^r + 28/3 L_8^r - 1/3 L_9^r \Big\}
\nonumber\\&&\hspace{-0.7cm}
+\overline{B}(m_\eta^2,m_K^2,t_\pi) \Big\{ m_K^2 m_\eta^2 
  \Big( 4 L_4^r + 8 L_5^r \Big)
   - 4 L_5^r m_\eta^4   \Big\}
\nonumber\\&&\hspace{-0.7cm}
  + \overline{B}(m_\eta^2,m_K^2,t_\pi) m_K^2/3   \Big\{ \Big(3 L_3^r + L_9^r \Big) \Big(s_\pi +  u_\pi\Big)
-t_\pi \Big( 47/3 L_3^r
+ 24 L_4^r - 4 L_5^r + 2 L_9^r\Big) \Big\}
\nonumber\\&&\hspace{-0.7cm}
  + \overline{B}(m_\eta^2,m_K^2,t_\pi)   \Big\{ 1/3 s_\pi t_\pi L_3^r + s_\pi t_\pi L_9^r + 7/3 t_\pi^2
          L_3^r + t_\pi^2 L_9^r + 1/3 t_\pi u_\pi L_3^r + t_\pi u_\pi L_9^r \Big\}
\nonumber\\&&\hspace{-0.7cm}
  + \overline{B}_1(m_\pi^2,m_K^2,t_\pi) m_\pi^4   \Big\{ 8 L_1^r + 20 L_2^r + 18 L_3^r + 46/3 L_4^r
          - 16/3 L_5^r + 4/3 L_9^r \Big\}
\nonumber\\&&\hspace{-0.7cm}
  + \overline{B}_1(m_\pi^2,m_K^2,t_\pi) m_\pi^2 m_K^2/3   \Big\{  - 16 L_1^r - 8 L_2^r + 2 L_3^r
          - 56 L_4^r - 36 L_5^r + 76 L_6^r + 48 L_8^r + 4 L_9^r \Big\}
\nonumber\\&&\hspace{-0.7cm}
  + \overline{B}_1(m_\pi^2,m_K^2,t_\pi) m_\pi^2/3   \Big\{  - 8 s_\pi L_1^r - 4 s_\pi L_2^r - 5 
s_\pi L_3^r - 2 s_\pi L_9^r + 24 t_\pi L_1^r - 12 t_\pi L_2^r 
\nonumber\\&&\hspace{-0.7cm}
- 39 t_\pi L_3^r 
- 22 t_\pi
L_4^r + 20 t_\pi L_5^r - 10 t_\pi L_9^r - 8 u_\pi L_1^r - 4 u_\pi L_2^r - 5 
         u_\pi L_3^r - 2 u_\pi L_9^r \Big\}
\nonumber\\&&\hspace{-0.7cm}
  + \overline{B}_1(m_\pi^2,m_K^2,t_\pi) m_K^4/3   \Big\{  - 8 L_1^r - 4 L_2^r + 4 L_3^r + 40
L_4^r + 20 L_5^r - 40 L_6^r - 20 L_8^r + L_9^r \Big\}
\nonumber\\&&\hspace{-0.7cm}
  + \overline{B}_1(m_\pi^2,m_K^2,t_\pi) m_K^2/3   \Big\{ 8 s_\pi L_1^r + 4 s_\pi L_2^r - s_\pi 
L_3^r - s_\pi L_9^r + 8 t_\pi L_1^r - 4 t_\pi L_2^r - 17 t_\pi L_3^r 
\nonumber\\&&\hspace{-0.7cm}
+ 4 
t_\pi L_4^r + 2 t_\pi L_5^r 
- 5 t_\pi L_9^r + 8 u_\pi L_1^r + 4 u_\pi L_2^r - 
          u_\pi L_3^r - u_\pi L_9^r \Big\}
\nonumber\\&&\hspace{-0.7cm}
  + \overline{B}_1(m_\pi^2,m_K^2,t_\pi)   \Big\{  
\Big(s_\pi+u_\pi\Big) t_\pi \Big( - 8 L_1^r - 4 L_2^r + 
L_3^r + 4 L_9^r\Big) + t_\pi^2 \Big(8 L_2^r + 13 L_3^r + 4
L_9^r\Big)
         \Big\}
\nonumber\\&&\hspace{-0.7cm}
  + \overline{B}_1(m_\pi^2,m_K^2,u_\pi) m_\pi^4   \Big\{  - 40 L_1^r - 20 L_2^r - 20 L_3^r - 56/3 
         L_4^r \Big\}
\nonumber\\&&\hspace{-0.7cm}
  + \overline{B}_1(m_\pi^2,m_K^2,u_\pi) m_\pi^2 m_K^2   \Big\{  - 32/3 L_1^r - 16/3 L_2^r - 16/3 L_3^r
          - 16/3 L_4^r \Big\}
\nonumber\\&&\hspace{-0.7cm}
  + \overline{B}_1(m_\pi^2,m_K^2,u_\pi) 4/3 m_\pi^2   \Big\{ 
\Big( 8 L_1^r + 4 L_2^r + 4 
L_3^r \Big) \Big( s_\pi +  t_\pi \Big) 
\nonumber\\&&\hspace{-0.7cm}
+ u_\pi \Big(12 L_1^r - 
12 L_2^r 
+ 6 L_3^r + 2 L_4^r- L_5^r+ L_9^r \Big) \Big\}
\nonumber\\&&\hspace{-0.7cm}
  + \overline{B}_1(m_\pi^2,m_K^2,u_\pi) m_K^4   \Big\{  - 40/3 L_1^r - 20/3 L_2^r - 20/3 L_3^r
     + 8 L_4^r \Big\}
\nonumber\\&&\hspace{-0.7cm}
  + \overline{B}_1(m_\pi^2,m_K^2,u_\pi) 8/3 m_K^2   \Big\{ 2 s_\pi L_1^r + s_\pi L_2^r + s_\pi 
L_3^r + 2 t_\pi L_1^r + t_\pi L_2^r + t_\pi L_3^r 
\nonumber\\&&\hspace{-0.7cm}
 + 2 u_\pi L_1^r - 4 
         u_\pi L_2^r + u_\pi L_3^r - 7 u_\pi L_4^r - 2 u_\pi L_5^r + 1/4 u_\pi L_9^r \Big\}
\nonumber\\&&\hspace{-0.7cm}
  + \overline{B}_1(m_\pi^2,m_K^2,u_\pi) 1/3  \Big\{  - 16 s_\pi u_\pi L_1^r - 8 s_\pi u_\pi L_2^r - 8
s_\pi u_\pi L_3^r - 2 s_\pi u_\pi L_9^r 
\nonumber\\&&\hspace{-0.7cm}
- 16 t_\pi u_\pi L_1^r - 8 t_\pi u_\pi L_2^r
          - 8 t_\pi u_\pi L_3^r - 2 t_\pi u_\pi L_9^r + 24 u_\pi^2 L_1^r + 52 u_\pi^2 L_2^r
     + 12 u_\pi^2 L_3^r - 2 u_\pi^2 L_9^r \Big\}
\nonumber\\&&\hspace{-0.7cm}
  + \overline{B}_1(m_K^2,m_\pi^2,t_\pi) 10/3 \Big\{ m_\pi^2 m_K^2 
  \Big( L_4^r - 2 L_6^r \Big)
  + m_K^4   \Big( 2 L_4^r + L_5^r - 4 L_6^r - 2
         L_8^r \Big) \Big\}
\nonumber\\&&\hspace{-0.7cm}
  + \overline{B}_1(m_K^2,m_\eta^2,t_\pi) m_\pi^4   \Big\{  - 1/3 L_4^r - 1/3 L_5^r + 8/3 L_6^r + 8/
         3 L_8^r \Big\}
\nonumber\\&&\hspace{-0.7cm}
  + \overline{B}_1(m_K^2,m_\eta^2,t_\pi) \Big\{ 
m_\pi^2 m_K^2   \Big(  - 25/3 L_4^r - 7 L_5^r + 20/3 L_6^r \Big)
  + 6  m_K^2  6 t_\pi L_4^r \Big\}
\nonumber\\&&\hspace{-0.7cm}
  + \overline{B}_1(m_K^2,m_\eta^2,t_\pi) \Big\{
3 m_\pi^2 t_\pi \Big( L_4^r +  L_5^r \Big)
  +  m_K^4/3   \Big(  - 46 L_4^r - 2 L_5^r + 8 L_6^r + 4
          L_8^r \Big) \Big\}
\nonumber\\&&\hspace{-0.7cm}
  + \overline{B}_1(m_\eta^2,m_K^2,t_\pi) m_\pi^4/3   \Big\{  - 25 L_3^r - 55 L_4^r - 16/3 L_5^r
     + 8 L_6^r + 32 L_7^r + 24 L_8^r - 17 L_9^r \Big\}
\nonumber\\&&\hspace{-0.7cm}
  + \overline{B}_1(m_\eta^2,m_K^2,t_\pi) m_\pi^2 m_K^2/3   \Big\{ 56/3 L_3^r - 73 L_4^r - 22/3 L_5^r + 
         20 L_6^r + 32 L_7^r + 16 L_8^r + 5/2 L_9^r \Big\}
\nonumber\\&&\hspace{-0.7cm}
  - \overline{B}_1(m_\eta^2,m_K^2,t_\pi) \Big\{ 
m_\pi^2 m_\eta^2   \Big(   2 L_4^r - 4/3 L_5^r \Big)
  + m_K^2 m_\eta^2   \Big(  4 L_4^r + 16/3 L_5^r \Big) 
  + 2  m_\eta^4   L_5^r \Big\}
\nonumber\\&&\hspace{-0.7cm}
  + \overline{B}_1(m_\eta^2,m_K^2,t_\pi) m_\pi^2   \Big\{ 19/9 s_\pi L_3^r + 17/6 s_\pi L_9^r + 79/9 
         t_\pi L_3^r + 3 t_\pi L_4^r - 2 t_\pi L_5^r 
\nonumber\\&&\hspace{-0.7cm}
+ 35/6 t_\pi L_9^r 
+ 19/9 u_\pi L_3^r + 17/6 
         u_\pi L_9^r \Big\}
\nonumber\\&&\hspace{-0.7cm}
  + \overline{B}_1(m_\eta^2,m_K^2,t_\pi) m_K^4/3   \Big\{ 7/3 L_3^r + 74 L_4^r + 38/3 L_5^r + 8
          L_6^r - 64 L_7^r - 28 L_8^r + 11/2 L_9^r \Big\}
\nonumber\\&&\hspace{-0.7cm}
  + \overline{B}_1(m_\eta^2,m_K^2,t_\pi) m_K^2   \Big\{  - 25/9 s_\pi L_3^r - 11/6 s_\pi L_9^r + 41/9
          t_\pi L_3^r + 6 t_\pi L_4^r - 4 t_\pi L_5^r 
\nonumber\\&&\hspace{-0.7cm}
- 1/3 t_\pi L_9^r 
- 25/9 u_\pi L_3^r - 11/6 
         u_\pi L_9^r \Big\}
\nonumber\\&&\hspace{-0.7cm}
  - \overline{B}_1(m_\eta^2,m_K^2,t_\pi)   \Big\{  
\Big(  1/3  L_3^r + 3/2 L_9^r \Big) \Big( s_\pi
+u_\pi \Big)t_\pi + 4 
t_\pi^2 L_3^r + 3/2 t_\pi^2 L_9^r \Big\}
\nonumber\\&&\hspace{-0.7cm}
  + \overline{B}_{21}(m_\pi^2,m_K^2,t_\pi) m_\pi^4   \Big\{  - 40 L_1^r - 20 L_2^r - 10 L_3^r - 16 
         L_4^r - 9 L_5^r + 3 L_9^r \Big\}
\nonumber\\&&\hspace{-0.7cm}
  + \overline{B}_{21}(m_\pi^2,m_K^2,t_\pi) m_\pi^2 m_K^2   \Big\{ 112/3 L_1^r + 56/3 L_2^r + 10/3 L_3^r
     + 8 L_4^r + 9 L_5^r - 3/2 L_9^r \Big\}
\nonumber\\&&\hspace{-0.7cm}
  + \overline{B}_{21}(m_\pi^2,m_K^2,t_\pi) m_\pi^2   \Big\{ 32/3 s_\pi L_1^r + 16/3 s_\pi L_2^r + 8/3 
         s_\pi L_3^r - 3/2 s_\pi L_9^r - 16 t_\pi L_2^r - 12 t_\pi L_3^r 
\nonumber\\&&\hspace{-0.7cm}
+ 52/3 t_\pi L_4^r + 23/3
          t_\pi L_5^r - 19/6 t_\pi L_9^r + 32/3 u_\pi L_1^r + 16/3 u_\pi L_2^r + 8/3 u_\pi L_3^r
          - 3/2 u_\pi L_9^r \Big\}
\nonumber\\&&\hspace{-0.7cm}
  + \overline{B}_{21}(m_\pi^2,m_K^2,t_\pi) m_K^4   \Big\{ 8/3 L_1^r + 4/3 L_2^r - 4/3 L_3^r + 8 
         L_4^r - 3/2 L_9^r \Big\}
\nonumber\\&&\hspace{-0.7cm}
  + \overline{B}_{21}(m_\pi^2,m_K^2,t_\pi) m_K^2   \Big\{  - 32/3 s_\pi L_1^r - 16/3 s_\pi L_2^r - 2/3
          s_\pi L_3^r + 3/2 s_\pi L_9^r + 16/3 t_\pi L_4^r 
\nonumber\\&&\hspace{-0.7cm}
+ 20/3 t_\pi L_5^r + 2/3 t_\pi L_9^r - 
         32/3 u_\pi L_1^r - 16/3 u_\pi L_2^r - 2/3 u_\pi L_3^r + 3/2 u_\pi L_9^r \Big\}
\nonumber\\&&\hspace{-0.7cm}
  + \overline{B}_{21}(m_\pi^2,m_K^2,t_\pi)   \Big\{ 5/2 s_\pi t_\pi L_9^r + 
t_\pi^2 \Big( 24 L_1^r + 44 
L_2^r + 34 L_3^r + 5/2 L_9^r\Big) + 5/2 t_\pi u_\pi L_9^r \Big\}/3
\nonumber\\&&\hspace{-0.7cm}
  + \overline{B}_{21}(m_\pi^2,m_K^2,u_\pi) \Big\{
\Big( 20 m_\pi^4 +16/3 m_\pi^2 m_K^2 \Big)  \Big( 2 L_1^r + L_2^r+ L_3^r \Big)
\Big\}
\nonumber\\&&\hspace{-0.7cm}
  + \overline{B}_{21}(m_\pi^2,m_K^2,u_\pi) 16/3 m_\pi^2   \Big\{  - 2 s_\pi L_1^r - s_\pi L_2^r - 
s_\pi L_3^r - 2 t_\pi L_1^r - t_\pi L_2^r - t_\pi L_3^r + u_\pi L_4^r
          \Big\}
\nonumber\\&&\hspace{-0.7cm}
  + \overline{B}_{21}(m_\pi^2,m_K^2,u_\pi) m_K^4   \Big\{ 40/3 L_1^r + 20/3 L_2^r + 20/3 L_3^r \Big\}
\nonumber\\&&\hspace{-0.7cm}
  - \overline{B}_{21}(m_\pi^2,m_K^2,u_\pi) 8/3 m_K^2   \Big\{2 s_\pi L_1^r + s_\pi L_2^r + 
s_\pi L_3^r + 2 t_\pi L_1^r +t_\pi L_2^r +t_\pi L_3^r \Big\}
\nonumber\\&&\hspace{-0.7cm}
  + \overline{B}_{21}(m_\pi^2,m_K^2,u_\pi)   \Big\{  - 8 u_\pi^2 L_1^r + 4/3 u_\pi^2 L_2^r - 4 u_\pi^2
          L_3^r \Big\}
\nonumber\\&&\hspace{-0.7cm}
  + \overline{B}_{21}(m_K^2,m_\eta^2,t_\pi) \Big\{
- 9 m_\pi^4   \Big(  L_4^r + L_5^r \Big)
  + 9 m_\pi^2 m_K^2   \Big(  - L_4^r + L_5^r \Big) 
  + 3 m_\pi^2 t_\pi  \Big( L_4^r + L_5^r \Big) \Big\}
\nonumber\\&&\hspace{-0.7cm}
  + \overline{B}_{21}(m_K^2,m_\eta^2,t_\pi) \Big\{
18 m_K^4   18 L_4^r 
  + 6 m_K^2   t_\pi L_4^r 
  + m_\pi^4   \Big( 13/3 L_3^r + 9 L_4^r + 3 L_9^r \Big) \Big\}
\nonumber\\&&\hspace{-0.7cm}
  + \overline{B}_{21}(m_\eta^2,m_K^2,t_\pi) m_\pi^2 m_K^2   \Big\{  - 56/9 L_3^r + 9 L_4^r - 3/2 L_9^r \Big\}
\nonumber\\&&\hspace{-0.7cm}
  - \overline{B}_{21}(m_\eta^2,m_K^2,t_\pi) m_\pi^2   \Big\{
 \Big( 10/9 L_3^r + 3/2 L_9^r \Big)\Big(s_\pi +u_\pi\Big)
- t_\pi\Big( - 4 
L_3^r + L_4^r + 4/3 _5^r - 5/2 L_9^r\Big)
          \Big\}
\nonumber\\&&\hspace{-0.7cm}
  + \overline{B}_{21}(m_\eta^2,m_K^2,t_\pi) m_K^4   \Big\{  - 7/9 L_3^r - 18 L_4^r - 3/2 L_9^r \Big\}
\nonumber\\&&\hspace{-0.7cm}
  + \overline{B}_{21}(m_\eta^2,m_K^2,t_\pi) m_K^2   \Big\{ 
\Big( 16/9  L_3^r + 3/2 L_9^r \Big) \Big(s_\pi + u_\pi\Big)
+ 2 t_\pi 
         L_4^r + 8/3 t_\pi L_5^r + t_\pi L_9^r \Big\}
\nonumber\\&&\hspace{-0.7cm}
  + \overline{B}_{21}(m_\eta^2,m_K^2,t_\pi)   \Big\{ 1/2 s_\pi t_\pi L_9^r + 5/3 t_\pi^2 L_3^r + 1/2 
         t_\pi^2 L_9^r + 1/2 t_\pi u_\pi L_9^r \Big\}
\nonumber\\&&\hspace{-0.7cm}
  + \overline{B}_{22}(m_\pi^2,m_\pi^2,s_\pi) \Big\{
m_\pi^2   \Big( 8 L_3^r - 32 L_4^r - 24 L_5^r + 8 L_9^r \Big)
  +  m_K^2   \Big( 8 L_3^r + 4 L_9^r \Big) \Big\}
\nonumber\\&&\hspace{-0.7cm}
  + \overline{B}_{22}(m_\pi^2,m_\pi^2,s_\pi) 4  \Big\{4 s_\pi L_1^r - 2 s_\pi L_2^r + 2 s_\pi L_3^r - 
s_\pi L_9^r - t_\pi L_3^r - t_\pi L_9^r - u_\pi L_3^r - u_\pi L_9^r \Big\}
\nonumber\\&&\hspace{-0.7cm}
 + \overline{B}_{22}(m_\pi^2,m_K^2,t_\pi) m_\pi^2   \Big\{  - 64/3 L_1^r - 80/3 L_2^r - 64/3 L_3^r + 
         4/3 L_4^r - 4/3 L_5^r + 4/3 L_9^r \Big\}
\nonumber\\&&\hspace{-0.7cm}
  + \overline{B}_{22}(m_\pi^2,m_K^2,t_\pi) m_K^2   \Big\{ 32 L_1^r + 4 L_3^r - 8/3 L_4^r + 20/3 L_5^r
     + 2/3 L_9^r \Big\}
\nonumber\\&&\hspace{-0.7cm}
  + \overline{B}_{22}(m_\pi^2,m_K^2,t_\pi)   \Big\{ 16/3 s_\pi L_1^r + 8/3 s_\pi L_2^r + 4/3 s_\pi L_3^r
          - 2/3 s_\pi L_9^r + 32/3 t_\pi L_1^r + 8 t_\pi L_2^r 
\nonumber\\&&\hspace{-0.7cm}
+ 16 t_\pi L_3^r - 2/3 t_\pi L_9^r
          - 32/3 u_\pi L_1^r + 8/3 u_\pi L_2^r - 8/3 u_\pi L_3^r - 2/3 u_\pi L_9^r \Big\}
\nonumber\\&&\hspace{-0.7cm}
  + \overline{B}_{22}(m_\pi^2,m_K^2,u_\pi) m_\pi^2   \Big\{ 160/3 L_1^r 
+ 80/3 L_2^r + 80/3 L_3^r + 16/
         3 L_4^r \Big\}
\nonumber\\&&\hspace{-0.7cm}
  + \overline{B}_{22}(m_\pi^2,m_K^2,u_\pi) m_K^2   \Big\{ 32 L_1^r + 16 L_2^r + 16 L_3^r \Big\}
\nonumber\\&&\hspace{-0.7cm}
  + \overline{B}_{22}(m_\pi^2,m_K^2,u_\pi)   \Big\{  - 16/3 s_\pi L_1^r - 8/3 s_\pi L_2^r - 8/3 s_\pi 
         L_3^r - 16/3 t_\pi L_1^r - 8/3 t_\pi L_2^r 
\nonumber\\&&\hspace{-0.7cm}
- 8/3 t_\pi L_3^r 
- 80/3 u_\pi L_1^r - 8 u_\pi
          L_2^r - 40/3 u_\pi L_3^r \Big\}
\nonumber\\&&\hspace{-0.7cm}
  + \overline{B}_{22}(m_K^2,m_K^2,s_\pi) m_\pi^2   \Big\{ 32 L_1^r + 12 L_3^r - 12 L_5^r + 4 L_9^r \Big\}
\nonumber\\&&\hspace{-0.7cm}
  + \overline{B}_{22}(m_K^2,m_K^2,s_\pi) m_K^2   \Big\{ 32 L_1^r + 12 L_3^r - 16 L_4^r + 2 L_9^r \Big\}
\nonumber\\&&\hspace{-0.7cm}
  + \overline{B}_{22}(m_K^2,m_K^2,s_\pi)  2  \Big\{ 2 s_\pi L_3^r - s_\pi L_9^r - 8 t_\pi L_1^r - 3 
t_\pi L_3^r -t_\pi L_9^r -8 u_\pi L_1^r - 3 u_\pi L_3^r - u_\pi L_9^r \Big\}
\nonumber\\&&\hspace{-0.7cm}
  + \overline{B}_{22}(m_K^2,m_\eta^2,t_\pi) 
\Big\{ m_\pi^2   \Big(  - 6 L_4^r - 6 L_5^r \Big)
  -12  m_K^2   L_4^r \Big\}
\nonumber\\&&\hspace{-0.7cm}
  + \overline{B}_{22}(m_\eta^2,m_K^2,t_\pi) m_\pi^2   \Big\{  - 16/9 L_3^r + 10 L_4^r + 4/3 L_5^r + 2 
         L_9^r \Big\}
\nonumber\\&&\hspace{-0.7cm}
  + \overline{B}_{22}(m_\eta^2,m_K^2,t_\pi) m_K^2   \Big\{ 28/9 L_3^r + 20 L_4^r + 8/3 L_5^r + L_9^r \Big\}
\nonumber\\&&\hspace{-0.7cm}
  + \overline{B}_{22}(m_\eta^2,m_K^2,t_\pi)   \Big\{  - 2/3 s_\pi L_3^r - s_\pi L_9^r + 16/3 t_\pi L_3^r
          - t_\pi L_9^r - 8/3 u_\pi L_3^r - u_\pi L_9^r \Big\}
\nonumber\\&&\hspace{-0.7cm}
  + \overline{B}_{31}(m_\pi^2,m_K^2,t_\pi) \Big\{
\Big(m_\pi^4 -m_\pi^2 m_K^2\Big)  \Big( 32 L_1^r + 16 L_2^r + 4 L_3^r \Big)
  \Big\}
\nonumber\\&&\hspace{-0.7cm}
  - \overline{B}_{31}(m_\pi^2,m_K^2,t_\pi) m_\pi^2   \Big\{ 
4 \Big(2 L_1^r +  L_2^r \Big) \Big(s_\pi + t_\pi+ u_\pi \Big) + s_\pi L_3^r
- t_\pi L_3^r +  u_\pi L_3^r
          \Big\}
\nonumber\\&&\hspace{-0.7cm}
  + \overline{B}_{31}(m_\pi^2,m_K^2,t_\pi) m_K^2   \Big\{ 
\Big( s_\pi +u_\pi\Big) \Big(8 L_1^r + 4 L_2^r + L_3^r \Big)- 8/
         3 t_\pi L_1^r - 4/3 t_\pi L_2^r + 1/3 t_\pi L_3^r  
         \Big\}
\nonumber\\&&\hspace{-0.7cm}
  + \overline{B}_{31}(m_\pi^2,m_K^2,t_\pi)   \Big\{ 
\Big(s_\pi+u_\pi\Big) t_\pi/3 \Big( 8 L_1^r + 4  L_2^r -  
L_3^r\Big)  - 8 t_\pi^2 L_1^r - 4 t_\pi^2 L_2^r - t_\pi^2 L_3^r 
\Big\}
\nonumber\\&&\hspace{-0.7cm}
  - \overline{B}_{31}(m_\pi^2,m_K^2,u_\pi) 8 u_\pi \Big\{
\Big(m_\pi^2+ m_K^2/3\Big) \Big(2 L_1^r + L_2^r + 
         L_3^r \Big) \Big\}
\nonumber\\&&\hspace{-0.7cm}
  + \overline{B}_{31}(m_\pi^2,m_K^2,u_\pi) 8/3  \Big\{ 2 s_\pi u_\pi L_1^r +  s_\pi u_\pi L_2^r + 
s_\pi u_\pi L_3^r+2 t_\pi u_\pi L_1^r+t_\pi u_\pi L_2^r+t_\pi u_\pi L_3^r \Big\}
\nonumber\\&&\hspace{-0.7cm}
  + \overline{B}_{31}(m_K^2,m_\eta^2,t_\pi) L_3^r \Big\{ 
4 \Big(- 
m_\pi^4+m_\pi^2 m_K^2
\Big) 
  - 1/3 s_\pi t_\pi  + t_\pi^2  - 1/3 t_\pi
          u_\pi  \Big\}
\nonumber\\&&\hspace{-0.7cm}
+ \overline{B}_{31}(m_K^2,m_\eta^2,t_\pi) L_3^r
\Big\{ m_\pi^2   \Big( s_\pi + t_\pi + u_\pi  \Big)
  +  m_K^2   \Big(  - s_\pi  + 1/3 t_\pi  - u_\pi \Big) \Big\}
\nonumber\\&&\hspace{-0.7cm}
  + \overline{B}_{32}(m_\pi^2,m_K^2,t_\pi) \Big\{
m_\pi^2   \Big( 64 L_1^r + 32 L_2^r + 12 L_3^r \Big)
  - m_K^2/3  \Big( 64 L_1^r + 32 L_2^r + 4 L_3^r \Big) \Big\}
\nonumber\\&&\hspace{-0.7cm}
  + \overline{B}_{32}(m_\pi^2,m_K^2,t_\pi)   \Big\{  - 32 s_\pi L_1^r - 16 s_\pi L_2^r - 8 s_\pi
 L_3^r - 112 t_\pi L_1^r - 56 t_\pi L_2^r 
\nonumber\\&&\hspace{-0.7cm}
- 10 t_\pi L_3^r + 16 u_\pi L_1^r 
+ 8
 u_\pi L_2^r - 2 u_\pi L_3^r \Big\}/3
\nonumber\\&&\hspace{-0.7cm}
 - \overline{B}_{32}(m_\pi^2,m_K^2,u_\pi) \Big\{
 16/3 \Big(3 m_\pi^2+ 3 m_K^2- s_\pi - t_\pi + u_\pi\Big) 
  \Big(  2 L_1^r + L_2^r + L_3^r \Big) \Big\}
\nonumber\\&&\hspace{-0.7cm}
  + \overline{B}_{32}(m_K^2,m_K^2,s_\pi) \Big(m_\pi^2 +m_K^2\Big)  \Big\{  - 32 L_1^r - 16 L_2^r - 16 L_3^r \Big\}
\nonumber\\&&\hspace{-0.7cm}
  + \overline{B}_{32}(m_K^2,m_K^2,s_\pi)   \Big\{ 16 t_\pi L_1^r + 8 t_\pi L_2^r + 8 t_\pi L_3^r + 16 
         u_\pi L_1^r + 8 u_\pi L_2^r + 8 u_\pi L_3^r \Big\}
\nonumber\\&&\hspace{-0.7cm}
  + \overline{B}_{32}(m_K^2,m_\eta^2,t_\pi) \Big\{
 8 \Big( - m_\pi^2 +m_K^2/3\Big)  
   L_3^r 
  +  4/3 s_\pi L_3^r + 14/3 t_\pi L_3^r - 2/3 u_\pi L_3^r
          \Big\}
\nonumber\\&&\hspace{-0.7cm}
+
\overline{A}(m_\pi^2) \Big(
   83/12 s_\pi L_9^r + 27 t_\pi  L_2^r +27/2 t_\pi L_3^r +83/12 t_\pi L_9^r
- 27 u_\pi L_2^r + 83/12 u_\pi  L_9^r\Big)
\nonumber\\&&\hspace{-0.7cm}
+
\overline{A}(m_K^2) \Big(
 19/6 s_\pi L_9^r 
+ 14/3 t_\pi 
L_2^r - 17/3 t_\pi L_3^r + 19/6 t_\pi L_9^r 
- 110/3 u_\pi L_2^r + 19/6 u_\pi L_9^r \Big)
\nonumber\\&&\hspace{-0.7cm}
+
\overline{A}(m_\eta^2) \Big(
- 1/4 s_\pi L_9^r + 27 t_\pi
+ 3 t_\pi L_2^r
- 11/6 t_\pi L_3^r - 1/4 t_\pi L_9^r 
- 3 u_\pi 
L_2^r - 1/4 u_\pi L_9^r\Big)
\end{eqnarray}
%%%%%%%%%%%%%%%%%%%%%%%%%%%%%%%%%%%%%%%%%%%%%%%%%%%%%%%%%%%%%%%%%
\subsection{${\cal L}_6$ Contribution}
\label{counter}
Here we collect the relevant contribution coming from the tree diagram
involving a 
${\cal O}(p^6)$ Lagrangian vertex \cite{BCElag}. 
Each $C^r_i$ is finite and as was
explained in Sect. \ref{estimates}, at present, for their numerical estimate one
has to resort to models.

For the $F$ form-factor we obtain
\noindent
\ba
F_{ct}&&\hspace{-0.5cm} =
m_\pi^4\Big\{  - 32 C^r_4 - 32 C^r_5 - 24 C^r_6 - 64 C^r_7 + 8 C^r_8 + 8 C^r_{10} + 16 
C^r_{11} - 80 C^r_{12} - 176 C^r_{13} + 32 C^r_{14} 
\nonumber\\&&\hspace{-0.5cm}
+ 8 C^r_{15} + 128C^r_{16}
-48C^r_{17} 
-32 C^r_{23} + 16 C^r_{25} + 32 C^r_{26} + 128 C^r_{28} + 8 C^r_{64} +8 C^r_{65}+8C^r_{66} - 
16 C^r_{67} 
\nonumber\\&&\hspace{-0.5cm}
+ 32 C^r_{68} + 16 C^r_{69} + 16 C^r_{83} + 32 C^r_{84} + 8 C^r_{90}\Big\}
\nonumber\\&&\hspace{-0.5cm}
+ m_\pi^2 m_K^2 \Big\{ -16 C^r_4 + 8 C^r_5 - 40 C^r_6 - 64 C^r_7 
- 24 C^r_8 + 16 C^r_{10} + 
48 C^r_{11}+24 C^r_{12} - 96 C^r_{13} - 16 C^r_{14} 
\nonumber\\&&\hspace{-0.5cm}
+ 80 C^r_{15} + 64 C^r_{17} 
- 16 C^r_{22}
-16 C^r_{23}
-8 C^r_{25} - 16 C^r_{26} - 32 C^r_{29} - 64 C^r_{30} - 32 C^r_{36} + 16 C^r_{63}
+ 20 C^r_{64}
\nonumber\\&&\hspace{-0.5cm}
+4 C^r_{65} - 4 C^r_{66} - 24 C^r_{67} + 16 C^r_{68} 
+ 8 C^r_{69}
+ 8 C^r_{83} + 
16 C^r_{84} + 12 C^r_{90} \Big\}
\nonumber\\&&\hspace{-0.5cm}
+ m_\pi^2 s_\pi \Big\{ 20 C^r_1 + 64 C^r_2 + 8 C^r_3 + 8 C^r_4 + 16 C^r_5 + 16 C^r_6 + 32 
C^r_7 + 16 C^r_{13} + 8 C^r_{22} 
+ 8 C^r_{25}- 4 C^r_{64} 
\nonumber\\&&\hspace{-0.5cm}
- 4 C^r_{65}-12C^r_{66}+8 C^r_{67}
- 32 C^r_{68} - 16 C^r_{69} - 8 C^r_{83} - 16 C^r_{84} - 4 C^r_{90} \Big\}
\nonumber\\&&\hspace{-0.5cm}
+ m_\pi^2 t_\pi \Big\{  - 20 C^r_1 - 64 C^r_2 
+ 8 C^r_3 - 4 C^r_4 - 8 C^r_6 - 8 C^r_8 + 8 
C^r_{12} + 32 C^r_{13} + 12 C^r_{22} + 16 C^r_{23} - 20 C^r_{25}
\nonumber\\&&\hspace{-0.5cm}
+ 4 C^r_{63} - 4 C^r_{64}
-4 
C^r_{65} + 6 C^r_{66} 
+ 16 C^r_{67} - 16 C^r_{68} - 10 C^r_{69} - 12 C^r_{83} - 16 C^r_{84}+8 
C^r_{88} - 10 C^r_{90} \Big\}
\nonumber\\&&\hspace{-0.5cm}
+ m_\pi^2 u_\pi \Big\{  - 12 C^r_1 - 64 C^r_2 - 8 C^r_3+36 C^r_4-8 C^r_{10} - 16 C^r_{11} 
+ 8 C^r_{12} + 80 C^r_{13} - 12 C^r_{22} + 16 C^r_{23} 
\nonumber\\&&\hspace{-0.5cm}
+ 4 C^r_{25}- 4 C^r_{63}-4 C^r_{64}
- 4 
 C^r_{65} - 6 C^r_{66} + 16 C^r_{67} - 16 C^r_{68} - 6 C^r_{69} - 4 C^r_{83} - 16 C^r_{84}-6 
 C^r_{90} \Big\}
\nonumber\\&&\hspace{-0.5cm}
+ m_K^4   \Big\{ 8 C^r_5 + 16 C^r_6 + 8 C^r_{10} + 32 C^r_{11} - 8 C^r_{12}+8 C^r_{22} + 16 
 C^r_{23} - 8 C^r_{34} + 8 C^r_{63} + 8 C^r_{64} 
\nonumber\\&&\hspace{-0.5cm}
- 4 C^r_{66} - 8 C^r_{67} + 4 C^r_{90} \Big\}
\nonumber\\&&\hspace{-0.5cm}
+ m_K^2 s_\pi \Big\{ 4 C^r_1 + 8 C^r_3 + 16 C^r_4 + 32 C^r_6 + 32 C^r_7 + 16 C^r_8 + 8 
 C^r_{12} + 32 C^r_{13}-16 C^r_{23}+8 C^r_{25}-8 C^r_{63}
\nonumber\\&&\hspace{-0.5cm}
-8 C^r_{64}+8 C^r_{67} 
- 8 C^r_{68}
  - 4 C^r_{69} - 4 C^r_{90} \Big\}
\nonumber\\&&\hspace{-0.5cm}
+ m_K^2 t_\pi\Big\{  - 4 C^r_1 + 8 C^r_3 + 4 C^r_4 - 8 C^r_5-16 C^r_6 - 24 C^r_{12} - 64 
 C^r_{13} - 4 C^r_{22} - 8 C^r_{23}-12 C^r_{63}
\nonumber\\&&\hspace{-0.5cm}
-8 C^r_{64}+6 C^r_{66}+12 C^r_{67} 
+ 2 C^r_{69}
  + 4 C^r_{83} + 4 C^r_{88} - 10 C^r_{90} \Big\}
\nonumber\\&&\hspace{-0.5cm}
+ m_K^2 u_\pi\Big\{ 4 C^r_1 - 8 C^r_3 - 4 C^r_4 - 8 C^r_{10} - 32 C^r_{11} + 16 C^r_{12} 
+ 32 
 C^r_{13} - 4 C^r_{22}-8 C^r_{23}-4 C^r_{63} - 8 C^r_{64} 
\nonumber\\&&\hspace{-0.5cm}
+ 6 C^r_{66} + 12 C^r_{67} 
- 2 C^r_{69}
  - 4 C^r_{83} - 2 C^r_{90} \Big\}
\nonumber\\&&\hspace{-0.5cm}
+ s_\pi^2 \Big\{  - 8 C^r_1 - 32 C^r_2 + 8 C^r_4 + 4 C^r_{66} 
+ 8 C^r_{68} + 4 C^r_{69} \Big\}
\nonumber\\&&\hspace{-0.5cm}
+ s_\pi t_\pi   \Big\{ 4 C^r_1 + 32 C^r_2 - 4 C^r_4 - 2 C^r_{66} - 4 C^r_{67} + 8 C^r_{68} + 6 
 C^r_{69} - 2 C^r_{88} + 2 C^r_{90} \Big\}
\nonumber\\&&\hspace{-0.5cm}
+ s_\pi u_\pi \Big\{ 8 C^r_1 + 32 C^r_2 - 8 C^r_3 - 20 C^r_4 + 6 C^r_{66} - 4 C^r_{67} + 8 
 C^r_{68} + 2 C^r_{69} - 2 C^r_{88} + 2 C^r_{90} \Big\}
\nonumber\\&&\hspace{-0.5cm}
+ t_\pi^2 \Big\{ 4 C^r_1 - 4 C^r_4 - 2 C^r_{66} - 2 C^r_{67} 
- 2 C^r_{69} - 6 C^r_{88} + 6 C^r_{90}
  \Big\}
\nonumber\\&&\hspace{-0.5cm}
+ t_\pi u_\pi \Big\{-4 C^r_1-8 C^r_3 + 4 C^r_4 - 6 C^r_{66} - 12 C^r_{67} + 2 C^r_{69} - 2 
 C^r_{88} + 2 C^r_{90}  \Big\}
\nonumber\\&&\hspace{-0.5cm}
+ u_\pi^2   \Big\{ 8 C^r_3 - 2 C^r_{67} \Big\},
\ea
\noindent
while for the $G$ form-factor the contribution is
\noindent
\ba
G_{ct}&&\hspace{-0.5cm} =
m_\pi^4 \Big\{  - 24 C^r_4 - 8 C^r_6 - 8 C^r_8 + 8 C^r_{10} + 16 C^r_{11} - 16 C^r_{13} + 8 
 C^r_{15} + 16 C^r_{17} + 8 C^r_{22} - 8 C^r_{25} 
\nonumber\\&&\hspace{-0.5cm}
+ 8 C^r_{63} + 8 C^r_{64} + 8 C^r_{65} 
- 4 C^r_{66}
  + 12 C^r_{69} + 8 C^r_{83} + 12 C^r_{90} \Big\}
\nonumber\\&&\hspace{-0.5cm}
+ m_\pi^2 m_K^2 \Big\{  - 16 C^r_4 - 8 C^r_5 - 24 C^r_6 - 8 C^r_8 + 16 C^r_{10} + 48 C^r_{11} - 
 24 C^r_{12} - 64 C^r_{13} + 16 C^r_{14}
\nonumber\\&&\hspace{-0.5cm}
+ 16 C^r_{15} - 16 C^r_{22} + 16 C^r_{25} + 16 C^r_{26}
  - 32 C^r_{29} + 16 C^r_{63} + 20 C^r_{64} + 4 C^r_{65} 
\nonumber\\&&\hspace{-0.5cm}
+ 4 C^r_{66} + 8 C^r_{69} + 8 C^r_{83} + 
 12 C^r_{90} \Big\}
\nonumber\\&&\hspace{-0.5cm}
+ m_\pi^2 s_\pi \Big\{  - 4 C^r_1 + 8 C^r_3 + 8 C^r_4 - 16 C^r_{12} - 16 C^r_{13} - 8 C^r_{63} - 4
  C^r_{64} - 4 C^r_{65} + 4 C^r_{66} 
\nonumber\\&&\hspace{-0.5cm}
- 8 C^r_{69} + 4 C^r_{88} - 12 C^r_{90} \Big\}
\nonumber\\&&\hspace{-0.5cm}
+ m_\pi^2 t_\pi\Big\{ 4 C^r_1 + 8 C^r_3 + 12 C^r_4 + 8 C^r_6 + 8 C^r_8 - 8 C^r_{12} + 4 C^r_{22}
  + 16 C^r_{23} + 4 C^r_{25} - 4 C^r_{63} - 4 C^r_{64} 
\nonumber\\&&\hspace{-0.5cm}
- 4 C^r_{65} - 2 C^r_{66} - 4 C^r_{67} 
- 10
  C^r_{69} - 4 C^r_{83} + 4 C^r_{88} - 10 C^r_{90} \Big\}
\nonumber\\&&\hspace{-0.5cm}
+ m_\pi^2 u_\pi \Big\{  - 4 C^r_1 - 8 C^r_3 + 20 C^r_4 - 8 C^r_{10} - 16 C^r_{11} + 8 C^r_{12} + 
 16 C^r_{13} - 12 C^r_{22} - 16 C^r_{23} 
\nonumber\\&&\hspace{-0.5cm}
+ 4 C^r_{25} - 4 C^r_{63} - 4 C^r_{64} - 4 C^r_{65} + 2 
 C^r_{66} + 4 C^r_{67} - 6 C^r_{69} - 4 C^r_{83} - 6 C^r_{90} \Big\}
\nonumber\\&&\hspace{-0.5cm}
+ m_K^4 \Big\{  - 8 C^r_4 - 8 C^r_5 - 16 C^r_6 + 8 C^r_{10} + 32 C^r_{11} 
- 40 C^r_{12} - 64
  C^r_{13} - 8 C^r_{34} + 8 C^r_{64} + 4 C^r_{69} + 8 C^r_{83} \Big\}
\nonumber\\&&\hspace{-0.5cm}
+ m_K^2 s_\pi   \Big\{  - 4 C^r_1 + 8 C^r_3 + 8 C^r_{12} - 32 C^r_{13} + 8 C^r_{22} - 8 C^r_{25} 
- 8
  C^r_{64} - 4 C^r_{69} - 8 C^r_{83} + 4 C^r_{88} - 4 C^r_{90} \Big\}
\nonumber\\&&\hspace{-0.5cm}
+ m_K^2 t_\pi \Big\{ 4 C^r_1 + 8 C^r_3 + 4 C^r_4 + 8 C^r_5 + 16 C^r_6 + 8 C^r_{12} + 4 C^r_{22}
  + 8 C^r_{23} - 4 C^r_{63} 
- 8 C^r_{64} - 2 C^r_{66} 
\nonumber\\&&\hspace{-0.5cm}
- 6 C^r_{69} - 4 C^r_{83} - 2 C^r_{90} \Big\}
\nonumber\\&&\hspace{-0.5cm}
+ m_K^2 u_\pi   \Big\{  - 4 C^r_1 - 8 C^r_3 + 12 C^r_4 - 8 C^r_{10} - 32 C^r_{11} + 16 C^r_{12} + 
 32 C^r_{13} - 4 C^r_{22}- 8 C^r_{23} 
\nonumber\\&&\hspace{-0.5cm}
- 8 C^r_{23} - 4 C^r_{63} - 8 C^r_{64} - 2 C^r_{66} - 2 C^r_{69} - 4 
 C^r_{83} - 2 C^r_{90} \Big\}
+ s_\pi^2   \Big\{  - 4 C^r_{88} + 4 C^r_{90} \Big\}
\nonumber\\&&\hspace{-0.5cm}
+ s_\pi t_\pi   \Big\{ 4 C^r_1 - 4 C^r_4 - 2 C^r_{66} + 6 C^r_{69} - 2 C^r_{88} + 2 C^r_{90} \Big\}
\nonumber\\&&\hspace{-0.5cm}
+ s_\pi u_\pi   \Big\{  - 8 C^r_3 - 4 C^r_4 - 2 C^r_{66} + 2 C^r_{69} - 2 C^r_{88}+2 C^r_{90} \Big\}
\nonumber\\&&\hspace{-0.5cm}
+ t_\pi^2   \Big\{  - 4 C^r_1 + 4 C^r_4 + 2 C^r_{66} + 2 C^r_{67} + 2 C^r_{69} - 2 C^r_{88} + 2 
 C^r_{90} \Big\}
\nonumber\\&&\hspace{-0.5cm}
+ t_\pi u_\pi \Big\{ 4 C^r_1 - 8 C^r_3 - 12 C^r_4 + 2 C^r_{66} + 2 C^r_{69} - 2 C^r_{88} + 2 
 C^r_{90} \Big\}
+ u_\pi^2   \Big\{ 8 C^r_3 - 2 C^r_{67} \Big\}.
\ea

\subsubsection{Resonance Contribution}
\label{Appresonance}

The aim of this section is to give an estimate, based on Resonance Saturation
as discussed in Sect. \ref{estimates},
of the previous combinations of ${\cal O}(p^6)$ constants. The main
drawback of this method  is that it does not specify
the scale at which it should be apply. A rough estimate of
this uncertainty is obtained varying the scale $\mu$, let's say
between $0.5$ and $1$ GeV, and comparing
the results with the scale at the $\rho$ mass.\\
Using the equations of motion
for integrating out the massive fields we get for the $F$ 
form-factor

\ba
\label{ref}
F_{RS}&&\hspace{-0.5cm} =
\frac{1}{F_0^2 M_S^4} \Big\{ c_m^2 \Big( -8 m_\pi^4+12 m_\pi^2 m_K^2
-4 m_K^4 \Big)
+c_m c_d \Big(4 m_\pi^4-4 m_\pi^2 m_K^2 +8 m_\pi^2 s_\pi -4 m_\pi^2 t_\pi \Big)
\nonumber\\&&\hspace{-0.5cm}
+c_d^2\Big(-8 m_\pi^2 s_\pi + 2 m_\pi^2 t_\pi + 2 m_K^2 t_\pi + 4 s_\pi^2
-2 t_\pi^2 \Big) \Big\}
\nonumber\\&&\hspace{-0.5cm}
+\frac{1}{F_0^2 \sqrt{2} M_V^4} \Big\{ f_V f_\chi \Big(
-4 m_\pi^4 + m_\pi^2 ( -6 m_K^2 +2 s_\pi + t_\pi + 3  u_\pi)
-2 m_K^4 + m_K^2 (2 s_\pi + 3 t_\pi + u_\pi) \Big)
\nonumber\\&&\hspace{-0.5cm}
+g_V f_\chi \Big( 12 m_\pi^4 + 10 m_\pi^2 m_K^2 -4 m_\pi^2 s_\pi
+10 m_\pi^2 t_\pi - 6 m_\pi^2 u_\pi +2 m_K^4 -4 m_K^2 s_\pi -2 m_K^2 u_\pi \Big)
\nonumber\\&&\hspace{-0.5cm}
+g_V \alpha_V \Big( 2 m_\pi^4 -m_\pi^2 (m_K^2 +s_\pi-t_\pi+u_\pi)
-m_K^4+m_K^2 (s_\pi+2 t_\pi + u_\pi)
+ s_\pi u_\pi 
-2 t_\pi (s_\pi+u_\pi)\Big)
\nonumber\\&&\hspace{-0.5cm}
+\sqrt{2} g_V f_V \Big(-2 m_\pi^2 t_\pi - m_K^2 t_\pi + (s_\pi t_\pi+s_\pi u_\pi
+3 t_\pi^2 + t_\pi u_\pi)/2 \Big)
\nonumber\\&&\hspace{-0.5cm}
+\sqrt{2} g_V^2 \Big( 5 m_\pi^2 t_\pi+m_K^2 t_\pi -s_\pi t_\pi - s_\pi u_\pi
-t_\pi u_\pi \Big)
+\sqrt{2} f_\chi^2 \Big( 32 m_\pi^4 + 16 m_\pi^2 m_K^2 \Big)
\nonumber\\&&\hspace{-0.5cm}
+\sqrt{2} f_\chi \alpha_V \Big( 16 m_\pi^4 + 8 m_\pi^2 m_K^2 -8 m_\pi^2 s_\pi
-12 m_\pi^2 t_\pi - 4 m_\pi^2 u_\pi + 4 m_K^2 t_\pi -4 m_K^2 u_\pi \Big) 
\nonumber\\&&\hspace{-0.5cm}
+\sqrt{2} f_\chi^2  \Big(32 m_\pi^4 + 16 m_\pi^2 m_K^2 \Big)
\Big\}
\nonumber\\&&\hspace{-0.5cm}
+\frac{\sqrt{2} s_\ell }{F_0^2 M_A^2} \Big\{ 
f_A \gamma_{A}^{(1)} \Big( -m_\pi^2 -m_K^2 +u_\pi\Big) 
+f_A \gamma_{A}^{(2)} \Big( 3 m_\pi^2-m_K^2-2 s_\pi + t_\pi \Big) \Big\}\,,
\ea

\noindent
and for the $G$ form-factor

\ba
\label{reg}
G_{RS}&&\hspace{-0.5cm} =
\frac{1}{F_0^2 M_S^4} \Big\{ 4 c_m^2 \Big( m_\pi^2 m_K^2
- m_K^4 \Big)
-4 c_m c_d m_\pi^2 \Big( m_\pi^2 -  m_K^2 -  t_\pi \Big)
- 2 t_\pi c_d^2 \Big( m_\pi^2 + m_K^2 - t_\pi \Big) \Big\}
\nonumber\\&&\hspace{-0.5cm}
+\frac{1}{F_0^2 \sqrt{2} M_V^4} \Big\{ f_V f_\chi \Big(
-6 m_\pi^4 -6 m_\pi^2 m_K^2 + 4 m_\pi^2 s_\pi + (3 m_\pi^2  
+ m_K^2)  (  u_\pi + t_\pi ) \Big)
\nonumber\\&&\hspace{-0.5cm}
+g_V f_\chi \Big( 8m_\pi^4 + 14m_\pi^2 m_K^2 -2 m_\pi^2 t_\pi -6 m_\pi^2 u_\pi
+2 m_K^4 + 4 m_K^2 s_\pi -2 m_K^2 u_\pi \Big)
\nonumber\\&&\hspace{-0.5cm}
+g_V \alpha_V \Big( -2 m_\pi^4 +m_\pi^2 ( m_K^2 + 3 s_\pi +  
t_\pi+u_\pi) + m_K^4 - m_K^2 (s_\pi+u_\pi)-2 s_\pi t_\pi -s_\pi u_\pi \Big)
\nonumber\\&&\hspace{-0.5cm}
+\sqrt{2} g_V f_V \Big(- m_\pi^2 (s_\pi +t_\pi) -m_K^2 s_\pi + s_\pi^2 +
(s_\pi t_\pi + s_\pi u_\pi + t_\pi^2 + t_\pi u_\pi)/2 \Big)
\nonumber\\&&\hspace{-0.5cm}
+\sqrt{2} g_V^2 \Big( ( 2 s_\pi + t_\pi ) (m_\pi^2 +m_K^2)
- s_\pi t_\pi - s_\pi u_\pi -t_\pi u_\pi \Big)
\nonumber\\&&\hspace{-0.5cm}
+\sqrt{2} f_\chi \alpha_V \Big( 8 m_\pi^4 + 8 m_\pi^2 m_K^2 -4 m_\pi^2 
(t_\pi+u_\pi) + 8 m_K^4 - 4 m_K^2 ( 2 s_\pi + t_\pi + u_\pi) \Big)
\nonumber\\&&\hspace{-0.5cm}
+48 \sqrt{2} f_\chi^2  m_\pi^2 m_K^2 \Big\}
\nonumber\\&&\hspace{-0.5cm}
+\frac{\sqrt{2} s_\ell }{F_0^2 M_A^2} \Big\{ 
f_A \gamma_{A}^{(1)} \Big( -m_\pi^2 -m_K^2 +u_\pi\Big) 
+f_A \gamma_{A}^{(2)} \Big(  m_\pi^2+m_K^2- t_\pi \Big) \Big\}\,.
\ea

\noindent
As can be seen from Eqs. (\ref{ref}), (\ref{reg}) the axial-vector contribution
vanishes at $s_\ell =0$ and thus has no influence on the values
quoted in Eq. (\ref{newfit}).

%%%%%%%%%%%%%%%%%%%%%%%%%%%%%%%%%%%%%%%%%%%%%%%%%%%%%%%%%%%%%%%%%
\setcounter{section}{0}
\setcounter{subsection}{0}

\renewcommand{\thesection}{\Alph{zahler}}
\renewcommand{\theequation}{\Alph{zahler}.\arabic{equation}}

\setcounter{equation}{0}
\addtocounter{zahler}{1}
\renewcommand{\thesection}{\Alph{zahler}}
\renewcommand{\theequation}{\Alph{zahler}.\arabic{equation}}
\newcommand{\overb}{\overline{b}}

\section{Scattering Lengths}
\label{scattering}
Using the definitions of Eq. (\ref{ex}) is rather straightforward to evaluate
the threshold parameters $a^I_l$ and $b^I_l$ not displayed in \cite{pipi2}.

\ba
b^0_2 =&&\hspace{-0.6cm}
\frac{m_\pi^{-2}}{80 \pi^3\, F_\pi^4 } \left\{
-\frac{481}{2520} + x \left[
-\frac{1849}{17010}
-\frac{1583 \pi^2}{22680}
-\frac{1}{84} \,\overb_1
-\frac{124}{315} \,\overb_2
-\frac{179}{378}\,\overb_3
-\frac{773}{210}\,\overb_4
-\overline{b}_5
+\frac{17}{3} \,\overb_6 \right] \right\} \,, \nonumber\\
b^2_2 =&&\hspace{-0.6cm}\frac{1}{240 \pi^3 \, F_\pi^4 m_\pi^2} \left\{
-\frac{277}{840}
+ x \left[ \frac{193 \pi^2}{7560}
-\frac{24218}{2835}
-\frac{1}{7}\,\overb_1
-\frac{337}{420}\,\overb_2
+\frac{157}{126}\,\overb_3
+\frac{87}{14}\,\overb_4
-3 \,\overline{b}_5 + 5 \,\overline{b}_6
\right] \right\} \,, \nonumber\\
a^1_3 =&&\hspace{-0.6cm} \frac{1}{560 \pi^3\, F_\pi^4 m_\pi^2} \left\{
\frac{11}{168}
+ x \left[
\frac{4111}{1134}
+\frac{37 \pi^2}{840}
+\frac{1}{14}\,\overb_1
+\frac{17}{84}\,\overb_2
-\frac{151}{126}\,\overb_3
-\frac{653}{126}\,\overb_4
+\overb_5
+\overb_6
\right] \right\} \,, \nonumber\\
b^1_3 =&&\hspace{-0.6cm} \frac{1}{35280 \pi^3\, F_\pi^4 m_\pi^4} \left\{
-\frac{47}{15}
+ x \left[
\frac{549221}{5400}
-\frac{41 \pi^2}{18}
-5 \, \overb_1
-\frac{169}{15} \,\overb_2
-\frac{958}{5} \,\overb_3
- 418 \, \overb_4 \right] \right\} \,.
\ea
where
\be
x=\frac{m_\pi^2}{16 \pi^2 F_\pi^2},\quad \overline{b}_{1,2,3,4}
= 16 \pi^2 b_{1,2,3,4}\,, \quad 
\overb_{5,6} = (16 \pi^2)^2 b_{5,6}\, .
\ee
For the rest of the notation and the expression for the $b_i$ coefficients,
we refer to App. D in \cite{pipi2}.

%%%%%%%%%%%%%%%%%%%%%%%%%%%%%%%%%%%%%%%%%%%%%%%%%%%%%%%%%%%%%%%%%
\setcounter{section}{0}
\setcounter{subsection}{0}

\renewcommand{\thesection}{\Alph{zahler}}
\renewcommand{\theequation}{\Alph{zahler}.\arabic{equation}}

\setcounter{equation}{0}
\addtocounter{zahler}{1}
\renewcommand{\thesection}{\Alph{zahler}}
\renewcommand{\theequation}{\Alph{zahler}.\arabic{equation}}

\section{Vacuum Expectation Values at ${\cal O}(p^6)$}
\label{vacuum}
Here we collect the relevant formulas for the vacuum condensate in the
limit $m_u=m_d$. As can be seen they can be cast in terms of the 
simplest one-loop integral $\overline{A}$ defined in App. \ref{1loop}.

For the non-strange current we find
\begin{eqnarray}
\label{eqvacuum1}
\langle 0 \vert \overline{q} q \vert 0 \rangle^{(4)} &&\hspace{-0.7cm} =
\frac{1}{F^2_\pi}  \Big\{ 3/2 \overline{A}(m_\pi^2)
  +\overline{A}(m_K^2)
  +\overline{A}(m_\eta^2)/6
  + 4 m_\pi^2 \Big( 4 L_6^r + 2 L_8^r + H_2^r\Big) + 32 m_K^2 L_6^r \Big\}\,,
\nonumber\\
\langle 0 \vert \overline{q} q \vert 0 \rangle^{(6)} &&\hspace{-0.7cm} =
  \frac{1}{F^4_\pi}  \Big\{
  \frac{1}{(16 \pi )^2}
\Big( m_\pi^4 (47/3888 \pi^2 + 47/648 ) -  m_\pi^2 m_K^2 (113/1944
 \pi^2 + 113/324 )
\nonumber\\&&\hspace{-0.7cm}
+ m_\pi^2 m_\eta^2 ( 19/432 \pi^2 + 19/72 )
+ m_K^4 (19/486
\pi^2 + 19/81 ) - m_K^2 m_\eta^2 ( 13/216 \pi^2 + 13/36 )
\nonumber\\&&\hspace{-0.7cm}
+
m_\eta^4 ( 5/216  \pi^2 + 5/36 ) \Big)
\nonumber\\&&\hspace{-0.7cm}
  + 21/8 \Big(\overline{A}(m_\pi^2)\Big)^2
  + 7/2 \overline{A}(m_\pi^2) \overline{A}(m_K^2)
  + 7/12 \overline{A}(m_\pi^2) \overline{A}(m_\eta^2)
\nonumber\\&&\hspace{-0.7cm}
  + \overline{A}(m_\pi^2)
   \Big(  m_\pi^2 ( - 24 L_4^r - 12 L_5^r + 112 L_6^r + 68 L_8^r
     +10  H_2^r ) + 112 m_K^2 L_6^r \Big)
  + \Big((\overline{A}(m_K^2)\Big)^2
\nonumber\\&&\hspace{-0.7cm}
   -\overline{A}(m_K^2) \overline{A}(m_\eta^2)/6
  + \overline{A}(m_\eta^2)^2/72
\nonumber\\&&\hspace{-0.7cm}
+ \overline{A}(m_K^2) \Big( m_\pi^2 ( 8 L_5^r + 32 L_6^r + 8 L_8^r + 4 H_2^r)
+ m_K^2(  - 32
L_4^r - 16 L_5^r + 128 L_6^r + 32 L_8^r ) \Big)
\nonumber\\&&\hspace{-0.7cm}
  +\overline{A}(m_\eta^2) \Big( m_\pi^2/3 ( 20/3 L_5^r - 16 L_6^r
+ 64 L_7^r + 12
L_8^r - 2 H_2^r)
\nonumber\\&&\hspace{-0.7cm}
- m_K^2/3 ( 32/3 L_5^r - 112 L_6^r + 64 L_7^r) - 8
    m_\eta^2 L_4^r \Big)
\nonumber\\&&\hspace{-0.7cm}
+  64 m_\pi^4 \Big( (L_4^r+L_5^r)( 4 L_6^r + 2 L_8^r + H_2^r)
- 4 (L_6^r)^2 - 6 L_6^r L_8^r - L_6^r H_2^r - 2 (L_8^r)^2 -
L_8^r H_2^r \Big)
\nonumber\\&&\hspace{-0.7cm}
+ 128 m_\pi^2 m_K^2 \Big(
(L_4^r - L_6^r) ( 8 L_6^r + 2 L_8^r +  H_2^r )
+ 2 L_5^r L_6^r
\Big)
\nonumber\\&&\hspace{-0.7cm}
+ m_K^4 \Big( 1024 L_4^r L_6^r + 256 L_5^r
         L_6^r - 1024 (L_6^r)^2 - 512 L_6^r L_8^r \Big) \Big\}
\nonumber\\&&\hspace{-0.7cm}
+\frac{4}{F^2_\pi}  \Big\{
 m_\pi^4 \Big( 12 C^r_{19} + 20  C^r_{20}
+ 12  C^r_{21} -
C^r_{94} \Big) +
m_\pi^2 m_K^2 \Big( 48 C^r_{21} + 2 C^r_{94} \Big)
+ m_K^4 \Big( 16 C^r_{20} + 48
C^r_{21} \Big) \Big\} \,,
\nonumber\\&&\hspace{-0.7cm}
\end{eqnarray}
\noindent
where $\langle 0 \vert \overline{q} q \vert 0 \rangle$ stands for
$\langle 0 \vert \overline{u} u \vert 0 \rangle$ either
$\langle 0 \vert \overline{d} d \vert 0 \rangle$. The results
for the strange quark reads
\begin{eqnarray}
\label{eqvacuum2}
\langle 0 \vert \overline{s} s \vert 0 \rangle^{(4)} &&\hspace{-0.7cm} =
\frac{2}{F_\pi^2} \Big\{ \overline{A}(m_K^2)
    +  \overline{A}(m_\eta^2)/3
     + m_\pi^2 ( 8 L_6^r - 4 L_8^r - 2
H_2^r ) + 4 m_K^2 ( 4 L_6^r + 2 L_8^r + H_2^r ) \Big\}
\nonumber\\
\langle 0 \vert \overline{s} s \vert 0 \rangle^{(6)} 
&&\hspace{-0.7cm} =
\frac{1}{F_\pi^4} \Big\{
\frac{1}{(16 \pi)^2}  (\pi^2/6+1) \Big( - 7/162 m_\pi^4
+ 13/81  m_\pi^2 m_K^2
+ 1/36 m_\pi^2 m_\eta^2 + 4/81 m_K^4
\nonumber\\&&\hspace{-0.7cm}
- 2/3 m_K^2 m_\eta^2
+ 17/36 m_\eta^4  \Big)
\nonumber\\&&\hspace{-0.7cm}
     + 4 \overline{A}(m_\pi^2) \overline{A}(m_K^2)
     + 4/3 \overline{A}(m_\pi^2) \overline{A}(m_\eta^2)
     + 2 \Big(\overline{A}(m_K^2)\Big)^2
     + 2/9 \Big(\overline{A}(m_\eta^2)\Big)^2
\nonumber\\&&\hspace{-0.7cm}
     + \overline{A}(m_\pi^2)   \Big(   m_\pi^2 ( - 24 L_4^r + 88
L_6^r - 20 L_8^r - 10 H_2^r) + 16 m_K^2 ( 4 L_6^r + 2 L_8^r
+ H_2^r ) \Big)
\nonumber\\&&\hspace{-0.7cm}
     + \overline{A}(m_K^2)   \Big( 4 m_\pi^2 ( 4 L_5^r + 12
L_6^r - 2 L_8^r - H_2^r)  + 8 m_K^2 ( - 4 L_4^r - 4 L_5^r + 20
L_6^r + 10 L_8^r + H_2^r ) \Big)
\nonumber\\&&\hspace{-0.7cm}
+\overline{A}(m_\eta^2)   \Big( m_\pi^2/3 ( 80/3 L_5^r + 8
L_6^r - 128 L_7^r - 60 L_8^r + 2 H_2^r)
\nonumber\\&&\hspace{-0.7cm}
+ m_K^2/3 (  - 128/3
L_5^r + 160 L_6^r + 128 L_7^r + 144 L_8^r
+ 8 H_2^r) - 8 m_\eta^2 L_4^r \Big)
\nonumber\\&&\hspace{-0.7cm}
+ m_\pi^4    \Big( (L_4^r+L_5^r-L_6^r) (256  L_6^r - 128
L_8^r - 64 H_2^r )
+ 128 (L_8^r)^2 + 64
L_8^r H_2^r \Big)
\nonumber\\&&\hspace{-0.7cm}
+ 64 m_\pi^2 m_K^2 \Big( 16 L_4^r L_6^r + 4 L_5^r L_6^r
+ 2 L_5^r L_8^r + L_5^r H_2^r - 16
   (L_6^r)^2 \Big)
\nonumber\\&&\hspace{-0.7cm}
+ 64 m_K^4 \Big( 4 ( L_4^r + L_5^r)( 4 L_6^r + 2 L_8^r +  H_2^r)
               - 4 L_6^r ( 4 L_6^r + 4 L_8^r +  H_2^r)
- 4 (L_8^r)^2 - 2 L_8^r H_2^r \Big) \Big\}
\nonumber\\&&\hspace{-0.7cm}
+ \frac{1}{F_\pi^2}   \Big\{ 4 m_\pi^4 \Big( 12 C^r_{19} + 4
 C^r_{20} + 12 C^r_{21} + C^r_{94}\Big)  - 64m_\pi^2 m_K^2 \Big( 3 C^r_{19} +
C^r_{20} - 3 C^r_{21}\Big)
\nonumber\\&&\hspace{-0.7cm}
+ 192 m_K^4 \Big( C^r_{19} +
  C^r_{20} + C^r_{21}\Big)  \Big\} \,.
\nonumber\\&&\hspace{-0.7cm}
\end{eqnarray}

%%%%%%%%%%%%%%%%%%%%%%%%%%%%%%%%%%%%%%%%%%%%%%%%%%%%%%%%%%%%%%%%%
\setcounter{section}{0}
\setcounter{subsection}{0}

\renewcommand{\thesection}{\Alph{zahler}}
\renewcommand{\theequation}{\Alph{zahler}.\arabic{equation}}

\setcounter{equation}{0}
\addtocounter{zahler}{1}
\renewcommand{\thesection}{\Alph{zahler}}
\renewcommand{\theequation}{\Alph{zahler}.\arabic{equation}}

\section{Loop Integrals}
\label{1loop}

In this Appendix we collect 
for completeness some familiar formulas for the
one-loop integrals.
The two-loop integrals that appear we discussed in
\cite{Amoros} and for the others we used the expressions
of \cite{Ghinculov}. We have not discussed these integrals
in more detail here since we do not present any formulas involving them.

Through the calculation one has to use one-loop integrals of one, two and three
point functions. The latter disappears after mass renormalization and the
use of some
recursion relation. All in all we only have to deal with the following set
of functions
-- in the remainder we use $d=4-2 \epsilon$.
\ba
\label{oneloop}
A(m_1^2)& =& \frac{1}{i}\int \frac{d^dq}{(2\pi)^d} \frac{1}{q^2-m_1^2}\,,
\nonumber\\
B(m_1^2,m_2^2,p^2) &=& \frac{1}{i}\int 
\frac{d^dq}{(2\pi)^d} \frac{1}{(q^2-m_1^2)((q-p)^2-m_2^2)}\,,\nonumber\\
B_\mu(m_1^2,m_2^2,p^2) &=& \frac{1}{i}\int 
\frac{d^dq}{(2\pi)^d} \frac{q_\mu}{(q^2-m_1^2)((q-p)^2-m_2^2)}\nonumber\\
&=& p_\mu B_1(m_1^2,m_2^2,p^2)\,,\nonumber\\
B_{\mu\nu}(m_1^2,m_2^2,p^2) &=& \frac{1}{i}\int 
\frac{d^dq}{(2\pi)^d} \frac{q_\mu q_\nu}{(q^2-m_1^2)((q-p)^2-m_2^2)}\nonumber\\
&=& p_\mu p_\nu B_{21}(m_1^2,m_2^2,p^2)+g_{\mu \nu} B_{22}(m_1^2,m_2^2,p^2)\,,
\nonumber\\
B_{\mu\nu\alpha}(m_1^2,m_2^2,p^2) &=& \frac{1}{i}\int 
\frac{d^dq}{(2\pi)^d} \frac{q_\mu q_\nu q_\alpha}{(q^2-m_1^2)((q-p)^2-m_2^2)}
\nonumber\\
&& \hspace{-2cm}
= p_\mu p_\nu p_\alpha B_{31}(m_1^2,m_2^2,p^2) + 
(p_\mu g_{\nu \alpha} + p_\nu g_{\mu \alpha} 
+ p_\alpha g_{\mu \nu})B_{32}(m_1^2,m_2^2,p^2)
\,.\nonumber\\
\ea
An expansion in $\epsilon$ leads to the following series
\ba
\label{expan}
A(m_1^2)& =& \frac{m_1^2}{16\pi^2}\lambda_0+\overline{A}(m_1^2)+\epsilon 
\overline{A}^\epsilon(m_1^2)+\ldots\,,
\nonumber\\
B_{ij}(m_1^2,m_2^2,p^2)& =& \frac{1}{16\pi^2} \mbox{pole}_{ij} 
+ \overline{B}_{ij}(m_1^2,m_2^2,p^2)
+\epsilon \overline{B}_{ij}^\epsilon(m_1^2,m_2^2,p^2) +\ldots\,,
\ea
with $\overline{A}$, $\overline{B}_{ij}$ defining finite quantities and
where "$\mbox{pole}_{ij}$" denotes the singular part of each of the $B_{ij}$
functions,
\ba
&&\mbox{pole} = \lambda_0\,,\quad
\mbox{pole}_{1} = \frac{\lambda_0}{2}\,,\quad 
\mbox{pole}_{21} = \frac{\lambda_0}{3}\,,\quad
\mbox{pole}_{22} = \frac{\lambda_0}{4}(m_1^2+m_2^2-\frac{p^2}{3})\,,\quad 
\nonumber\\&&
\mbox{pole}_{31} = \frac{\lambda_0}{4}\,,\quad
\mbox{pole}_{32} = \frac{\lambda_0}{24}(2m_1+4m_2^2-p^2)\,,
\ea
with
\be
\label{pole}
\lambda_0 = \frac{1}{\epsilon}+\ln(4\pi)+1-\gamma\,.
\ee

After some simpler algebraic manipulation, the functions defined in
Eq. (\ref{oneloop})
can be related to the basic integrals $A(m_1^2)$ and $B_1(m_1^2,m_2^2,p^2)$
through the identities
\ba
\label{relations}
B_{31}(m_1^2,m_2^2,p^2)& =& \frac{1}{2p^2} \Big(
A(m_2^2)
     -(m_2^2-m_1^2-p^2)B_{21}(m_1^2,m_2^2,p^2)
-4B_{32}(m_1^2,m_2^2,p^2) \Big) \,, \nonumber\\ 
B_{32}(m_1^2,m_2^2,p^2)& =& \frac{1}{2p^2}
\Big( -\frac{m_1^2}{d}A(m_1^2)+\frac{m_2^2}{d}A(m_2^2)
-(m_2^2-m_1^2-p^2)B_{22}(m_1^2,m_2^2,p^2) \Big) \,,\nonumber\\
B_{21}(m_1^2,m_2^2,p^2)& =& 
\frac{1}{p^2}\Big( A(m_2^2) 
+m_1^2 B(m_1^2,m_2^2,p^2)
-d B_{22}(m_1^2,m_2^2,p^2)\Big) \,, \nonumber\\
B_{22}(m_1^2,m_2^2,p^2)& =& \frac{1}{2(d-1)} \Big(
A(m_2^2)+2m_1^2B(m_1^2,m_2^2,p^2)
\nonumber\\ &&
-(p^2+m_1^2-m_2^2)B_1(m_1^2,m_2^2,p^2) \Big)\,,  \nonumber\\
B_{1}(m_1^2,m_2^2,p^2)& =& \frac{1}{2p^2} \Big(
A(m_2^2)-A(m_1^2) 
   + ( m_1^2-m_2^2+ p^2 ) B(m_1^2,m_2^2,p^2) \Big)\,, \nonumber\\
B(m_1^2,m_1^2,0)& = & \frac{(d-2)}{2m_1^2}A(m_1^2)\,. 
\ea
Notice that the inclusion of the previous identities
is only done at the final (numerical) level in order to avoid
cancellations between different terms occurring in the form-factors. Then 
the one-loop contribution is reduced to 
\ba
\overline{A}(m_1^2) &=& - \frac{m_1^2}{16\pi^2} \ln(m_1^2)\,,\nonumber\\
\overline{B}(m_1^2,m_2^2,p^2) &=&
-\frac{1}{16\pi^2}
\frac{m_1^2\ln(m_1^2)-m_2^2\ln(m_2^2)}{m_1^2-m_2^2}
\nonumber\\&&
+\frac{1}{(32\pi^2)} \left(
2+\left(-\frac{\Delta}{p^2}+\frac{\Sigma}{\Delta}\right)\ln\frac{m_1^2}{m_2^2}
-\frac{\nu}{p^2}\ln\frac{(p^2+\nu)^2-\Delta^2}{(p^2-\nu)^2-\Delta^2} \right)\,,
\ea
with $\Delta=m_1^2-m_2^2$, $\Sigma=m_1^2+m_2^2$
and $\nu^2 = [p^2-(m_1+m_2)^2][p^2-(m_1-m_2)^2]$.
Similarly combining Eqs. (\ref{expan}) and (\ref{relations}) one can obtain 
the expressions of the  $\epsilon$ terms in the expansion as functions of
$\overline{A}^\epsilon$ and $\overline{B}^\epsilon$.
A straightforward calculation leads to
\ba
16\pi^2 \overline{A}^\epsilon(m_1^2)&=& m_1^2[\frac{C^2}{2}+\frac{1}{2}
+\frac{\pi^2}{12}+\frac{1}{2}
\ln^2(m_1^2)-C\ln(m_1^2)] \nonumber\\
16\pi^2 \overline{B}^\epsilon(m_1^2,m_2^2,p^2) &=& \frac{C^2}{2}-\frac{1}{2}
+\frac{\pi^2}{12}
+(C-1) \overline{B}(m_1^2,m_2^2,p^2) +\frac{1}{2} \int_0^1 dx\, 
\ln^2(m^2)\,,\nonumber\\
\ea
with $C=\ln(4\pi)+1-\gamma$ and $m^2=(1-x)m_1^2+x m_2^2-x(1-x)p^2$.

%%%%%%%%%%%%%%%%%%%%%%%%%%%%%%%%%%%%%%%%%%%%%%%%%%%%%%%%%%%%%%%%%


\begin{thebibliography}{99}
\bibitem{CHPT} For a series of lectures and more references see
A.~Pich, lectures given at Les Houches Summer School in Theoretical Physics,
Session 68: Probing the Standard Model of Particle Interactions,
Les Houches, France, 28
Jul - 5 Sep 1997,
%``Effective field theory,''
hep-ph/9806303.
%%CITATION = HEP-PH 9806303;%%
\bibitem{GL1}
J.~Gasser and H.~Leutwyler,
%``Chiral Perturbation Theory: Expansions In The Mass Of The Strange Quark,''
Nucl.\ Phys.\ {\bf B250} (1985) 465.
%%CITATION = NUPHA,B250,465;%%

\bibitem{Chounet}
L.~Chounet, J.~Gaillard and M.K.~Gaillard,
%``Leptonic Decays Of Hadrons,''
Phys.\ Rept.\ {\bf 4} (1972) 199.
%%CITATION = PRPLC,4,199;%%

\bibitem{Kl4oneloop}
J.~Bijnens,
%``K(L4) Decays And The Low-Energy Expansion,''
Nucl.\ Phys.\ {\bf B337} (1990) 635;
%%CITATION = NUPHA,B337,635;%%
C.~Riggenbach {\it et al.},
%``Chiral symmetry and the large N(c) limit in K(L4) decays,''
Phys.\ Rev.\ {\bf D43} (1991) 127.
%%CITATION = PHRVA,D43,127;%%

\bibitem{BCG}
J.~Bijnens, G.~Colangelo and J.~Gasser,
%``K(l4) decays beyond one loop,''
Nucl.\ Phys.\ {\bf B427} (1994) 427
hep-ph/9403390.
%%CITATION = NUPHA,B427,427;%%

\bibitem{pipi}
J.~Bijnens {\it et al.},
%``Elastic $\pi\pi$ scattering to two loops,''
Phys.\ Lett.\ {\bf B374} (1996) 210
hep-ph/9511397.
%%CITATION = PHLTA,B374,210;%%

\bibitem{pipi2}
%``Pion pion scattering at low energy,''
J.~Bijnens {\it et al.},
Nucl.\ Phys.\ {\bf B508} (1997) 263
hep-ph/9707291.
%%CITATION = NUPHA,B508,263;%%

\bibitem{Knecht}
M.~Knecht {\it et al.},
%``The Low-energy pi pi amplitude to one and two loops,''
Nucl.\ Phys.\ {\bf B457} (1995) 513
hep-ph/9507319.
%%CITATION = NUPHA,B457,513;%%

\bibitem{Amoros}
G.~Amor\'os, J.~Bijnens and P.~Talavera,
%``Two-point functions at two loops in three flavour
%chiral perturbation  theory,''
Nucl.\ Phys.\ in press, hep-ph/9907264.
%%CITATION = HEP-PH 9907264;%%

\bibitem{2loop}
S.~Bellucci, J.~Gasser and M.~E.~Sainio,
%``Low-energy photon-photon collisions to two loop order,''
Nucl.\ Phys.\  {\bf B423} (1994) 80
hep-ph/9401206;
%%CITATION = HEP-PH 9401206;%%
U.~Burgi,
%``Pion polarizabilities and charged pion pair production to two loops,''
Nucl.\ Phys.\  {\bf B479} (1996) 392
hep-ph/9602429;
%%CITATION = HEP-PH 9602429;%%
J.~Bijnens and P.~Talavera,
%``pi --> l nu gamma form factors at two-loop,''
Nucl.\ Phys.\ {\bf B489}, (1997) 387
hep-ph/9610269;
%%CITATION = NUPHA,B489,387;%%
P.~Post and K.~Schilcher,
%''Higher-Order Corrections to Sirlin's
%Theorem in O(p^6) Chial Perturbation Theory,''
Phys.\ Rev.\ Lett.\ {\bf 79} (1997) 4088;
%``Two loop analysis of vector current propagators in chiral perturbation theory,''
E.~Golowich and J.~Kambor,
Nucl.\ Phys.\  {\bf B447} (1995) 373
hep-ph/9501318;
%%CITATION = HEP-PH 9501318;%%
%``Two-loop analysis of axialvector current propagators
%in chiral  perturbation theory,''
E.~Golowich and J.~Kambor, 
Phys.\ Rev.\  {\bf D58} (1998) 036004
hep-ph/9710214;
%%CITATION = HEP-PH 9710214;%%
K.Maltman,
%'' THE VECTOR CURRENT CORRELATOR <O|T(V(3)(MU)V(8)(NU))|O> TO
%TWO LOOPS IN CHIRAL PERTURBATION THEORY,''
Phys.\ Rev.\ {\bf D53} (1996) 2573.
%%CITATION = PHRVA,D53,2573;%%

\bibitem{BCT}
J.~Bijnens, G.~Colangelo and P.~Talavera,
%``The vector and scalar form factors of the pion to two loops,''
JHEP {\bf 9805} (1998) 014
hep-ph/9805389.
%%CITATION = HEP-PH 9805389;%%

\bibitem{LEC}
G.~Amor\'os, J.~Bijnens and P.~Talavera,
%''Low Energy Constants from Kl4 Form-Factors,''
hep-ph/9912398.

\bibitem{cabibbo} 
N. Cabibbo and A. Maksymowicz,
Phys.\ Rev.\ {\bf 137} (1965) B438.
%%CITATION = PHRVA,137,B438;%%

\bibitem{Hanomaly}
Ll.~Ametller {\it et al.},
%``Semileptonic pi and K decays and the chiral anomaly at one loop,''
Phys.\ Lett.\ {\bf B303} (1993) 140
hep-ph/9302219.
%%CITATION = PHLTA,B303,140;%%

\bibitem{PT}
A.~Pais and S.B.~Treiman,
Phys.\ Rev.\ {\bf 168} (1968) 1858;
%%CITATION = PHRVA,168,1858;%%
F.A.~Berends, A.~Donnachie and G.C.~Oades, 
Phys.\ Lett.\ {\bf B26} (1967) 109;
%%CITATION = PHLTA,B26,109;%%
Phys.\ Rev.\ {\bf 171} (1968) 1457.
%%CITATION = PHRVA,171,1457;%%

\bibitem{BCElag}
J.~Bijnens, G.~Colangelo and G.~Ecker,
%``The mesonic chiral Lagrangian of order p**6,''
JHEP {\bf 02} (1999) 020
hep-ph/9902437.
%%CITATION = JHEPA,9902,020;%%

\bibitem{Rosselet}
L.~Rosselet {\it et al.},
%``Experimental Study Of 30,000 K(E4) Decays,''
Phys.\ Rev.\ {\bf D15} (1977) 574.
%%CITATION = PHRVA,D15,574;%%

\bibitem{Makoff}
G.~Makoff {\it et al.},
%``Study of the decay K(L) $\to$ pi+- pi0 e-+ anti-neutrino (neutrino),''
Phys.\ Rev.\ Lett.\ {\bf 70} (1993) 1591; Erratum {\bf 75} (1995) 2069. 
%%CITATION = PRLTA,70,1591;%%

\bibitem{PDG}
Particle Data Group,
C.~Caso {\it et al.},
%``Review of particle physics,''
Eur.\ Phys.\ J.\  {\bf C3} (1998) 1.
%%CITATION = EPHJA,C3,1;%%

\bibitem{weinberg}
S. Weinberg, Phys.\ Rev.\ Lett.\ {\bf 17} (1966) 336;
Erratum {\bf 18} (1967) 1178.

\bibitem{BCEdble}
J.~Bijnens, G.~Colangelo and G.~Ecker,
%``Double chiral logs,''
Phys.\ Lett.\ {\bf B441} (1998) 437
hep-ph/9808421.
%%CITATION = PHLTA,B441,437;%%

\bibitem{Buttiker}
B.~Ananthanarayan and P.~Buttiker,
%``The Chiral Coupling Constants $\lb{1}$ and $\lb{2}$
%from \pipi Phase Shifts,''
Phys.\ Rev.\ {\bf D54} (1996) 1125
hep-ph/9601285.
%%CITATION = PHRVA,D54,1125;%%

\bibitem{Girlanda:1997ed}
L.~Girlanda {\it et al.},
%``Comment on the prediction of two-loop standard chiral
%perturbation  theory for low-energy pi pi scattering,''
Phys.\ Lett.\ {\bf B409} (1997) 461
hep-ph/9703448.
%%CITATION = PHLTA,B409,461;%%

\bibitem{BCEinf}
J.~Bijnens, G.~Colangelo and G.~Ecker,
%``Renormalization of chiral perturbation theory to order p**6,''
Ann.\ Phys.\ (NY) {\bf 280} (2000) 100 hep-ph/9907333.
%%CITATION = HEP-PH 9907333;%%

\bibitem{DT}
A.I.~Davydychev and J.B.~Tausk,
Nucl.\ Phys.\ {\bf B397} (1993) 123.
%%CITATION = NUPHA,B397,123;%%

\bibitem{GS}
J.~Gasser and M.~E.~Sainio,
%``Two-loop integrals in chiral perturbation theory,''
Eur.\ Phys.\ J.\  {\bf C6} (1999) 297
hep-ph/9803251.
%%CITATION = HEP-PH 9803251;%%

\bibitem{Ghinculov}
A.~Ghinculov and J.J.~van der Bij,
%``Massive two loop diagrams: The Higgs propagator,''
Nucl.\ Phys.\ {\bf B436} (1995) 30
hep-ph/9405418;
%%CITATION = NUPHA,B436,30;%%
A.~Ghinculov and Y.~Yao,
%``Massive two-loop integrals in renormalizable theories,''
Nucl.\ Phys.\ {\bf B516} (1998) 385
hep-ph/9702266.
%%CITATION = NUPHA,B516,385;%%

\bibitem{resonance} 
G.~Ecker {\it et al.},
%``The Role Of Resonances In Chiral Perturbation Theory,''
Nucl.\ Phys.\ {\bf B321} (1989) 311;
%%CITATION = NUPHA,B321,311;%%
G.~Ecker {\it et al.},
%``Chiral Lagrangians For Massive Spin 1 Fields,''
Phys.\ Lett.\ {\bf B223} (1989) 425.
%%CITATION = PHLTA,B223,425;%%

\bibitem{kamborsumrules}
E.~Golowich and J.~Kambor,
%``The DMO sum rule revisited,''
Phys.\ Lett.\  {\bf B421} (1998) 319
hep-ph/9711256;
%%CITATION = HEP-PH 9711256;%%
%\cite{Golowich:1997kn}
%``Chiral sum rules to second order in quark mass,''
Phys.\ Rev.\ Lett.\  {\bf 79} (1997) 4092
hep-ph/9707341;
%%CITATION = HEP-PH 9707341;%%
%``Inverse-Moment Chiral Sum Rules,''
Phys.\ Rev.\  {\bf D53} (1996) 2651
hep-ph/9509304;
%%CITATION = HEP-PH 9509304;%%
and E.~Golowich and J.~Kambor in \cite{2loop}.

\bibitem{Prades}
J.~Prades,
%``Massive spin 1 field chiral Lagrangian from an extended
%Nambu-Jona-Lasinio model of QCD,''
Z.\ Phys.\ {\bf C63} (1994) 491
hep-ph/9302246; Erratum Eur.\ Phys.\ J.\ {\bf C11} (1999) 571.
%%CITATION = ZEPYA,C63,491;%%

\bibitem{etap}
P.~Herrera-Siklody,
%''ETA AND ETA-PRIME HADRONIC DECAYS IN U(L)(3) X U(R)(3) CHIRAL PERTURBATION THEORY,''
hep-ph/9902446; 
%%CITATION = ;%%
P.~Herrera-Siklody et al.,
%''ETA - ETA-PRIME MIXING FROM U(3)(L) X U(3)(R) CHIRAL PERTURBATION THEORY,''
Phys.\ Lett.\ {\bf B419} (1998) 326
hep-ph/9710268.
%%CITATION = PHLTA,B419,326;%%

\bibitem{Dashen}
J.F.~Donoghue, B.R.~Holstein and D.~Wyler,
%``Electromagnetic selfenergies of pseudoscalar mesons and Dashen's theorem,''
Phys.\ Rev.\ {\bf D47} (1993) 2089;
%%CITATION = PHRVA,D47,2089;%%
J.~Bijnens,
%``Violations of Dashen's theorem,''
Phys.\ Lett.\ {\bf B306} (1993) 343
hep-ph/9302217;
%%CITATION = PHLTA,B306,343;%%
J.~Bijnens and J.~Prades,
%``Electromagnetic corrections for pions and kaons: 
%Masses and  polarizabilities,''
Nucl.\ Phys.\ {\bf B490} (1997) 239
hep-ph/9610360;
%%CITATION = NUPHA,B490,239;%%
B.~Moussallam,
%``A sum rule approach to the violation of Dashen's theorem,''
Nucl.\ Phys.\ {\bf B504} (1997) 381
hep-ph/9701400.
%%CITATION = NUPHA,B504,381;%%

\bibitem{Stern}
N.~H.~Fuchs, M.~Knecht and J.~Stern,
%``Contributions of order O(m**2(quark) to K(l3) form factors and  unitarity of the CKM matrix,''
hep-ph/0001188.
%%CITATION = HEP-PH 0001188;%%

\bibitem{domenech}
D.~Espriu, E.~de Rafael and J.~Taron,
%``The QCD Effective Action At Long Distances,''
Nucl.\ Phys.\  {\bf B345} (1990) 22.
%%CITATION = NUPHA,B345,22;%%

\bibitem{michelle}
S.~Peris, M.~Perrottet and E.~de Rafael,
%``Matching long and short distances in large-N(c) {QCD},''
JHEP {\bf 9805} (1998) 011
hep-ph/9805442.
%%CITATION = HEP-PH 9805442;%%

\bibitem{descotes}
S.~Descotes, L.~Girlanda and J.~Stern,
%``Paramagnetic effect of light quark loops on chiral symmetry breaking,''
JHEP {\bf 0001} (2000) 041
hep-ph/9910537.
%%CITATION = HEP-PH 9910537;%%

\bibitem{enjl}
J.~Bijnens, C.~Bruno and E.~de Rafael,
%``Nambu-Jona-Lasinio like models and the low-energy effective action of QCD,''
Nucl.\ Phys.\  {\bf B390} (1993) 501
hep-ph/9206236.
%%CITATION = HEP-PH 9206236;%%

\bibitem{peris}
M.~F.~Golterman and S.~Peris,
%``The 7/11 rule: An estimate of m(rho)/f(pi),''
Phys.\ Rev.\  {\bf D61} (2000) 034018
hep-ph/9908252.
%%CITATION = HEP-PH 9908252;%%

\bibitem{AB}
G.~Amor\'os and J.~Bijnens,
%``A parametrization for K+ --> pi+ pi- e+ nu,''
J.\ Phys.\ {\bf G25} (1999) 1607
hep-ph/9902463.
%%CITATION = JPHGB,G25,1607;%%

\bibitem{Dumbrajs:1983jd}
O.~Dumbrajs {\it et al.},
%``Compilation Of Coupling Constants And Low-Energy Parameters. 1982 Edition,''
Nucl.\ Phys.\  {\bf B216} (1983) 277.
%%CITATION = NUPHA,B216,277;%%

\bibitem{Hoogland}
W.~Hoogland {\it et al.},
Nucl.\ Phys.\  {\bf B126} (1977) 109.
%%CITATION = NUPHA,B126,109;%%

\bibitem{novikov}
V.~A.~Novikov {\it et al.},y
%``Are All Hadrons Alike? Technical Appendices,''
Nucl.\ Phys.\  {\bf B191} (1981) 301.
%%CITATION = NUPHA,B191,301;%%

\bibitem{shifman}
M.~A.~Shifman, A.~I.~Vainshtein and V.~I.~Zakharov,
%``QCD And Resonance Physics. Sum Rules,''
Nucl.\ Phys.\  {\bf B147} (1979) 385.
%%CITATION = NUPHA,B147,385;%%

\bibitem{nc}
G.~'t Hooft,
%``A Planar Diagram Theory For Strong Interactions,''
Nucl.\ Phys.\  {\bf B72} (1974) 461;
%%CITATION = NUPHA,B72,461;%%
G.~Veneziano,
%``Some Aspects Of A Unified Approach To Gauge, Dual And Gribov Theories,''
Nucl.\ Phys.\  {\bf B117} (1976) 519;
%%CITATION = NUPHA,B117,519;%%
E.~Witten,
%``Baryons In The 1/N Expansion,''
Nucl.\ Phys.\  {\bf B160} (1979) 57.
%%CITATION = NUPHA,B160,57;%%'t Hooft, Veneciano, Witten, Manohar.

\bibitem{eduard}
S.~Peris and E.~de Rafael,
%``On the large N(c) behaviour of the L(7) coupling in chi(PT),''
Phys.\ Lett.\  {\bf B348} (1995) 539
hep-ph/9412343.

\bibitem{bachir}
B.~Moussallam,
%``N(f) dependence of the quark condensate from a chiral sum rule,''
hep-ph/9909292.
%%CITATION = HEP-PH 9909292;%%

\end{thebibliography}
\end{document}